\documentclass{jfm}

\usepackage{graphicx}
\usepackage{newtxtext}
\usepackage{newtxmath}
\usepackage{natbib}
\usepackage{hyperref}
\usepackage{nth}
\usepackage{xcolor}
\usepackage{ar}

\hypersetup{
    colorlinks = true,
    urlcolor   = blue,
    citecolor  = black,
}

\newcommand{\RomanNumeralCaps}[1]
\newcommand{\AoA}[1]{\ensuremath{\alpha={#1}^{\circ}}}
\newcommand{\myAR}[1]{\ensuremath{\AR={#1}}}

\title{High-fidelity study of three-dimensional turbulent transonic buffet on wide-span infinite wings}

\author{David J. Lusher\aff{1}
  \corresp{\email{lusher.david@jaxa.jp}},
  Andrea Sansica\aff{1}, Markus Zauner\aff{1},
 \and Atsushi Hashimoto\aff{1}}

\affiliation{\aff{1}Japan Aerospace Exploration Agency (JAXA), Chofu Aerospace Center, 7-44-1 Jindaiji Higashi-machi, Chofu-shi, Tokyo 182-8522, Japan}

\begin{document}
\maketitle

\begin{abstract}
% All papers should feature a single-paragraph abstract of no more than 250 words which must not spill onto the seond page of the manuscript.
Turbulent transonic buffet is an aerodynamic instability causing periodic oscillations of lift/drag in aerospace applications. Involving complex coupling between inviscid and viscous effects, buffet is characterised by shock-wave oscillations and flow separation/reattachment. Previous studies have identified both 2D chordwise shock-oscillation and 3D buffet/stall-cell modes. While the 2D instability has been studied extensively, investigations of 3D buffet have been limited to only low-fidelity simulations or experiments. Due to computational costs, almost all high-fidelity studies to date have been limited to narrow span-widths around 5\% of aerofoil chord length (aspect ratio, $\AR = 0.05$), which is insufficiently wide to observe large-scale three-dimensionality. In this work, high-fidelity simulations are performed up to $\myAR{3}$, on infinite unswept NASA-CRM wing profiles at $Re=5\times 10^{5}$. At $\AR \geq 1$, intermittent 3D separation bubbles are observed at buffet conditions. While previous RANS/stability-based studies predict simultaneous onset of 2D- and 3D-buffet, a case with buffet that remains essentially-2D despite span-widths up to $\myAR{2}$ is identified here. Strongest three-dimensionality was observed near the low-lift phases of the buffet cycle at maximum flow separation, reverting to essentially-2D behaviour during high-lift phases. Buffet was found to become three-dimensional when extensive mean flow separation was present. At $\AR \geq 2$, multiple 3D separation bubbles form, in a wavelength range of $\lambda=\left[1c-1.5c\right]$. SPOD and cross-correlations were applied to analyse the spatio/temporal structure of 3D buffet-cells. In addition to the 2D chordwise shock-oscillation mode (Strouhal number $St \approx 0.07-0.1$), 3D modal structures were found in the shocked region of the flow at $St \approx 0.002-0.004$.
\end{abstract}

\begin{keywords}
Authors should not enter keywords on the manuscript, as these must be chosen by the author during the online submission process.
\end{keywords}

{\bf MSC Codes }  {\it(Optional)} Please enter your MSC Codes here

\newpage
\section{Introduction}\label{sec:introduction}
\subsection{Overview of transonic buffet and its physical origins}
Transonic shock buffet is an aerodynamic instability commonly found in a wide range of industry-relevant aerospace applications. Buffet is comprised of certain types of shockwave/boundary-layer interactions (SBLI) \citep{D2001}, and is characterized by periodic (albeit, often irregular) self-sustained shock oscillations, and phase-dependent boundary-layer separation and reattachment \citep{L1990,L2001}. Often investigated via a combination of flight tests, wind tunnel experiments, and Computational Fluid Dynamics (CFD) simulations, transonic buffet is a high-speed instability with onset criteria that, for a given aerofoil of chord length $c$ and Reynolds number $Re = \frac{\rho_{\infty} U_{\infty}c}{\mu_{\infty}}$, depends on certain combinations of freestream Mach number $\left(M_{\infty}\right)$ and angle of incidence $\left(\alpha\right)$. Buffet has important physical ramifications for aircraft design and efficiency, motivating the need for a complete understanding of its physical mechanisms. It is for example relevant at the boundaries of the flight envelope of commercial aircraft, namely for high speeds and high Angles of Attack (AoA). Transonic shock buffet can cause large amplitude oscillations in lift and drag, leading to structural vibrations, deteriorated control, and, subsequently, increased fatigue and failure rates. An extensive review of the buffet instability was provided by \citet{Giannelis_buffet_review}.

Buffet is known to consist of both a two-dimensional (2D) chord-wise shock oscillation instability, and three-dimensional (3D) cross-flow outboard propagating cellular separation patterns known as `buffet-cells' \citep{IR2015}. Two-dimensional buffet on turbulent aerofoils occurs in a frequency range of Strouhal numbers around $St = \left[0.06, 0.1\right]$, while the three-dimensional instability for swept wings is found to have broadband energy content at three to ten times higher \citep{plante2020towards} frequencies. Previous RANS/stability-based studies predict simultaneous onset of 2D- and 3D-buffet \citep{PBDSR2019,CGS2019}. Furthermore, it was shown by \citet{PDL2020} that, at least in the context of low-fidelity simulations, the frequency of the 3D modes tend towards lower-frequencies for unswept wings. Whether the same behaviour is found in high-fidelity simulations without the influence of approximate turbulence models remains to be seen. Recent comparisons have been drawn between buffet-cells and the qualitatively similar `stall-cell' \citep{rodriguez2011birth} phenomenon observed at low-speed and high AoA largely-separated flow conditions \citep{plante2020towards,PDL2020}. While transonic buffet typically occurs at angles of attack far below those seen in aerofoil stall applications, the adverse-pressure gradient imposed by the SBLI at transonic conditions can result in similarly large regions of flow separation on wings. While the 2D buffet instability has been studied extensively by numerical simulations in recent years \citep{FK2018,MarkusPRF_2020,SLKHR2022,Nguyen2022,ZMS2022,Moise2023_AIAAJ,LongWong2024_Laminar_buffet}, the picture is less clear for the 3D instability which forms the focus of the present work.

In an attempt to explain the observed 2D shock oscillations on aerofoils at transonic conditions, earlier studies proposed feedback loop models based on upstream and downstream travelling waves from the trailing edge and shock foot regions \citep{L1990}. More recent studies have linked the origins of transonic buffet to a global instability \citep{CGMT2009,SMS2015}. In the case of a global instability, the onset of the unsteadiness is the result of a Hopf bifurcation \citep{CGMT2009}, and the instability is localized to the region around the shock and partially in the separated shear layer \citep{SMS2015}. Despite further developments on the feedback loop model \citep{D2005,JMDMS2009,HFS2013}, the explanation based on global stability remains the most pervasive, but does not fully clarify the mechanisms that lead to self-sustained shock oscillations. 
In this regard, \citet{KAWAI_RESOLVENT2023} recently used a resolvent analysis approach to argue that the shock-induced separation height and pressure dynamics around the shock-wave both contribute to, and maintain, the self-sustained oscillations.

\subsection{Categorization of buffet types and geometrical complexity}
% Type I buffet, Type II buffet, laminar vs turbulent briefly
Transonic buffet is broadly characterised into two types: (Type I) buffet, identified as phase-locked shockwaves propagating on both sides of symmetric aerofoils at zero angle of attack, and (Type II) buffet, characterised by shock oscillations and periodic separation/reattachment on the suction side of aerofoils at non-zero angles of attack \citep{Giannelis_buffet_review}. In this work we limit our discussion to Type II buffet, as it is commonly found on the asymmetric supercritical aerofoils widely-used in practical applications such as commercial airliners. Further categorization can be made based on the state of the boundary-layer upstream of the main SBLI. Based on this definition, transonic buffet can be further separated into laminar buffet \citep{dandois_mary_brion_2018,MarkusPRF_2020,moise_zauner_sandham_2022,LongWong2024_Laminar_buffet}, and turbulent buffet \citep{FK2018,Nguyen2022}, with comparable low-frequency 2D energy content found between the two \citep{Moise2023_AIAAJ}. In this study we limit our analysis to fully-turbulent buffet, as it is the most representative of the higher Reynolds numbers found in practical applications \citep{Giannelis_buffet_review}.

\begin{figure}
\begin{center}
\includegraphics[width=0.8\textwidth]{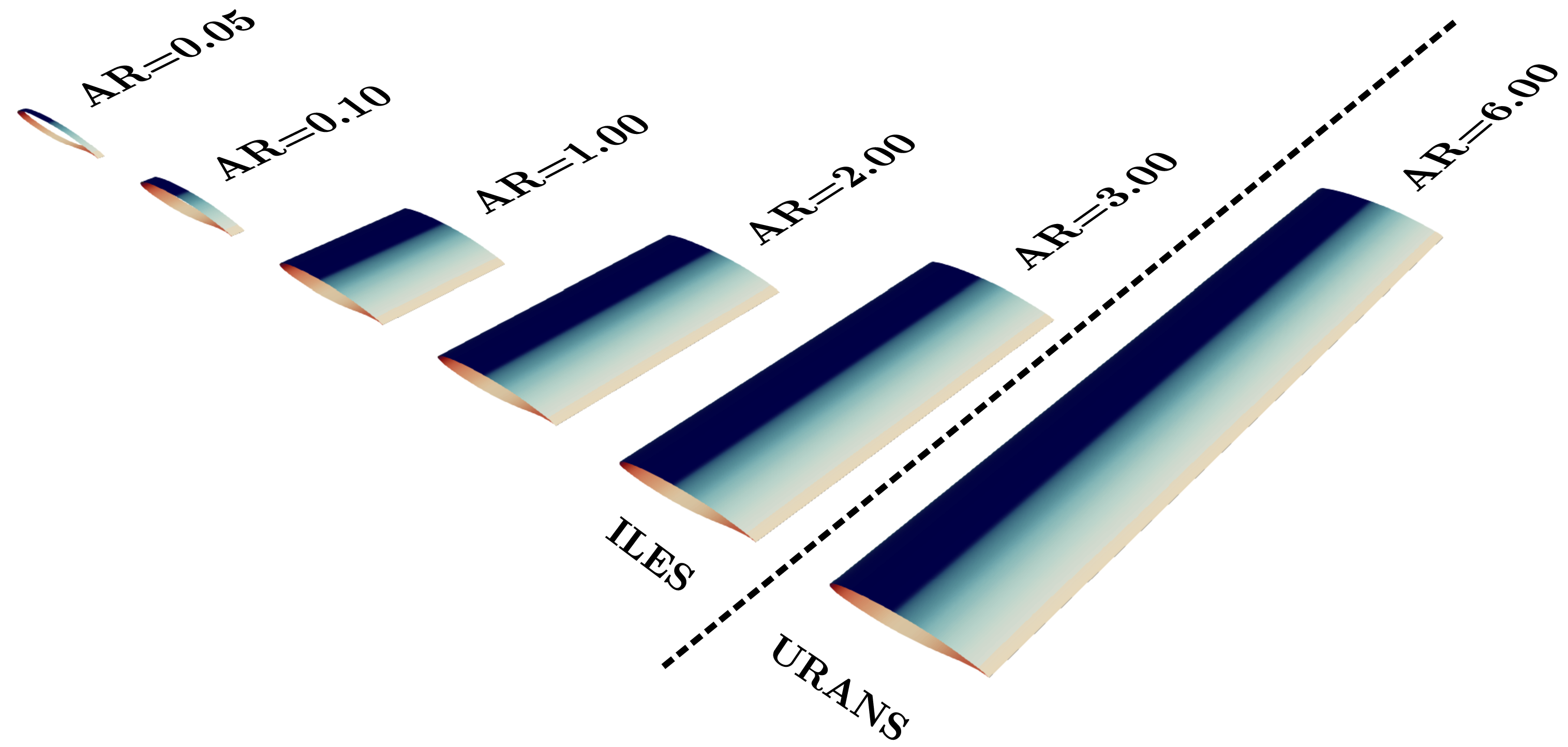}
\caption{Schematic of aspect ratio wings considered in this work. Most high-fidelity (ILES/DNS) computational studies of buffet are limited to narrow $\AR=0.05-0.1$ configurations unable to capture three-dimensional buffet. Up to $\myAR{3}$ is simulated in this work with ILES. Low-fidelity URANS is also used at $\myAR{6}$ to check limiting behaviour.}
\label{fig:intro}
\end{center}
\end{figure}

% 2D vs 3D introduction
Within the scope of aerofoil buffet studies, distinction must also be made between the level of geometrical complexity for the model used in the investigation. The complexity can range from purely 2D aerofoil profiles \citep{CGMT2009,SMS2015,IIHAT2016,PRD2019,SLKHR2022}, to 3D simulations of 2D aerofoils extruded in the third dimension \citep{D2005,GD2010,FK2018,MBP2018,CGS2019,MarkusPRF_2020,Moise2023_AIAAJ} (typically with infinite/periodic boundary conditions applied), finite-wings \citep{IR2015,ST2017,MTP2020,Houtman_Timme_Sharma_2023}, and full aircraft configurations \citep{SH2023,TamakiKawai_2024_CRM} with fuselage, tails, nacelles, and high-lift devices \citep{T2019dpw,dpw-summ1}. Each has potential trade-offs in terms of cost, ability to capture physically meaningful phenomena, and differing levels of relevance to real-world applications. The geometrical complexity also largely determines the level of fidelity of the simulation methods that can be feasibly applied to it. In this study we perform high-fidelity Implicit Large-Eddy Simulations (ILES) of infinite 3D aerofoils, at significantly wider span-widths than previously simulated (Figure~\ref{fig:intro}). High-fidelity in this instance is defined relative to low-fidelity Steady/Unsteady Reynolds-Averaged Navier-Stokes (RANS/URANS) methods used ubiquitously as standard throughout the relevant engineering applications.

A defining property of infinite wings is the Aspect Ratio ($\AR=L_z / c$, Figure~\ref{fig:intro}) selected, for an aerofoil chord length $c$ and spanwise length $L_z$. For high-fidelity aerofoil simulations at both low \citep{AK2023} and high \citep{GD2010} speeds, the size of the flow separation has been shown to be sensitive to the span-width ($L_z$), that therefore needs to be appropriately selected to avoid overly constraining the flow. Due to computational cost, high-fidelity simulations of periodic wings have typically been limited to narrow aspect ratios. Some relevant examples of 2D buffet studies on narrow domains used $\AR=0.0365 - 0.073$ \citep{GD2010}, $\AR=0.065$ \citep{FK2018,Nguyen2022}, $\AR=0.05$ \citep{zauner2019direct,moise_zauner_sandham_2022,Moise2023_AIAAJ}, and $\AR=0.25$ \citep{LongWong2024_Laminar_buffet}. Figure~\ref{fig:intro} provides a visual comparison of these types of narrow domain infinite wings, relative to the $\AR=3$ (ILES) and $\AR = 6$ (URANS) cases presented in this work. In the study of \citet{GD2010} the spanwise width of their simulations was raised from 3.65\% to 7.3\%, which led to a significant reduction of the pressure fluctuations at the trailing edge \citep{Giannelis_buffet_review}. The wider domain `better captures trailing edge pressures by allowing three-dimensional coherent structures to develop' \citep{Giannelis_buffet_review}. Our recent work \citep{LSH2024_narrow_buffet} applied ILES to assess $\AR$ sensitivity for the 2D buffet instability on domains between $0.025 \leq \AR \leq 0.5$. Domain widths at least as wide as the height of the separated boundary-layer near the trailing-edge ($\AR \geq 0.1$ for the cases considered) were required to avoid aspect ratio sensitivity for the 2D shock oscillations. Beyond the span-width sensitivity of the 2D instability, a more severe limitation to be addressed is the inability of narrow-span simulations to capture three-dimensional buffet.

% 3D buffet intro
\subsection{Characteristics of three-dimensional buffet}
Three-dimensional buffet has been investigated both computationally \citep{ST2017,T2020, SH2023}, and with experiments \citep{MTP2020, SKNNNA2018, SNKNNA2021,SKK2022}. \cite{plante2020towards} compiled a comprehensive summary of two- and three-dimensional buffet studies and their main features (Tables 1.3-1.6). One of the earlier studies of 3D buffet effects was performed by \citet{IR2015} via URANS on swept infinite- and finite-wing configurations of the RA16SC1 aerofoil at transonic conditions. For low sweep angles the buffet was found to be largely similar to 2D buffet, dominated by chord-wise shock oscillations. As the sweep angle was increased, 3D cellular separation patterns were observed, which the authors termed `buffet-cells'. Other comparable studies applied Delayed Detached–Eddy Simulation (DDES) methods to half finite-wing body geometries \citep{ST2017}, finding similar three-dimensional buffet features. The work of \citet{hashimoto2018current} applied a Zonal DES method to simulate 3D buffet on the NASA Common Research Model (CRM) aircraft geometry, with good agreement found for the shock position when compared to experimental pressure sensitive paint data. Buffet-cells were also observed, which convected in the spanwise direction. Subsequent work \citep{OIH2018} applied modal decomposition methods to identify a broadband peak associated to the 3D buffet-cell mode in the range $St=\left[0.2, 0.6\right]$. A low-frequency mode corresponding to the main shock-oscillation was also found to be present at $St=0.06$.
% more recent 3D buffet 
\citet{PBDSR2019} applied Global Stability Analysis (GSA) to the OAT15A aerofoil geometry, comparing transonic buffet on configurations ranging from 2D profiles to 3D swept wings. The authors found both a 2D buffet mode consistent with that of \citet{CGMT2009}, and a low-wavenumber 3D one. The 3D mode was found to have zero frequency on unswept configurations, becoming unsteady at non-zero sweep angles. Predictions of the wavenumber and convection velocity of the 3D buffet cells agreed well with numerical and experimental results \citep{SKNNNA2018}.

% buffet and stall cells
More recently, \citet{PDL2020} performed URANS investigations of 3D buffet on infinite swept wings, highlighting similarities between the cellular separation buffet-cells, and those found at low-speed stall. At transonic conditions, analysis of the frequency content showed a superposition of both the 2D buffet mode and a spanwise convecting 3D mode consistent with the 3D buffet-cell phenomenon. The three-dimensionality occurred on the infinite periodic wings without the introduction of any 3D disturbance from the physical setup. The frequency of the buffet-cells was observed to depend on the applied sweep angle. While buffet-cells have typically been reported within the intermediate Strouhal number range of $St=\left[0.2, 0.6\right]$ \citep{Giannelis_buffet_review,OIH2018}, \citet{PDL2020} showed that, in the context of buffet on infinite wings with minor sweep angles ($\Lambda \leq 10^{\circ}$), the 3D mode can appear at frequencies below that of the 2D shock oscillation. At zero-sweep ($\Lambda = 0^{\circ}$) the flow was observed to be unsteady but irregular, with three-dimensional cellular perturbations on the surface streamlines and at the shock front. As the sweep angle was incrementally raised, the buffet-cells became regular, with a spanwise convection speed proportional to the applied tangential freestream speed. Compared to the broadband buffet spectra observed on full aircraft wings \citep{MTP2020}, the infinite-wing configuration had a well-defined convection frequency. Complementary simulations at low-speed stall conditions showed similar cellular separation patterns across the span. For non-zero sweep angles the stall-cells convected in a similar manner to those found during buffet. However, for zero sweep ($\Lambda = 0^{\circ}$), the low-speed flow was observed to be steady, in contrast to the behaviour at buffet conditions. A follow-up study \citep{PDBLS2021} expanded on the previous findings with the aid of global stability analysis. GSA predicted an unstable mode for both transonic buffet and low-speed stall, with a null frequency found at zero sweep. The mode became unsteady for increased sweep angle. While URANS and GSA predicted consistent wavelengths and frequencies in the context of stall-cells, discrepancies were found between the methods for buffet. Furthermore for buffet, the 3D mode was identified at angles of attack below those required for onset of the 2D instability, suggesting the 3D features can occur without 2D buffet being present, arguing that buffet/stall-cells share the same origin.

Similar stability-based studies of 3D buffet include those of \citet{T2020} and \citet{HT2021}. \citet{HT2021} applied tri-global stability analysis to infinite wings at high Reynolds number with aspect ratios ranging from $\AR = 1$ to 10. In addition to the 2D spanwise-uniform oscillatory mode, a group of spatially periodic stationary shock-distortion modes were found for unswept flow with a wavenumber dependent on the aspect ratio of the wing. The modes became travelling waves for non-zero sweep over a broadband range of frequencies. In \cite{T2020}, GSA was applied to the wing-body-tail geometry of the NASA CRM at high-Reynolds-number for turbulent transonic flow. In contrast to previous findings on infinite straight and swept wings, \cite{T2020} did not observe the same essentially two-dimensional long-wavelength mode on this more complex geometry. Instead, a single three-dimensional unstable oscillatory mode was observed, with outboard-propagating shock oscillations. 

Applying resolvent analysis, \citet{Houtman_Timme_Sharma_2023} identified a group of weakly damped modes in addition to the dominant unstable global buffet mode. The additional modes were found to be significantly amplified by external forcing, and were classified as `wing-tip modes', a `wake mode' and a `long-wavelength mode'. Wave-maker analysis showed that buffet is highly sensitive at the shock-foot and separated region immediately downstream of the SBLI. However, no sensitivity was reported around the base-flow shock-wave position itself. Other notable examples of buffet on complex configurations includes \cite{SH2023} and \cite{TamakiKawai_2024_CRM}. In \cite{SH2023}, GSA was performed on a full aircraft configuration at flight Reynolds numbers for the first time. The authors demonstrated the effectiveness of GSA for predicting buffet onset at flight-relevant flow conditions. In addition to a buffet-cell mode localized to the wing outboard region, side-of-body separation effects were also noted. 
Finally, with the aim of increasing the level of simulation fidelity that can be applied, \cite{TamakiKawai_2024_CRM} carried out the first Wall-Modelled Large-Eddy Simulation (WM-LES) of the NASA-CRM geometry at buffet conditions. The wavy shock-wave structure associated with outboard-propagating buffet-cells was observed.

% increased fidelity, comparison to narrow span simulations
The main limitation of the existing literature on 3D buffet is the widespread use of low-fidelity RANS-based methods and the associated difficulty they have in accurately modelling the kinds of unsteady SBLIs and highly-separated flow that characterizes transonic buffet. As shown by \citet{TC2006}, predictions obtained by RANS-based solvers can be sensitive to different turbulence models. \citet{Giannelis_buffet_review} further commented that URANS simulations exhibit high sensitivity to simulation parameters, turbulence model, and both the spatial and temporal discretisation methods used. \citet{MBP2018} compared URANS and DDES methods to scale-resolving Implicit LES (ILES), for the V2C supercritical laminar wing. ILES \citep{grinstein2007implicit} (or, alternatively, under-resolved Direct Numerical Simulation (DNS)) is the approach taken in the current work. ILES does not require additional turbulence modelling, as the governing equations are solved directly, with the numerical dissipation from the shock-capturing scheme and filters acting as a sub-grid scale model \citep{Garnier_ILES_1999, grinstein2007implicit, MBP2018,fu2023review}.

Owing to the extreme computational costs, examples of high-fidelity simulations of 3D buffet are extremely sparse within the available literature. Additionally, they are usually limited to low Reynolds numbers and very narrow domains ($\AR \sim 0.05-0.25$), which are insufficiently wide to observe 3D buffet. A couple of notable exceptions include \cite{MarkusPRF_2020} and \cite{moise_zauner_sandham_2022}, who simulated moderate Reynolds number buffet ($Re = 5 \times 10^5$) up to $\AR=1$. These examples, however, were only performed for fully-untripped laminar buffet, with a limited exploration of the parameter space. No 3D buffet effects were observed. The present contribution extends the literature by performing high-fidelity ILES buffet on aspect ratios up to $\AR \leq 3$ for the first time. A moderate Reynolds number of $Re = 5 \times 10^5$ is selected with numerical boundary-layer tripping applied to obtain the fully-turbulent conditions which are more relevant to the high-Reynolds-number real-world limit.

% Organisation of the paper
\subsection{Structure of the present study}
The present contribution is organised as follows. Section \ref{sec:opensbli} and Section \ref{sec:fastar} give brief overviews of the OpenSBLI \citep{LUSHER201817,OpenSBLI_2021_CPC} and FaSTAR \citep{FaSTAR_Hashimoto_2012,IIHAT2016} CFD solvers used to perform the simulations in this work. Section \ref{sec:mesh} describes the NASA-CRM infinite wing geometry used and associated grid metrics. Section \ref{sec:flow_conditions} completes the problem specification with details of the flow conditions and numerical tripping method. Section~\ref{sec:correlation_setup} and Section~\ref{sec:SPOD_setup} outline the methodology used for cross-correlations and modal Spectral Proper Orthogonal Decompositions (SPOD). For the results in Section \ref{sec:deg5_intro}, baseline high-fidelity (ILES) simulations of transonic buffet on wide-span ($\AR=1, \AR=2$) aerofoils are presented close to buffet onset conditions at an AoA of $\AoA{5}$. The initial results are cross-validated to URANS solutions to demonstrate agreement between the two solvers and simulation methods applied. A final case of $\myAR{6}$ is shown via URANS to observe the limiting behaviour of buffet on very-wide domains. The ILES configuration at aspect ratio of $\myAR{2}$ is then extended to higher angles of incidence of $\AoA{6}$ and $\AoA{7}$ in Section~\ref{sec:deg6_deg7_AR2}, to observe the effect of AoA on three-dimensional buffet. Aspect ratio effects are investigated in Section~\ref{sec:deg6_AR01_AR2_AR3}, contrasting buffet on narrow ($\AR=0.1$) to very wide ($\AR=2,3$) domains at a fixed angle of attack. Section~\ref{sec:sectional} investigates sectional evaluation of aerodynamic quantities at $\myAR{2}$ and $\myAR{3}$ to demonstrate deviation from span-averaged quantities due to three-dimensional buffet. Section~\ref{sec:cross_correlations} applies cross-correlations to further analyse the time-dependence and spanwise coherence of the three-dimensional structures. Finally, Section~\ref{sec:SPOD} performs SPOD-based modal decompositions at $\myAR{2}$ and $\myAR{3}$ to identify dominant modes and their relation to 2D- and 3D-buffet. Further discussion and conclusions are given in Section~\ref{sec:conclusions}.

\section{Computational Method}\label{sec:computational_method}

\subsection{OpenSBLI high-fidelity (ILES/DNS) solver}\label{sec:opensbli}

All high-fidelity simulations in this work were performed in OpenSBLI \citep{OpenSBLI_2021_CPC}, an open-source high-order compressible multi-block flow solver on structured curvilinear meshes. OpenSBLI was developed at the University of Southampton \citep{LUSHER201817,OpenSBLI_2021_CPC} and the Japan Aerospace Exploration Agency (JAXA) \citep{LZSH2023} to perform high-speed aerospace research, with a focus on fluid flows involving Shock-wave/Boundary-Layer Interactions (SBLI) \citep{lusher_sandham_2020,Lusher2020_FTAC}. Written in Python, OpenSBLI utilises symbolic algebra to automatically generate a complete finite-difference CFD solver in the Oxford Parallel Structured (OPS) \citep{Reguly_2014_OPSC,Istvan_OPS_Stencils2017} Domain-Specific Language (DSL). The OPS library is embedded in C/C++ code, enabling massively-parallel execution of the code on a variety of high-performance-computing architectures via source-to-source translation, including GPUs. OpenSBLI was recently cross-validated against six other independently-developed flow solvers using a range of different numerical methodologies by \citet{NATO2024_TGV}.

The base governing equations are the non-dimensional compressible Navier-Stokes equations for an ideal fluid. Applying conservation of mass, momentum, and energy, in the three spatial directions $x_i$ $\left(i=0, 1, 2\right)$, results in a system of five partial differential equations to solve. These equations are defined for a density $\rho$, pressure $p$, temperature $T$, total energy $\rho E$, and velocity components $u_k$ as
\begin{align}\label{ns_eqn}
\frac{\partial \rho}{\partial t} &+ \frac{\partial}{\partial x_k} \left(\rho u_k \right) = 0,\\
\frac{\partial}{\partial t}\left(\rho u_i\right) &+ \frac{\partial}{\partial x_k} \left(\rho u_i u_k + p \delta_{ik} - \tau_{ik}\right) = 0,\\
\frac{\partial}{\partial t}\left(\rho E\right) &+ \frac{\partial}{\partial x_k} \left(\rho u_k \left(E + \frac{p}{\rho}\right) + q_k - u_i \tau_{ik}\right) = 0,
\end{align}
with heat flux $q_k$ and stress tensor $\tau_{ij}$ defined as 
\begin{equation}\label{heat_flux}
q_k = \frac{-\mu}{\left(\gamma - 1\right) M_{\infty}^{2} Pr Re}\frac{\partial T}{\partial x_k},
\end{equation}
\begin{equation}\label{stress_tensor}
\tau_{ik} = \frac{\mu}{Re} \left(\frac{\partial u_i}{\partial x_k} + \frac{\partial u_k}{\partial x_i} - \frac{2}{3}\frac{\partial u_j}{\partial x_j} \delta_{ik}\right).
\end{equation}
$Pr$, $Re$, and $\gamma=1.4$ are the Prandtl number, Reynolds number and ratio of specific heat capacities for an ideal gas, respectively. Support for curvilinear meshes is provided by using body-fitted meshes with a coordinate transformation. The equations are non-dimensionalised by a reference velocity, density and temperature $\left(U^{*}_{\infty}, \rho^{*}_{\infty}, T^{*}_{\infty}\right)$. In this work, the reference conditions are taken as the freestream quantities. For a reference Mach number $M_{\infty}$, the pressure is defined as
\begin{equation}\label{pressure_eqn}
p = \left(\gamma - 1\right) \left(\rho E - \frac{1}{2} \rho u_i u_i\right) = \frac{1}{\gamma M^{2}_{\infty}} \rho T.
\end{equation}
Temperature dependent shear viscosity is evaluated with Sutherland's law such that
\begin{equation}
\mu(T) = T^{\frac{3}{2}}\frac{1 + C_\textrm{Suth}}{T + C_\textrm{Suth}},
\end{equation}
with $C_\textrm{Suth} = T_{S}^{*} / T_{\infty}^{*}$, for $T_{S}^{*} = 110.4K$ and reference temperature of $T_\infty^{*} = 273.15$. Skin-friction is defined for a wall shear stress $\tau_w$ as
\begin{equation}
C_f = \frac{\tau_w}{0.5 \rho_{\infty}^{*} U_{\infty}^{*2}}.
\end{equation}
The lift coefficient is evaluated over the aerofoil surface with arc-length $s_{total}$ as
\begin{equation}
C_L = \frac{1}{0.5\rho_{\infty}^{*} U_{\infty}^{*2}}\int_{s=0}^{s=s_{total}} -S(p_w - p_{\infty}) \left|\cos(\theta)  \right| ds
\end{equation}
where $\theta$ is the inclination angle at the surface, $p_w$ and $p_{\infty}$ are the wall and freestream pressures, and $S$ switches the sign for contributions on the pressure/suction side of the aerofoil. Aerodynamic quantities are averaged in time and the span-wise direction, unless otherwise stated.

OpenSBLI is explicit in both space and time, with a range of different discretisation options available to users. Spatial discretisation is performed in this work by \nth{4} order central differences recast in a cubic split form \citep{Coppola_CubicSplit_2019} to boost numerical stability. Time-advancement is performed by a \nth{4}-order 5-stage low-storage Runge-Kutta scheme \citep{carpenter_kennedy_1994}. Dispersion Relation Preserving (DRP) filters \citep{BOGEY2004194} are applied to the freestream using a targeted filter approach which turns the filter off in well-resolved regions to further reduce numerical dissipation \citep{LZSH2023}. The DRP filters are also only applied once every 25 iterations. Shock-capturing is performed via Weighted Essentially Non-Oscillatory (WENO) schemes, specifically the \nth{5}-order WENO-Z variant by \cite{Borges2008}. The effectiveness and resolution of the underlying shock-capturing schemes in OpenSBLI was assessed for the compressible Taylor-Green vortex case involving shock-waves and transition to turbulence in \citep{lusher2021assessment} and for compressible wall-bounded turbulence in \citep{hamzehloo2021_IJNMF}. The shock-capturing scheme is applied within a characteristic-based filter framework \citep{Yee1999_filter_schemes,Yee2018}. The dissipative part of the WENO-Z reconstruction is applied at the end of the full time-step to capture shocks based on a modified version of the Ducros sensor \citep{DUCROS1999517,Bhagatwala2009}. In addition to the validation and verification cases contained within the code releases \citep{OpenSBLI_2021_CPC}, the numerical methods in OpenSBLI were also recently validated for turbulent channel- and counter-flows in \citet{LC2022_TCF_HS,HLS2023_IJHFF_SPOD}, laminar-transitional buffet cases on the V2C aerofoil geometry in \citet{LZSH2023}, and against URANS and Global Stability Analysis (GSA) on the NASA-CRM aerofoil geometry in \citet{LSH2024_narrow_buffet}.

\subsection{FaSTAR low-fidelity (URANS) solver}\label{sec:fastar}
Comparisons to the high-fidelity ILES data are provided by the FaSTAR unstructured mesh CFD solver \citep{FaSTAR_Hashimoto_2012,IIHAT2016} developed at JAXA. A cell-centered finite volume method is used for the spatial discretisation of the compressible 3D RANS equations. The numerical fluxes are computed by the Harten-Lax-van Leer-Einfeldt-Wada (HLLEW) scheme \cite{O1995} and the weighted Green-Gauss method is used for the gradient computation \cite{M2003}. For the mean flow and transport equations, the spatial accuracy is set to the second and first order, respectively. The turbulence model selected for the present simulations is the Spalart-Allmaras turbulence model \cite{SA1992} without $f_{t2}$ term (SA-noft2) and with rotation/curvature corrections (SA-noft2-RC) \cite{SSTS2000}. No-slip velocity and adiabatic temperature boundary conditions are imposed on the wing walls; far-field boundary conditions are employed at the outer boundaries, and the angle of attack is applied to the incoming flow. 

For the unsteady RANS calculations, dual-time stepping \cite{Vetal2000} is used to improve accuracy of the implicit time integration method. The Lower/Upper Symmetric Gauss-Seidel (LU-SGS) scheme \cite{Setail1998} is used for the pseudo-time sub-iterations and the physical time derivative is approximated by the three-point backward difference. All URANS calculations are advanced in time with a global time integration step of $\Delta t = 0.005$ (corresponding to a dimensional time step $\Delta t^*= 1.48 \times 10^{-3}\: s$) and 40 sub-iterations for the pseudo-time integration of the dual time stepping method.

\subsection{Geometry and mesh configuration}\label{sec:mesh}
\begin{figure}
\begin{center}
\includegraphics[width=0.497\textwidth]{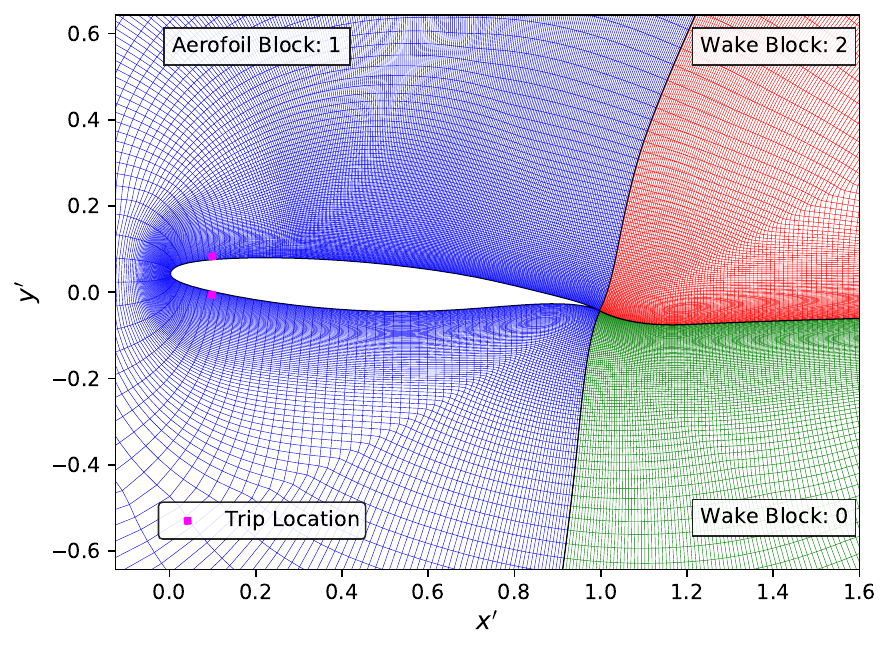}
\includegraphics[width=0.497\textwidth]{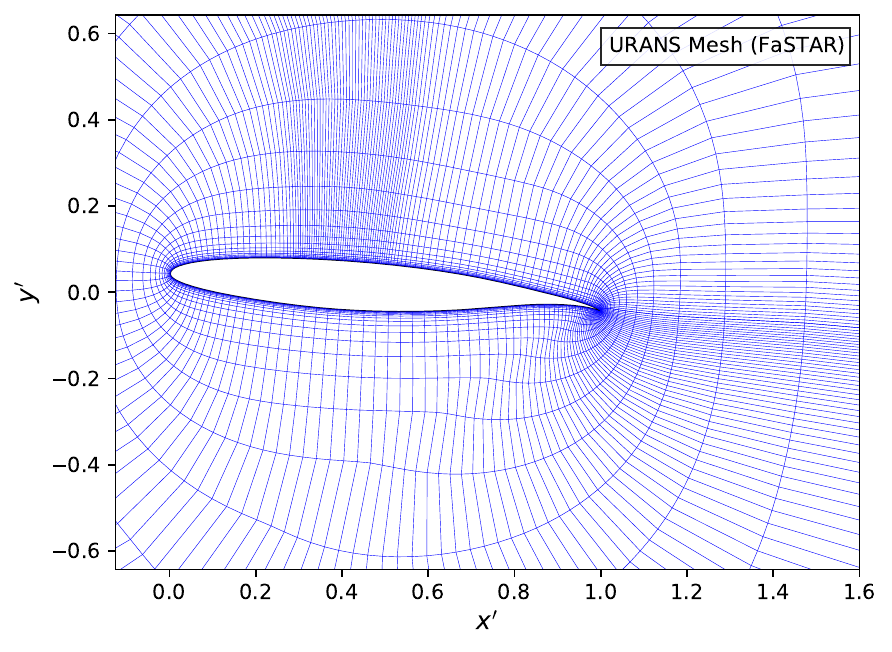}
\caption{Visualisation of the NASA-CRM-65 \citep{CRM} aerofoil mesh configurations used in this study at $\AoA{5}$. The ILES grid is comprised of a C-mesh and two wake blocks, with boundary-layer trips placed at $0.1c$. The URANS grid is comprised of a single block O-mesh. For illustrative purposes, the ILES and URANS grids are plotted at every \nth{7} and \nth{5} grid line respectively.}
\label{fig:mesh_blocks}
\end{center}
\end{figure}

The selected geometry is the 65\% semi-span station of the NASA Common Research Model (CRM) wing, commonly used for turbulent transonic buffet research. Two-dimensional body-fitted structured meshes are created in Pointwise\texttrademark. The three-dimensional mesh is generated by extruding the two-dimensional grid in the span-wise direction with uniform spacing. Figure~\ref{fig:mesh_blocks} shows the meshes used by the (a) ILES simulations in OpenSBLI, and (b) URANS simulations in FaSTAR, plotted at every \nth{7} and \nth{5} line respectively for visualization purposes. The 10\% chord trip location for the ILES is marked on the figure, whereas the URANS-based solutions are considered to be fully-turbulent with no fixed transition location.

In the case of OpenSBLI, an aerofoil C-mesh is connected to two wake blocks with a sharp trailing edge configuration. The in-flow boundary is set at a distance of $25c$ with the outlet $5c$ downstream of the aerofoil. The inflow is set to be uniform $U_{\infty}=1$, with the angle of attack prescribed by rotating the aerofoil within the mesh. For each case, the near-wake mesh is also slightly modified to take into account the deflection of the wake based on the AoA. In the $\xi$ and $\eta$ directions clockwise around the aerofoil and normal to the surface, the aerofoil and wake blocks have $\left(2249, 681\right)$ and $\left(701, 681\right)$ points respectively. Around the aerofoil, the pressure and suction sides have 500 and 1749 points in the $\xi$ direction, respectively. The $\xi$ distribution is refined between $0.3 < x < 0.6$ on the suction side to improve the resolution at the main shock-wave and SBLI. A span-wise grid study at buffet conditions was presented on the same CRM configuration as used here in \cite{LSH2024_narrow_buffet}, with the medium span-wise resolution of $\Delta z = 0.001$ selected for the wide-span cases in this work to make aspect ratios of $\AR \leq 3$ computationally feasible. Upstream of the main shock-wave at $x=0.4$, the grid has wall units of $\left(\Delta x^{+} , \Delta y^{+}, \Delta z^{+} \right) = \left(6.1, 2.2, 14.8\right)$, and in the attached turbulent region downstream of the shock reaches a maximum at $x=0.7$ of $\left(\Delta x^{+} , \Delta y^{+}, \Delta z^{+} \right) = \left(3.9, 1.1, 7.6\right)$. In addition to wall criteria, it is important to maintain good resolution throughout the entire boundary-layer by applying only weak grid stretching. At $x=0.4$ and $x=0.7$ there are 80 and 195 points in the boundary-layer respectively. Additional sensitivity tests to outlet length and $(x-y)$ mesh resolution were given in the appendix of \cite{LSH2024_narrow_buffet}. The results were found to be insensitive to outlets between $5c$ and $20c$ in length. For the $(x-y)$ mesh sensitivity, the buffet characteristics and aerodynamic quantities were found to be consistent with those on meshes three times coarser than those used here. Aerodynamic coefficients, pressure distributions and skin-friction are all time- and span-averaged in this work unless otherwise mentioned.

In the case of the cell-centred finite volume FaSTAR solver, the blunt trailing-edge version of the CRM wing is used. The numerical mesh is obtained by first defining the distribution of cells around the aerofoil and then by normal extrusion to obtain a single block O-grid. The number of cells in the $\xi$ and $\eta$ directions is $\left(1050, 162\right)$. The distribution around the aerofoil consists of $600$ and $400$ cells on suction and pressure sides respectively, and $50$ cells are used to discretize the blunt trailing edge. A region of chord-wise width equal to $0.2 c$ is refined around the shock and counts $200$ cells. To account for different shock locations, this refinement region changes chord-wise position depending on the angle of attack. The domain boundaries extend to about $100$ chords from the aerofoil in all directions. The O-grid is extruded in the spanwise direction to a target aspect ratio ($\AR = 1$ or $6$), that is discretized by using 20 cells per chord.

\subsection{Flow parameters, computational setup and initial conditions}\label{sec:flow_conditions}
\begin{table}
  \begin{center}
\def~{\hphantom{0}}
  \begin{tabular}{cccccccccc}
  % \hline
      Case       & Method  & $\alpha$    & $AR$ & $\overline{C_L}$ & $\overline{C_{Dp}}$ & $\overline{C_{Df}}$ & $\overline{C_D}$ &  $C_{L}^{\rm{RMS}}$ & $St_{\textrm{2D buffet}}$\\ \hline
      AR100-AoA5 & ILES    & $5^{\circ}$ & 1.00 &     0.999       &       0.0508             &         0.0079           &        0.0588        &   0.070 & 0.0775 \\
      AR200-AoA5 & ILES    & $5^{\circ}$ & 2.00 &     0.999       &       0.0507             &         0.0079           &        0.0587        &  0.069 &     0.0775 \\ \hline
      AR010-AoA6 & ILES    & $6^{\circ}$ & 0.10 &     0.990        &       0.0695             &         0.0073           &        0.0768        &   0.079 &    0.0858 \\
      AR200-AoA6 & ILES    & $6^{\circ}$ & 2.00 &     0.996        &       0.0695             &         0.0073           &        0.0767        &  0.073 &     0.0858 \\
      AR300-AoA6 & ILES    & $6^{\circ}$ & 3.00 &     0.993        &       0.0692             &         0.0072           &        0.0764        &   0.052 &    0.0858 \\ \hline
      AR100-AoA7 & ILES    & $7^{\circ}$ & 1.00 &     0.984        &       0.0880             &         0.0066           &        0.0947        &   0.034 &   0.1086 \\
      AR200-AoA7 & ILES    & $7^{\circ}$ & 2.00 &     0.979        &       0.0868             &         0.0066           &        0.0934        &    0.027 &   0.1086 \\\hline

  \end{tabular}
  \caption{Summary of wide-span buffet ILES cases at post buffet onset AoA ($\AoA{5}$), moderate AoA ($\AoA{6}$), and high AoA ($\AoA{7}$) conditions. For each case, Reynolds and Mach numbers are fixed at $Re=5 \times 10^{5}$ and $M_{\infty}=0.72$, with tripping amplitude $A=7.5\times 10^{-1}$ to obtain turbulent conditions. Aspect ratio wings between $\AR=0.1$ and $\myAR{3}$ are considered with ILES. Mean aerodynamic coefficients, RMS of lift oscillations, and two-dimensional buffet Strouhal number are shown for each case.}
  \label{tab:domain_cases}
  \end{center}
\end{table}

All simulations were performed at a moderate Reynolds number of $Re=500,000$ based on aerofoil chord length and freestream Mach number of $M_\infty=0.72$. The non-dimensional time-step is set as $\Delta t = 5\times 10^{-5}$ for all OpenSBLI ILES cases. The ILES simulations are advanced from uniform flow conditions for 20 time units until the boundary-layer is fully turbulent and the buffet unsteadiness fully develops.

In order to investigate turbulent transonic buffet, numerical tripping must be applied to the oncoming boundary-layer to promote a fast transition to turbulence upstream of the shock-wave. This is achieved by forcing a set of unstable modes as a time-varying blowing/suction strip near the leading edge of the aerofoil. This type of forcing is commonly used in CFD research as a method to mimic arrays of tripping dots used in experiments \citep{SKNNNA2018,SKK2022}. The forcing strip is centred around the $0.1c$ location on both the suction and pressure sides of the aerofoil. The forcing is applied to the wall-normal velocity component, which is then used to set the momentum and total energy on the wall. Outside of the forcing strip the wall is a standard isothermal no-slip viscous boundary condition. The forcing is taken to be a modified form of the one given in \cite{Moise2023_AIAAJ} as
\begin{equation}\label{eq:tripping_eqn} 
\rho v_w=\sum_{i=1}^3 A \exp \left(-\frac{\left(x-x_t\right)^2}{2 \sigma^2}\right) \sin \left(\frac{k_i z}{0.05c}\right) \sin \left(\omega_i t+\Phi_i\right),
\end{equation}
for simulation time $t$, trip location $x_t$, and Gaussian scaling factor $\sigma=0.00833$. The three modes $\left(0, 1, 2\right)$ have spatial wavenumbers of $k_i = \left(6\pi, 8\pi, 8\pi\right)$, phases $\Phi_i = \left(0, \pi, -\pi/2\right)$, and temporal frequencies of $\omega_i = \left(26, 88, 200\right)$. The tripping strength is set to $7.5\%$ of the freestream ($A=0.075$), to initiate the transition to turbulence. The sensitivity of the 2D buffet instability to this tripping strength parameter was investigated in our recent previous work \citep{LSH2024_narrow_buffet} over a range of $0.5\%$ to $10\%$ of freestream velocity. It was found that for $5\%$ and above, fully-turbulent interactions were obtained, with identical buffet frequencies observed in the range of $A=\left[5\%, 7.5\%, 10\%\right]$ and only minor variation in mean $C_L$. For weaker tripping, transitional and laminar buffet interactions were observed. In the context of the present work, the $A=7.5\%$ tripping is used throughout to produce fully-turbulent conditions for the investigation of wide-span 3D buffet effects.

\subsection{Cross correlation methodology}\label{sec:correlation_setup}
\begin{figure}
\begin{center}
\includegraphics[width=0.90\textwidth]{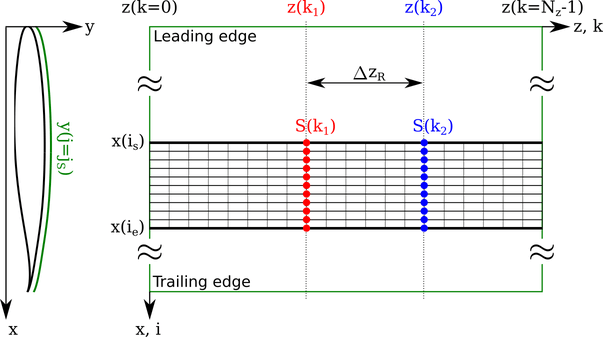}
  \caption{Schematic of cross-correlation approach applied to the aerofoil data in Section~\ref{sec:cross_correlations}.}
\label{fig:schematic_CC}
\end{center}
\end{figure}
To analyse the appearance and frequency of potential three-dimensional buffet structures, a cross correlation approach is implemented. Figure \ref{fig:schematic_CC} shows a schematic of the cross-correlation procedure used in Section~\ref{sec:cross_correlations}. The left-hand-side plot shows the $x/y$ coordinates of the NASA-CRM profile from Figure~\ref{fig:mesh_blocks}. The green curve indicates the position of the $x/z$ plane containing the data of interest, which for the present approach is the streamwise velocity component at the first grid point off the wall ($j_S=1$). At each spanwise location $z(k)$, stencils of flow data are extracted within a range of $x(i_s)<x<x(i_e)$, where each stencil contains $N_s=i_e-i_s$ points. 
At each time instance the span-wise average is subtracted from each stencil $\hat{S}$ to obtain the fluctuations of the quantities according to
\begin{align}
    S(t_n,k)=\hat{S}(t_n,k)- \frac{1}{N_z} \sum_{k=0}^{N_z}\hat{S}(t_n,k).
\end{align}
The streamwise velocity profiles enable us to identify important flow structures, their coherence, and the relation to aerodynamic coefficients such as lift and drag. Stencils containing flow structures with similar characteristics show high levels of correlation. In essentially two-dimensional regions of the flow, the stencils are only correlated within the range of turbulent length scales.
Let us now consider two stencils $S(t_n,k_1)$ and $S(t_n,k_2)$, separated by a distance of $\Delta z_R = z(k_2)-z(k_1)$. We can compute the averaged cross-correlation at a time instant $t_n$ according to:
\begin{align}
    R(t_n, \Delta z_R) = \frac{1}{N_z} \sum_{k_1=0}^{N_z} \frac{<S(t_n,k_1),S(t_n,k_2)>}{<S(t_n,k_1),S(t_n,k_1)><S(t_n,k_2),S(t_n,k_2)>}
\end{align}
where $<\cdot>$ denotes the inner product and $k_2$ corresponds to the index where $z(k_2)=z(k_1)+\Delta z_R$. 
If $z(k_2)$ exceeds the (periodic) domain width $L_z$, we correct it by wrapping around with $z(k_2)=z(k_1)-(L_z-\Delta z_R)$.
It should be noted that due to the normalisation, the cross-correlation does not comment on the amplitude of the corresponding fluctuations. Furthermore, the statistical quantities (e.g. root-mean-square) may be hard to interpret if the appearance of certain structures with high correlation are rare across the span. However, using the present cross-correlation strategy, we obtain an illustrative measure of the prominence of the 3D characteristics as a function of time, which allows us to (a) identify specific time instants of interest for analysis of the 3D data, and (b) correlate the occurrence of 3D buffet phenomena with aerodynamic coefficients associated to the wing.

\subsection{Spectral Proper Orthogonal Decomposition (SPOD) methodology}\label{sec:SPOD_setup}
Modal decomposition methods are widely used analysis techniques that have seen ever-increasing application to fluid flow problems in recent years \citep{ModalDecomp_Taira2017}. These methods extract a variety of representative flow structures (or, modes) that can be used for the identification/extraction of dominant physical mechanisms, or, for the construction of reduced-order models to represent the complex flow field \citep{ModalDecomp_Taira2017}. One popular example of a modal decomposition method is the frequency-resolved Spectral Proper Orthogonal Decomposition (SPOD) \citep{towne_schmidt_colonius_2018}, which decomposes the flow into a series of orthogonal modes ranked by their importance in the frequency domain. The SPOD algorithm has recently been applied to OpenSBLI data in \cite{HLS2023_IJHFF_SPOD}.

In this work, the open-source Python-based SPOD library, PySPOD \citep{mengaldo2021pyspod} has been coupled to OpenSBLI and used for the SPOD analyses presented in Section~\ref{sec:SPOD}. The flow fields extracted from OpenSBLI during unsteady calculations have been formatted and provided as input to the PySPOD library. For each case, side $x$-$y$ plane (at $z=L_z/2$) and $x$-$z$ surface or near surface (at the first point off the wall) data sets are processed independently. While side $x$-$y$ plane data include near and off body regions, the $x$-$z$ surface or near surface data only consider contributions from pressure and suction sides of the aerofoil. For each data set, the SPOD analysis is performed on different flow variables separately. The flow variables selected are pressure/wall-pressure (for both side plan and surface data) and $w$-velocity component (for near surface data only). Initial transients are removed from each data set and for all cases presented here, the flow field sampling period is $\Delta T_{SPOD,sampling} = 1000\Delta t$. The data sets are all divided into three segments with 50\% overlap. To enforce periodicity in each segment, a Hanning function is applied. When analysing the results, the SPOD eigenvalue frequency spectra are plotted only for the first SPOD mode and compared with the PSD of the lift-coefficient fluctuations. The SPOD modes selected for visualization and discussions are chosen based on considerations on the SPOD spectra and their relevance with respect to the lift-coefficient fluctuations PSD. These will be specified in the dedicated sections for each case. For visualization purposes, real and imaginary parts of the SPOD modes have been used to reconstruct the mode temporal evolution. The visualized modes correspond to the time instance within the corresponding period for which a positive maximum value of the mode is reached at $(x,y,z)=(0.5, y_{\rm{wall}}, L_z/2)$.

\section{Investigating wide-span transonic buffet close to 2D onset conditions ($\AoA{5}$)}\label{sec:deg5_intro}

\begin{figure}
\begin{center}
\includegraphics[width=0.8\textwidth]{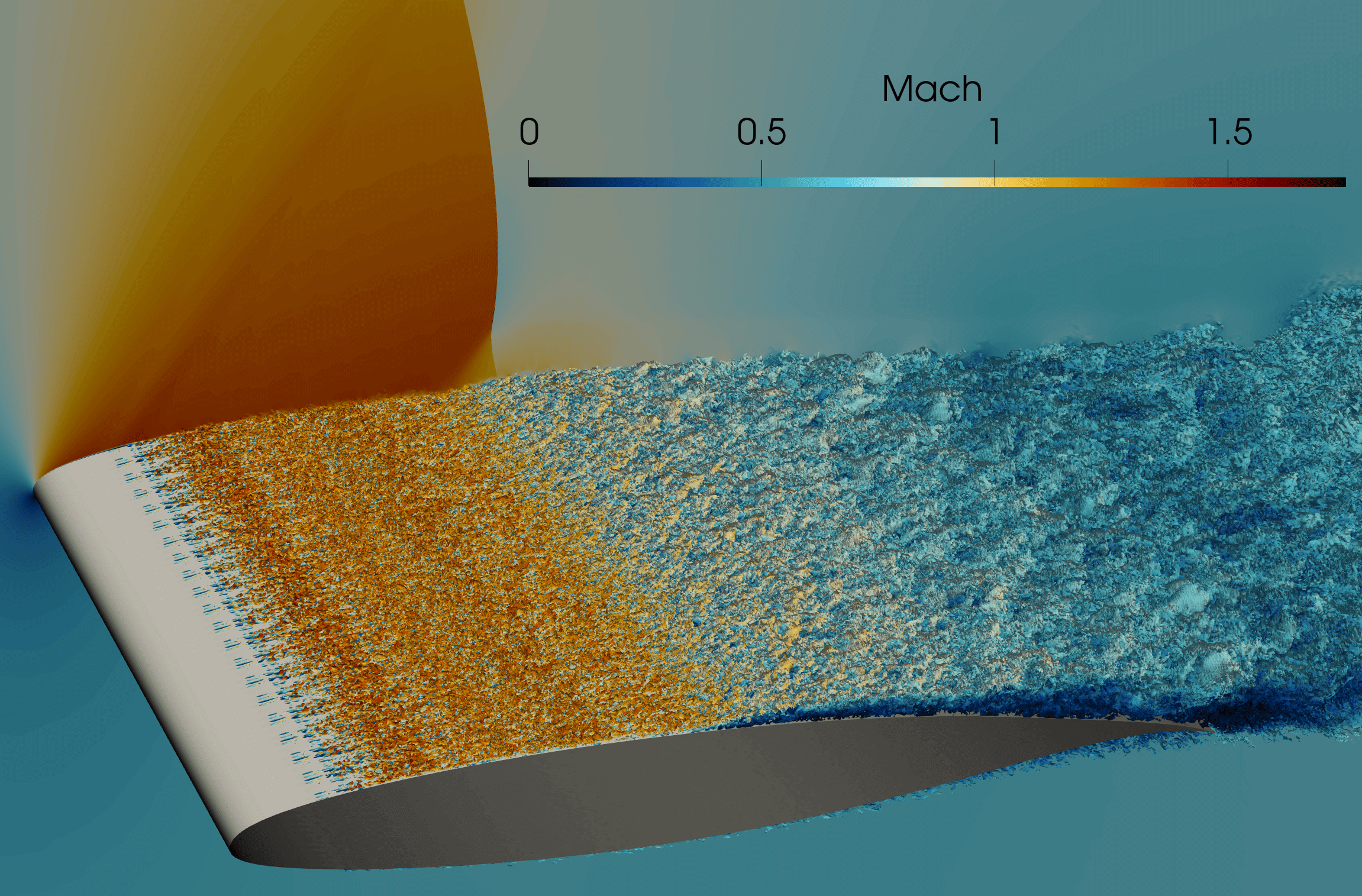}
\caption{Example instantaneous ILES flow field for the NASA-CRM wing geometry at an aspect ratio of $\myAR{1}$, for the baseline case of $\AoA{5}$. Showing spanwise velocity contours coloured by Mach number. Transition to turbulence is observed downstream of the numerical trip location at $10\%$ of chord length ($x=0.1c$). The boundary-layer thickens as a result of the adverse pressure gradient imposed by the main shock-wave.}
\label{fig:3D_picture}
\end{center}
\end{figure}

In our recent work \citep{LSH2024_narrow_buffet}, turbulent transonic buffet was investigated on the same NASA-CRM configuration used here at $Re=5\times 10^5$, albeit for narrow to medium span-widths in the range $\AR=\left[0.025, 0.5\right]$. Domain sensitivity was observed for $AR < 0.1$, for which the flow was shown to be overly-constrained for the narrowest domains. The main sensitivities were observed at the main shock location, and in the pressure fluctuations at the trailing edge which were overestimated compared to the wider domains. It was found that while there were differences in lift amplitudes, pressure distributions and skin-friction, all domain widths still reproduced the same low-frequency buffet oscillation of $St=0.0778$ when the AoA was moderate ($\AoA{5}$) and close to onset ($\alpha = 4.5^{\circ}$). In this section, the previous work is first extended to a baseline wide-span high-fidelity ILES case of $\AoA{5}$ and $\myAR{1}$, to search for three-dimensional buffet effects. Comparison is made between ILES and URANS to cross-validate the two solvers. The baseline ILES width is then doubled to $\myAR{2}$. This aspect ratio is around 40 times wider than commonly used in previous high-fidelity buffet studies (e.g. $\AR=0.0365 - 0.073$ \cite{GD2010}, $\AR=0.05$ \cite{Moise2023_AIAAJ}, and $\AR=0.065$ \cite{FK2018,Nguyen2022}). Comparison is made to URANS at $\myAR{6}$ in Appendix \ref{sec:URANS_AR6}, to check the limiting behaviour of buffet at these conditions on an extremely wide domain.

Figure~\ref{fig:3D_picture} shows an instantaneous snapshot of the NASA-CRM wing profile to be investigated. The plot shows span-wise $w$-velocity contours coloured by Mach number, for the baseline ILES case of $\myAR{1}$ and $\AoA{5}$. A well-captured terminating normal shock-wave is observed in the Mach number contours on the back panel. The numerical trip \eqref{eq:tripping_eqn} at $x=0.1c$ can be seen to cause a rapid transition to turbulence far upstream of the main SBLI. As in many experimental campaigns, the numerical tripping enables us to investigate buffet interactions at turbulent conditions despite the moderate Reynolds numbers used. Small-scale coherent structures introduced at the forcing location break down to turbulence rapidly and become uncorrelated, as will be shown later in Section~\ref{sec:cross_correlations}. Thickening of the boundary-layer is observed due to the adverse pressure gradient imposed by the main shock-wave. The shock-wave position is unsteady and, as we will see, oscillates at low-frequency along the suction side of the aerofoil. Previous computational studies of wide-span buffet have been limited to low-fidelity URANS/DDES methods with the well-known issues associated to these methods. While limited to infinite wing configurations and a moderate Reynolds number, the current contribution is the first set of high-fidelity scale-resolving simulations we are aware of targeting wide-span three-dimensional buffet.

\begin{figure}
\begin{center}
\includegraphics[width=1\textwidth]{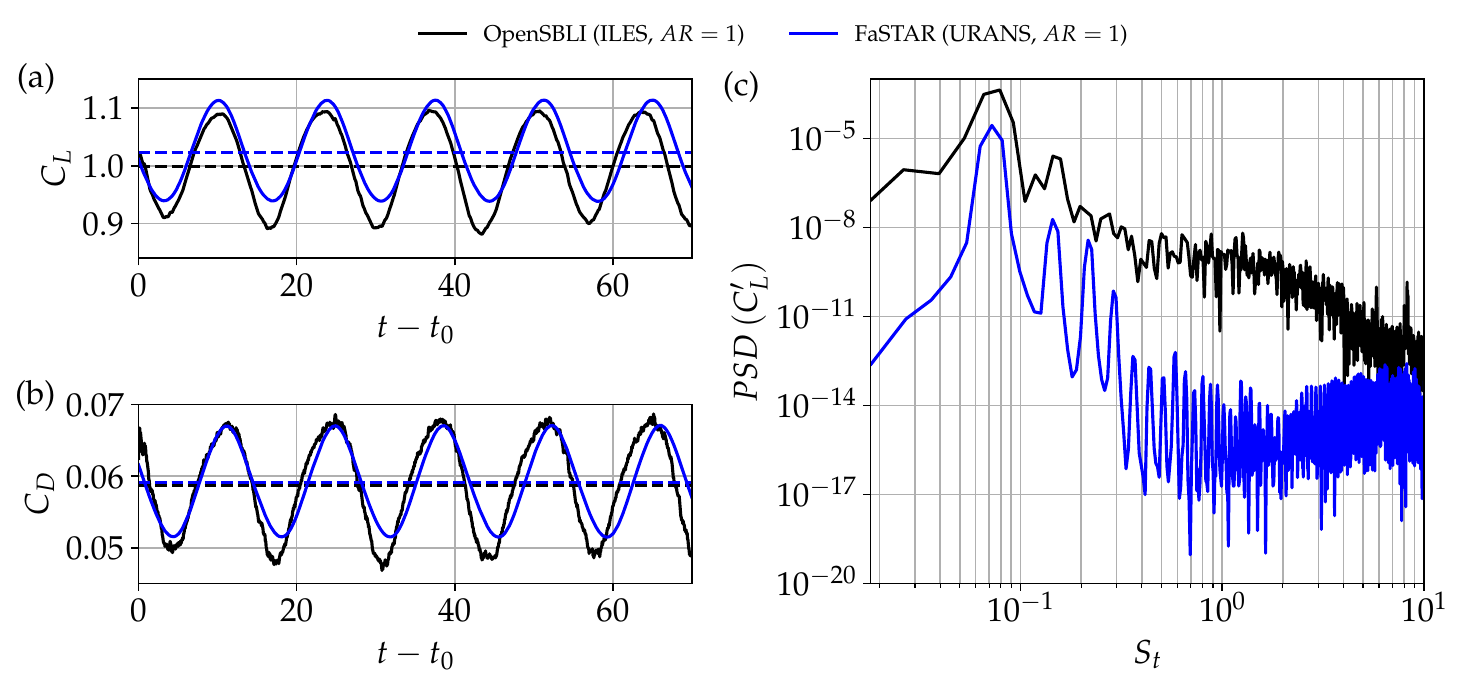}
\caption{Low- to high-fidelity cross-validation of buffet characteristics for simulations using two different solvers and methods. Showing strong agreement for the unsteady (a) lift coefficient, (b) drag coefficient, and (c) PSD of of lift fluctuations between the ILES and URANS solution methods.}
\label{fig:URANS_validation}
\end{center}
\end{figure}

Before proceeding to the main ILES results, it is important to first cross-validate the two solvers and methods. Our previous work \citep{LSH2024_narrow_buffet} performed Global Stability Analysis (GSA) to determine an onset criteria for the two-dimensional buffet shock oscillations of $\alpha = 4.5^{\circ}$. This GSA prediction was found to agree very well with both subsequent ILES and URANS cases, which simulated flow conditions of $\alpha=4^{\circ}$ (no buffet observed) and $\AoA{5}$ (buffet observed). We begin the present work at the same angle of incidence of $\AoA{5}$, where strong buffet is observed close to its onset. Figure~\ref{fig:URANS_validation} shows (a) unsteady lift coefficient, (b) drag coefficient, and (c) Power Spectral Density (PSD) of lift fluctuations between the two methods at $\myAR{1}$. Excellent agreement is found between the two solvers, with both reproducing the low-frequency buffet phenomena despite the differences between the fully-turbulent URANS modelling and tripped transition ILES approaches. Mean lift and drag from ILES and URANS match very well, with only 2.4\% and 0.5\% relative error, respectively.

\begin{figure}
\begin{center}
\includegraphics[width=0.9\textwidth]{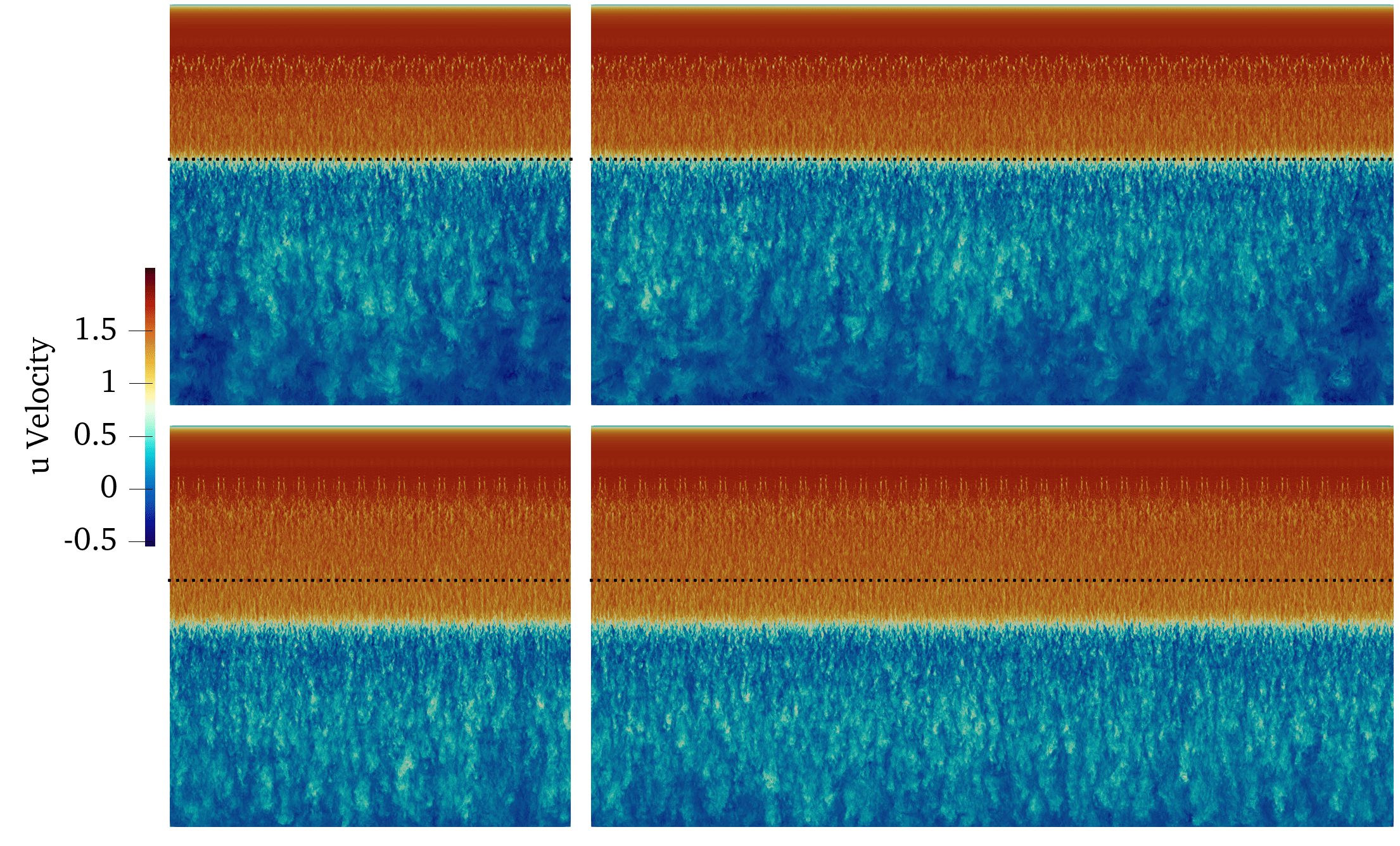}
\caption{Instantaneous streamwise velocity contours within the boundary-layer on the suction side of the aerofoil. Showing the $\AoA{5}$ flow during the (top) upstream low-lift shock-wave position and (bottom) downstream high-lift shock-wave position, at (left) $\myAR{1}$ and (right) $\myAR{2}$. The dashed black line represents the low-lift shock position.}
\label{fig:deg5_contours}
\end{center}
\end{figure}

\begin{figure}
\begin{center}
\includegraphics[width=1\textwidth]{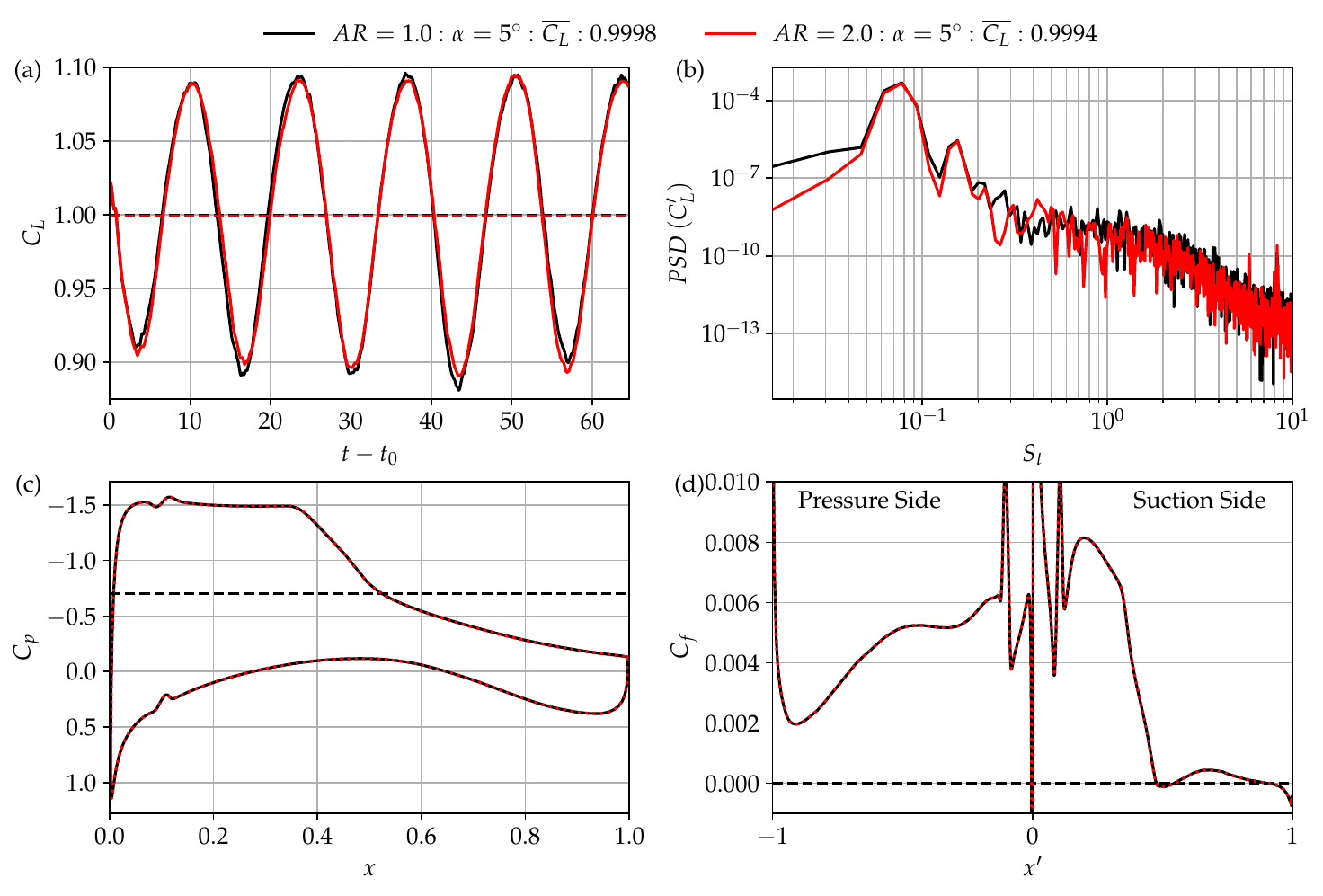}
\caption{ILES cases of wide-span buffet at $\myAR{1}$ and $\myAR{2}$ for a moderate angle of attack of $\AoA{5}$. Showing (a) lift coefficient, (b) PSD of lift fluctuations, (c) time- and span-averaged pressure coefficient, and (d) time- and span-averaged skin-friction distributions.}
\label{fig:AR1_AR2_deg5}
\end{center}
\end{figure}

The baseline AR100-AoA5 case pictured in Figure~\ref{fig:3D_picture} was first monitored at different stages of the buffet cycle and was observed to remain essentially two-dimensional throughout. While the turbulent boundary-layer is certainly three-dimensional, no significant large-scale variations were observed across the spanwise width. As in the narrow-to-medium domain width cases $(\AR = 0.05 - 0.5)$ presented in \citet{LSH2024_narrow_buffet} on the same configuration and angle of incidence ($\AoA{5}$), the spanwise shock-front remains perpendicular to the freestream and no buffet/stall-cells are observed. Although buffet is present, it is limited to only the two-dimensional chord-wise shock oscillations. To assess whether this is simply due to an insufficiently wide span, a second case was performed at $\myAR{2}$. The purpose of this is to investigate whether the lack of three-dimensional effects at this AoA is simply due to an AoA dependence on the wavelength of the span-wise perturbation, with potentially wider aspect ratios required to see its onset at $\AoA{5}$.

Figure~\ref{fig:deg5_contours} shows instantaneous streamwise velocity contours within the boundary-layer on the suction side of the aerofoil. The AR100-AoA5 and AR200-AoA5 cases are shown alongside one another. The lighter colouring of the velocity contours in the bottom panels show the acceleration of the flow to higher speeds as the shock moves farther back on the aerofoil at the point of maximum lift generation (Figure~\ref{fig:AR1_AR2_deg5} (a)). The flow separates at the low-lift phase as the shock-wave moves upstream. The dashed black line indicates the shock position during the low-lift phase for reference between the instantaneous snapshots which are separated by a phase of $t_\textrm{buffet}/2$. The flow is observed to still be essentially two-dimensional, with no span-wise variation of the shock position nor cellular structures present. Despite the wider spans of $\myAR{1}$ and $\myAR{2}$ in this work, the flow at this AoA is still visibly similar to the two-dimensional structure observed at lower aspect ratios ($0.025 \leq AR \leq 0.5$, \citep{LSH2024_narrow_buffet}). The shock-wave traverses only in the streamwise direction at the low-frequency buffet condition, with no discernible three-dimensional effects across the span. The terminating shock position remains perpendicular to the freestream, with the buffet phenomenon remaining essentially 2D at these flow conditions.

Figure~\ref{fig:AR1_AR2_deg5} shows a comparison of aerodynamic coefficients for the cases at $\myAR{1}$ and $\myAR{2}$. The plots show (a) unsteady lift coefficient, (b) PSD of lift fluctuations, (c) time- and span-averaged pressure coefficient, and (d) time- and span-averaged skin-friction. The mean lift varies by no more than 0.04\% with the doubling of the aspect ratio, with almost identical spectra seen between the cases. There are very minor differences between the unsteady lift curves at the extrema of high- and low-lift. These cycle-to-cycle variations are far smaller than commonly observed between buffet periods in other high-fidelity studies \citep{MarkusPRF_2020,Moise2023_AIAAJ,LongWong2024_Laminar_buffet}. Furthermore, essentially perfect agreement is observed for the span-averaged pressure and skin-friction distributions in Figure~\ref{fig:AR1_AR2_deg5} (c) and (d) despite the wider aspect ratio. The first two ILES cases at $\AoA{5}$ have shown that despite simulating aspect ratios in excess of the wavelength previously seen for buffet/stall-cell phenomena ($\lambda = 1-1.5$, \citep{Giannelis_buffet_review,PBDSR2019,plante2020towards,PDL2020}), we are able to isolate a wide-span transonic aerofoil case that possesses clear two-dimensional chord-wise buffet shock- and lift-oscillations, while showing no evidence of span-width sensitivity nor buffet/stall-cells. To further check the two-dimensionality of the $\AoA{5}$ solution, the $\myAR{1}$ URANS (Figure \ref{fig:URANS_validation}) is extended in Appendix \ref{sec:URANS_AR6} to assess whether the $\myAR{2}$ ILES domain is simply still too narrow to accommodate 3D buffet effects. However, the $\AoA{5}$ condition still remains essentially-2D up to $\myAR{6}$. Therefore, having identified several wide-span cases at $\AoA{5}$ that possess 2D-buffet but show no 3D effects, the next section increases the angle of incidence with ILES to $\AoA{6}$ and $\AoA{7}$ for a fixed aspect ratio of $\myAR{2}$.
 
\section{Sensitivity to increased angle of attack for wide-span transonic buffet at $\myAR{2}$}\label{sec:deg6_deg7_AR2}

\begin{figure}
\begin{center}
\includegraphics[width=1\textwidth]{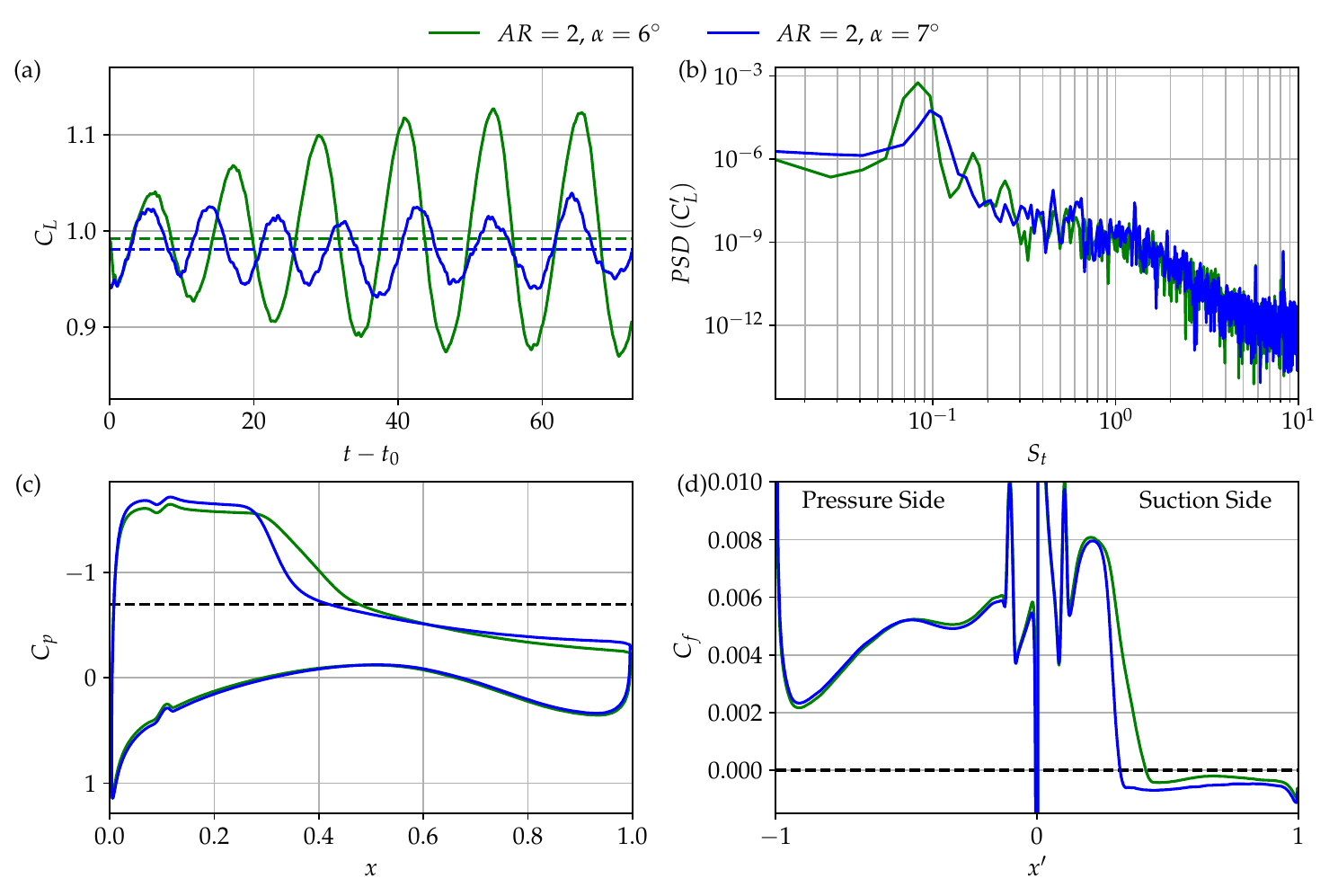}
\caption{ILES cases of wide-span buffet at $\myAR{2}$ for angles of attack of $\AoA{6}$ and $\AoA{7}$. Showing (a) lift coefficient, (b) PSD of lift fluctuations, (c) time- and span-averaged pressure coefficient, and (d) time- and span-averaged skin-friction distributions.}
\label{fig:AR2_deg6_deg7_LINES}
\end{center}
\end{figure}

\begin{figure}
\begin{center}
\includegraphics[width=0.7\textwidth]{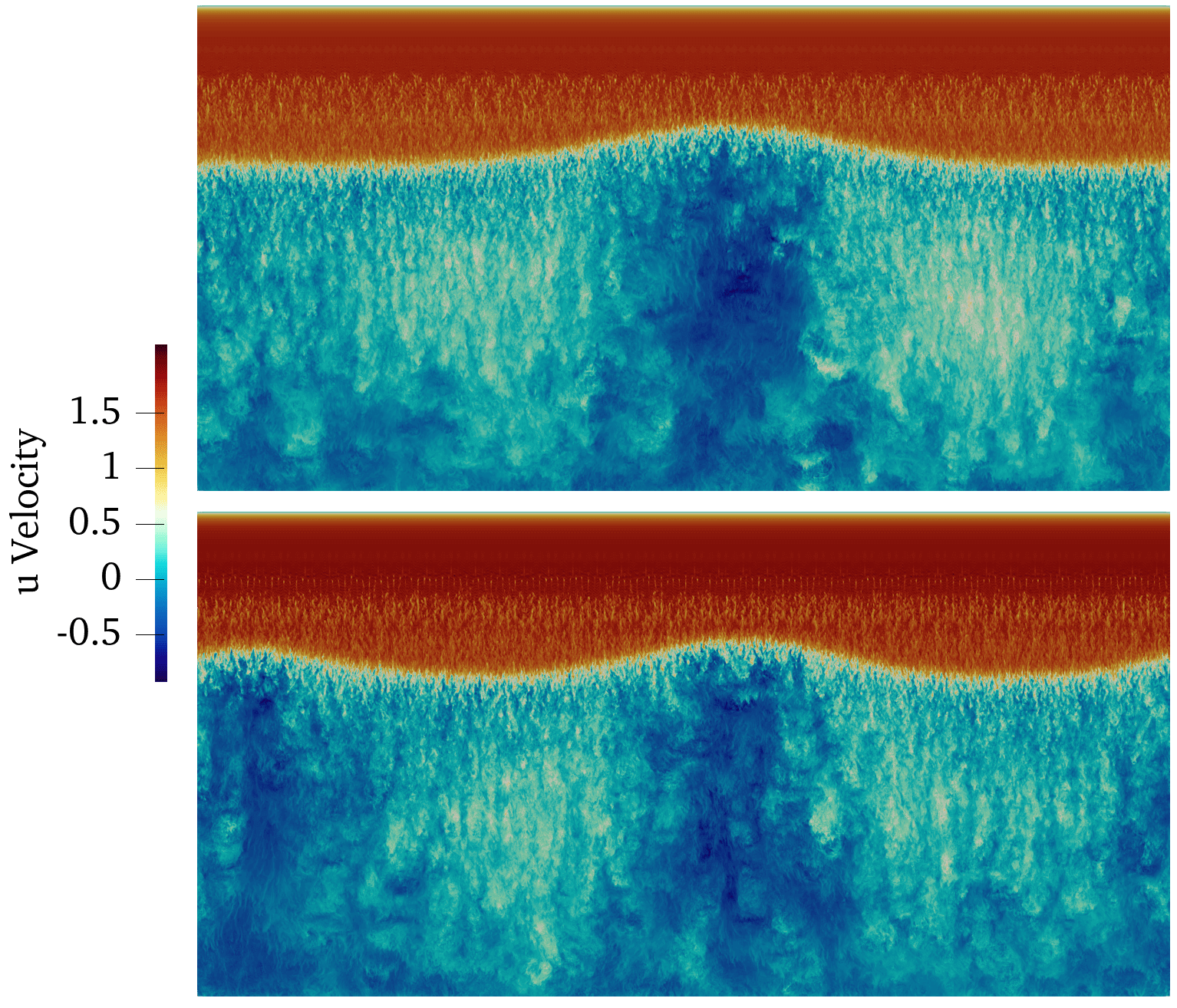}
\caption{Instantaneous streamwise velocity contours within the boundary-layer on the suction side of the aerofoil. Showing 3D buffet effects for the $\myAR{2}$ flow for angles of incidence of (top) $\AoA{6}$ and (bottom) $\AoA{7}$. Similar time instances are selected when the 3D effects are comparable between the two simulations. The dark blue colouring shows cellular structures as regions of strong flow recirculation.}
\label{fig:deg6_deg7_contours}
\end{center}
\end{figure}

In this section the effect of increased angle of attack is investigated at $\myAR{2}$ with ILES. An additional case of $\AoA{7}$ and $\myAR{1}$ is also shown in Appendix~\ref{sec:appendix:deg7_AR1} for completeness, whereas this section focuses on $\myAR{2}$. Figure~\ref{fig:AR2_deg6_deg7_LINES} shows the aerodynamic coefficients for cases AR200-AoA6 and AR200-AoA7. The first feature to note is that, in contrast to the regular periodic oscillations of the $\AoA{5}$ case in Figure~\ref{fig:AR1_AR2_deg5} (a), the higher AoAs begin to show irregularity in the buffet amplitudes and phase from period to period. Both cases were initialized by extruding fully-developed narrow-span solutions across the span at their respective angles of attack. We note that, while the buffet oscillations have the same period at $\AoA{6}$ (green line), there is a noticeable initial transient in the buffet amplitudes at this AoA which saturates after a few cycles. The same initialization process of fully-developed narrow-span solutions being extruded to wide-span was also used at $\AoA{5}$, however, the lower AoA did not show the same transient behaviour. Similarly, the $\AoA{7}$ case shows a similar peak-to-peak amplitude at each cycle without a long transient. The $\AoA{7}$ case, however, shows decreased regularity from period-to-period than the lower AoAs. The PSD of lift fluctuations in Figure~\ref{fig:AR2_deg6_deg7_LINES} (b), with tabulated values in Table~\ref{tab:domain_cases}, show an increase in the buffet frequency as the AoA is increased ($\AoA{5}$ : $St=0.0775$, $\AoA{6}$ : $St=0.0858$, and $\AoA{7}$ : $St=0.1086$). 

The mean pressure distributions in Figure~\ref{fig:AR2_deg6_deg7_LINES} (c) show that the higher AoA has a mean shock position farther upstream as expected, but both cases still consist of a smeared out pressure gradient as a result of the streamwise shock oscillations. Turning to the skin-friction in Figure~\ref{fig:AR2_deg6_deg7_LINES} (d), both cases now consist of large regions of time-averaged flow separation ($C_f < 0$) downstream of the shock position. This is in contrast to the moderate AoA case ($\AoA{5}$, Figure~\ref{fig:AR1_AR2_deg5} (d)), which had only small regions of time-averaged flow separation at the shock location and near the trailing edge. The flow at the moderate AoA was otherwise attached in a time-averaged sense. The pressure side of the aerofoil is observed to be less sensitive to the change in AoA, with a small shift in $C_p$ and $C_f$ visible. The decreased period-to-period regularity in the buffet oscillations in Figure~\ref{fig:AR1_AR2_deg5} (a) motivate us to inspect the spanwise flow fields for potential three-dimensional effects.

Figure~\ref{fig:deg6_deg7_contours} shows instantaneous streamwise velocity contours within the boundary-layer on the suction side of the aerofoil for cases AR200-AoA6 and AR200-AoA7. At these higher angles of incidence, the flow now exhibits large-scale three-dimensionality across the span. Compared to the flow fields for the more moderate AoA ($\AoA{5}$, Figure~\ref{fig:deg5_contours}), which had shock fronts aligned entirely parallel to the spanwise width, at $\alpha=6^{\circ},7^{\circ}$ we begin to see spanwise perturbations similar in structure to buffet/stall-cells \citep{IR2015,Giannelis_buffet_review,PDBLS2021}. The three-dimensional cellular features are observed as dark blue regions where the flow recirculation is at its strongest. After a short transient when initializing the wide-span simulation with the fully-developed narrow solution, the three-dimensionality develops naturally within the flow without any additional forcing of long wavelengths. We note that the wavelengths forced in the boundary-layer tripping from equation \eqref{eq:tripping_eqn} which are visualised in Figure~\ref{fig:3D_picture}, are over two orders of magnitude smaller than the observed three-dimensional cellular buffet effects. The scale separation between the boundary-layer tripping can also clearly be seen later in the spanwise velocity contours shown in Figure~\ref{fig:deg6_surface_w_contours}. The approximate location of the shock-wave can be identified in the white terminating region separating the supersonic (red) and subsonic (blue) regions of the flow. The cellular structures lead to curvature of the shock-wave orientation, which is no longer normal to the freestream in the streamwise direction. 

The snapshots shown in Figure~\ref{fig:deg6_deg7_contours} were selected for comparison between different angles of attack when the cellular structures were observed to be in a similar location across the span. We note that these three-dimensional buffet effects are present persistently over numerous low-frequency cycles (Figure~\ref{fig:AR2_deg6_deg7_LINES} (a)), and, as we will see, were observed to be strongest during the switch from high- to low-lift phases of the buffet cycle as the shock-wave propagates upstream. However, the spanwise arrangement of the cellular patterns are found to be intermittent in the nature and location of their appearance. Due to the zero sweep angle imposed in this unswept study, there is no preferred convection direction for these effects unlike those observed for cases with non-zero sweep \citep{IR2015,PDL2020,PDBLS2021}. For cases with non-zero sweep, the cells propagate at a set convection velocity based on the sweep angle and spanwise velocity component. At the time instance shown in Figure~\ref{fig:deg6_deg7_contours}, the main qualitative difference observed in our case is the reduction of wavelength as the AoA is increased. The higher AoA of $\AoA{7}$ exhibits two separation cells compared to the single cell visible at $\AoA{6}$.

\begin{figure}
\begin{center}
\includegraphics[width=1\textwidth]{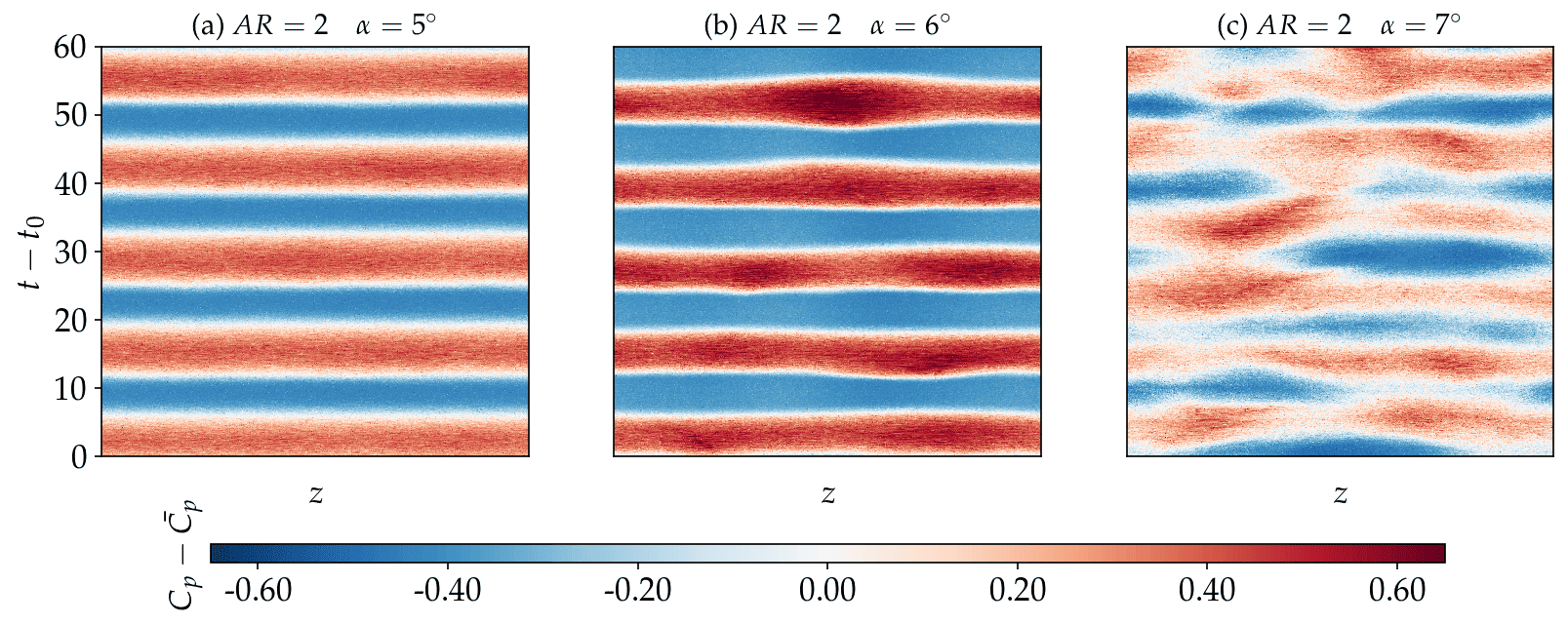}
\caption{Span-wise variation of pressure coefficient in time ($z-t$) at $\myAR{2}$, evaluated at the mean chord-wise shock location of (a) $x=0.45$ ($\AoA{5}$), (a) $x=0.375$ ($\AoA{6}$), and (b) $x=0.325$ ($\AoA{7}$) on the suction side of the aerofoil. }
\label{fig:ZT_deg5_deg6_deg7_AR2}
\end{center}
\end{figure}

To further demonstrate the three-dimensional effects, Figure~\ref{fig:ZT_deg5_deg6_deg7_AR2} shows the span-wise variation of $C_p - \overline{C_p}$ at a single chord-wise location on the suction side of the aerofoil, as it evolves in time ($z - t$ plot, for the span-wise coordinate $z$). The quantity is evaluated at the time-averaged mean shock positions of $x=0.45$, $0.375$, and $0.325$, for the $\myAR{2}$ cases at $\AoA{5}$, $\AoA{6}$, and $\AoA{7}$, respectively. Different chord-wise positions of the probe line were tested, however, centring the probe location at the mid-point of the shock oscillations at each AoA was deemed the fairest comparison. For a purely two-dimensional interaction, the pressure distribution should vary uniformly across the span-wise coordinate as the shock oscillates about its mean position in a streamwise-manner over the probe line. This is exactly what is observed for the moderate AoA case of $\AoA{5}$. Over multiple low-frequency buffet periods (corresponding to the lift history in Figure~\ref{fig:AR1_AR2_deg5} (a)), no span-wise variation is observed. The pressure oscillates symmetrically about the mean $\left(C_p - \overline{C_p} = 0\right)$ in the repeating red and blue bands, with a constant band thickness across the span. As the AoA is increased, the two-dimensionality of the $z-t$ signals begins to breakdown. At $\AoA{6}$, while similar low-frequency alternating red and blue bands show the same two-dimensional buffet shock oscillations exist about the measurement point as in the lower AoA, at this higher AoA they are no longer in phase across the span. At any given time instance, the non-constant band thickness across the span demonstrates the modulation of the shock front observed in the instantaneous flow visualisations in Figure~\ref{fig:deg6_deg7_contours}. At $\AoA{7}$ the three-dimensionality becomes more severe, and, while the streamwise shock oscillations are still present, the bands now intersect as the spanwise shock position shifts in time relative to the probe location.

We have identified configurations where the two dimensional (shock-oscillation) can occur either in isolation ($\AoA{5}$), or with three-dimensional separation effects superimposed ($\alpha = 6^{\circ}, 7^{\circ}$). These results highlight that, in the case of periodic wings for the flow conditions tested here, turbulent shock buffet can exist both in an essentially quasi-2D manner at moderate angles of attack with minimal flow separation, and as three-dimensional buffet with span-wise modulations when the angle of attack is raised. Even when using domain widths wide enough to capture the long wavelength structures typically attributed to being representative of buffet cells ($\lambda = 1-1.5$ \citep{plante2020towards}), we are able to isolate quasi-2D streamwise shock oscillations without any three-dimensional effects (Section \ref{sec:deg5_intro}). When the angle of attack is raised by $1-2^{\circ}$ from this initial 2D buffet state, onset of the three-dimensionality of the buffet phenomena is observed.

\section{Aspect ratio effects on buffet at $\AR=0.1$, $\myAR{2}$ and $\myAR{3}$ for $\AoA{6}$}\label{sec:deg6_AR01_AR2_AR3}
\begin{figure}
\begin{center}
\includegraphics[width=1\textwidth]{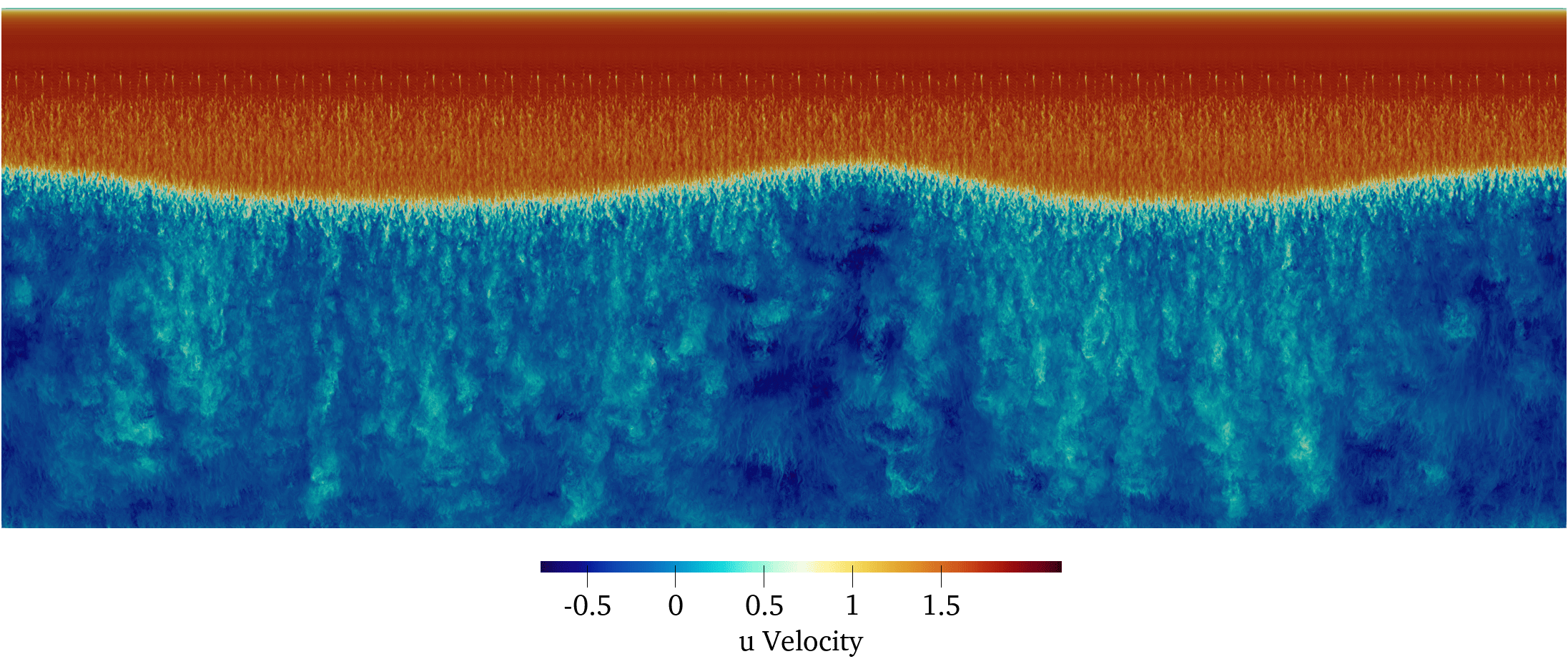}
\caption{Instantaneous streamwise velocity contours within the boundary-layer on the suction side of the aerofoil. Showing 3D buffet effects for the $\myAR{3}$ flow for an angle of incidence of $\AoA{6}$. The dark blue regions show strong flow recirculation.}
\label{fig:deg6_AR3_contours}
\end{center}
\end{figure}

Having identified three-dimensional buffet effects at $\alpha = 6^{\circ},7^{\circ}$ but not at $\AoA{5}$, this section investigates the effect of increasing aspect ratio further to $\myAR{3}$ with ILES, at a fixed AoA of $\AoA{6}$. This section also contrasts buffet behaviour between configurations of both very narrow ($\AR=0.1$) and very wide ($\AR=2, AR=3$) aerofoils. As previously mentioned, trends between buffet on narrow to intermediate domain widths ($\AR = \left[0.05, 0.5\right]$) were reported in \cite{LSH2024_narrow_buffet}. The narrow AR010-AoA6 case in this section is selected simply as a reference of two-dimensional buffet for comparison to the very wide ($\AR=2, AR=3$) domain cases.

Figure~\ref{fig:deg6_AR3_contours} plots instantaneous streamwise velocity contours within the boundary-layer on the suction side of the aerofoil at $\AoA{6}$ and $\myAR{3}$, at a time instance where three-dimensional effects are visible. Compared to the previously shown $\myAR{2}$ in Figure~\ref{fig:deg6_deg7_contours}, the wider domain at the same angle of incidence exhibits two clear peaks in the shock front instead of one. Figure~\ref{fig:deg6_deg7_contours} at $\alpha=6^{\circ}, 7^{\circ}$ suggests that the wavelength of the buffet/stall-cells decreases with increasing AoA, and the $\myAR{2}$ domain is overly narrow to support two buffet/stall-cells at $\AoA{6}$, but not at $\alpha=7^{\circ}$. Widening the domain at $\AoA{6}$ from $\myAR{2}$ to $\myAR{3}$ allows two stall/buffet cells to develop.

\begin{figure}
\begin{center}
\includegraphics[width=1\textwidth]{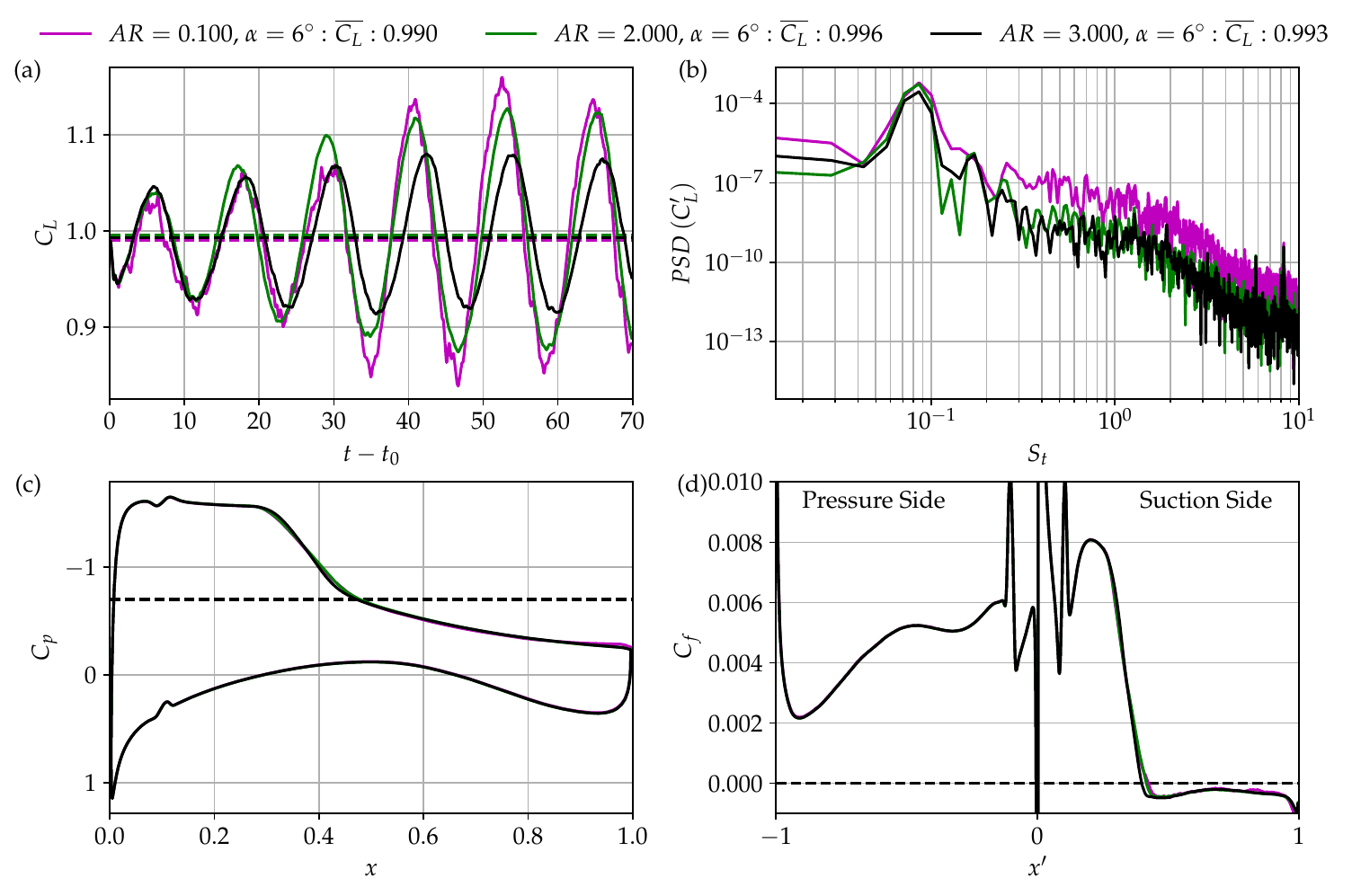}
\caption{ILES cases of narrow- and wide-span buffet at $\AR=0.1, 2, 3$ for an angle of attack of $\AoA{6}$. Showing (a) lift coefficient history (b) PSD of fluctuating lift component (c) time-averaged pressure coefficient and (d) time-averaged skin-friction.}
\label{fig:deg6_lines_ALL}
\end{center}
\end{figure}

Figure~\ref{fig:deg6_lines_ALL} shows the span-averaged aerodynamic coefficients at $\AoA{6}$ for $\AR=0.1, 2, 3$. We note in the lift coefficient that a similar initial transient in peak-to-peak amplitude observed at $\myAR{2}$ (shown previously in Figure \ref{fig:AR2_deg6_deg7_LINES} (a)) is also present at $\AR=0.1$. At $\myAR{3}$ it is also there but is much weaker and saturates early on. Despite the presence of strong three-dimensional effects at wider aspect ratio (Figure~\ref{fig:deg6_deg7_contours}, Figure~\ref{fig:deg6_AR3_contours}), the mean lift shows minimal variation between aspect ratios, differing only to the third decimal point. Each case has a clear low-frequency oscillation which becomes more regular with increasing aspect ratio. In the PSD of lift fluctuations in Figure~\ref{fig:deg6_lines_ALL} (b), the narrow domain predicts higher amplitude mid-to-high-frequency energy content, which reduces with increasing aspect ratio. These additional frequencies are in the Strouhal number range of $St \approx 0.5$ and above, which are commonly associated to vortical wake modes \citep{Moise2023_AIAAJ,LongWong2024_Laminar_buffet}. This suggests that the narrower domains have a stronger wake component compared to the wide-span cases. All cases predict the same low-frequency two-dimensional buffet peak of $St = 0.0858$.

Figures~\ref{fig:deg6_lines_ALL} (c,d) show the pressure coefficient and skin-friction distributions for the $\AR=0.1, 2, 3$ cases at $\AoA{6}$. As for the other angles of attack considered, the span-averaged profiles do not show a large sensitivity to aspect ratio. This is especially true for the leading edge, transition region, and pressure side of the aerofoil which are entirely insensitive to aspect ratio effects in the span-averaged sense. The narrowest case shows a slight deviation from the wide-span cases near the trailing edge ($x > 0.9$), which was reported in \cite{LSH2024_narrow_buffet} as one of the markers for an overly-narrow domain relative to the size of the separated boundary-layer and subsequent over-prediction of the wake component. Both the $\myAR{2}$ and $\myAR{3}$ cases converge in this region near the trailing edge, as, along with the essentially two-dimensional cases presented at $\myAR{1}$ and $\myAR{2}$ in Figure \ref{fig:AR1_AR2_deg5} (c), the wide aspect ratios considered in this work are far wider than the thickness of any separated boundary-layers encountered. The other region where sensitivity to aspect ratio is observed in Figures~\ref{fig:deg6_lines_ALL} (c,d) is at the main shock position, due to three-dimensional effects. The deviation in the line plots between aspect ratios is very minor due to the time- and span-averaging applied, but this can be viewed as evidence that a sufficiently long time signal was averaged over. Although there are three-dimensional buffet/stall-cells present, due to the zero sweep angle there is no preferential span-wise location for their occurrence nor direction for their convection, and, consequently, an extremely long time integration would provide results consistent with two-dimensional/narrow predictions. To obtain a clearer picture of the three-dimensionality, sectional evaluation of aerodynamic coefficients at discrete spanwise probe locations is shown in Section~\ref{sec:sectional}.

\begin{figure}
\begin{center}
\includegraphics[width=1\textwidth]{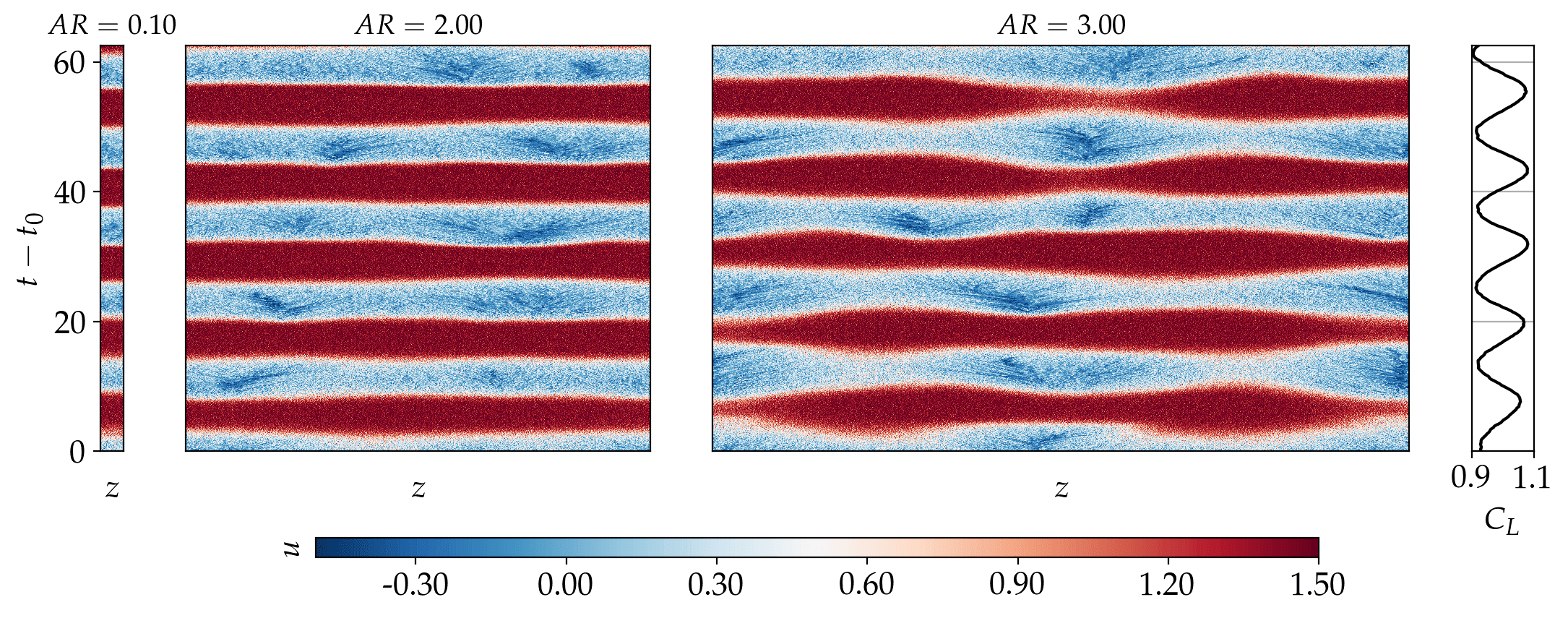}
\caption{$z-t$ diagrams showing the time evolution of $u$ velocity within the boundary-layer at the mean shock position of $x=0.375$. Sensitivity to aspect ratio at $\AoA{6}$ is observed for $\AR=0.1, 2, 3$. The $\myAR{3}$ lift coefficient history is shown on the right side to reference low/high-lift phases of the buffet cycle.}
\label{fig:deg6_ZT}
\end{center}
\end{figure}

Figure~\ref{fig:deg6_ZT} shows $z-t$ diagrams of streamwise velocity within the boundary-layer on the suction side of the aerofoil. The monitor line is again taken as the mean shock-wave position at this AoA of $x=0.375$. The lift-coefficient for $\myAR{3}$ is plotted on the side to relate the $z-t$ oscillations to the varying lift during the buffet cycle. In a similar fashion to the aerodynamic coefficients shown in Figure~\ref{fig:deg6_lines_ALL} (a), the three $z-t$ signals are in phase with one another despite the increase from $\myAR{0.1}$ to $\myAR{3}$. The flow velocity decreases periodically for each of the low-lift phases, as the shock-wave moves upstream and passes over the time-averaged mean shock location. At both $\myAR{2,3}$, dark blue patches of strong flow recirculation are visible during the low-lift phases. They are persistently appearing at each cycle, albeit with a shifted span-wise location. Similar to the $z-t$ progression shown with increasing AoA in Figure~\ref{fig:ZT_deg5_deg6_deg7_AR2}, the two-dimensionality of the alternating colour bands also breaks down with increasing aspect ratio. While the two-dimensional chord-wise shock oscillations still dominate at $\myAR{2}$, the onset of three-dimensionality is already apparent. At $\myAR{3}$ the three-dimensional effect grows stronger in amplitude, with strong spanwise perturbations superimposed on the chord-wise shock oscillation, visible as a spanwise warping of the $z-t$ signal which affects the entire buffet cycle. These results demonstrate that in the context of un-swept infinite wings, buffet becomes three-dimensional across the span when a critical angle of attack is reached. However, there exists lower angle of attack cases for which only the chord-wise shock oscillation can be present, without any significant three-dimensional effects (Section~\ref{sec:deg5_intro}). For the cases containing three-dimensional features, the amplitude of the buffet/stall-cells can be increased by widening the domain at a fixed angle of attack. Similarly, the wavelength of the instability can be shortened with increasing angle of attack for a fixed aspect ratio.

\begin{figure}
\begin{center}
\includegraphics[width=0.75\textwidth]{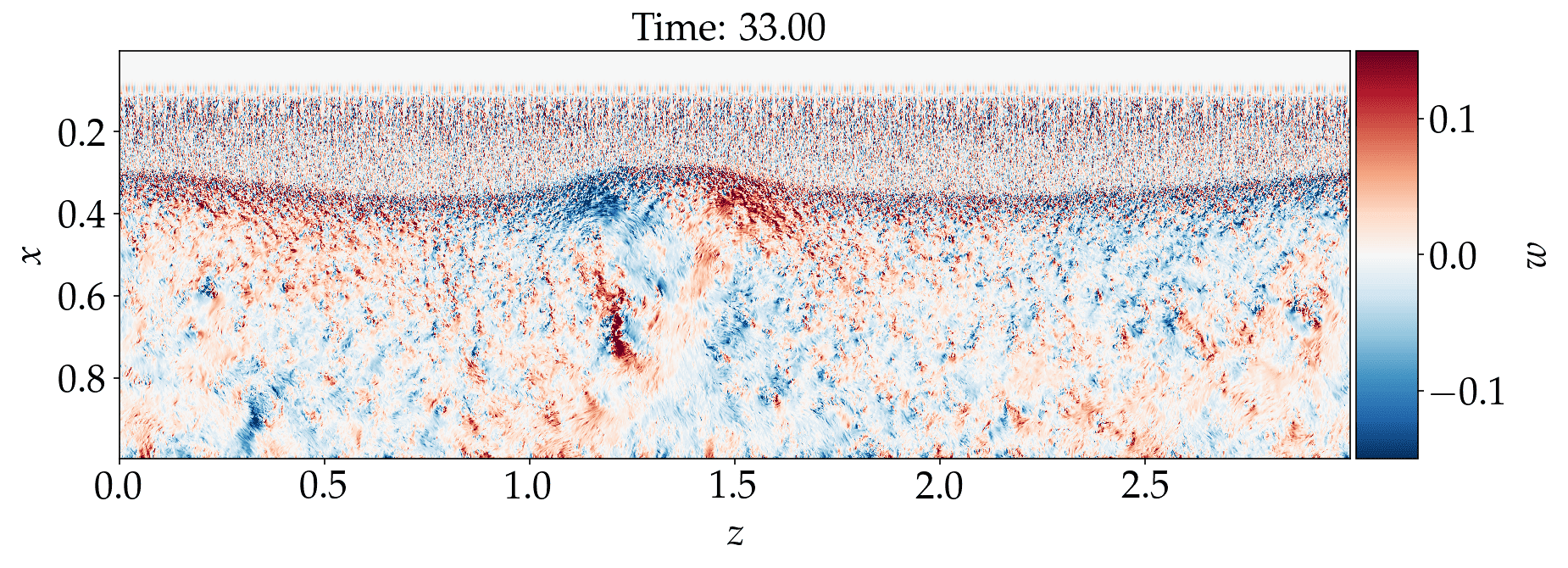}
\includegraphics[width=0.75\textwidth]{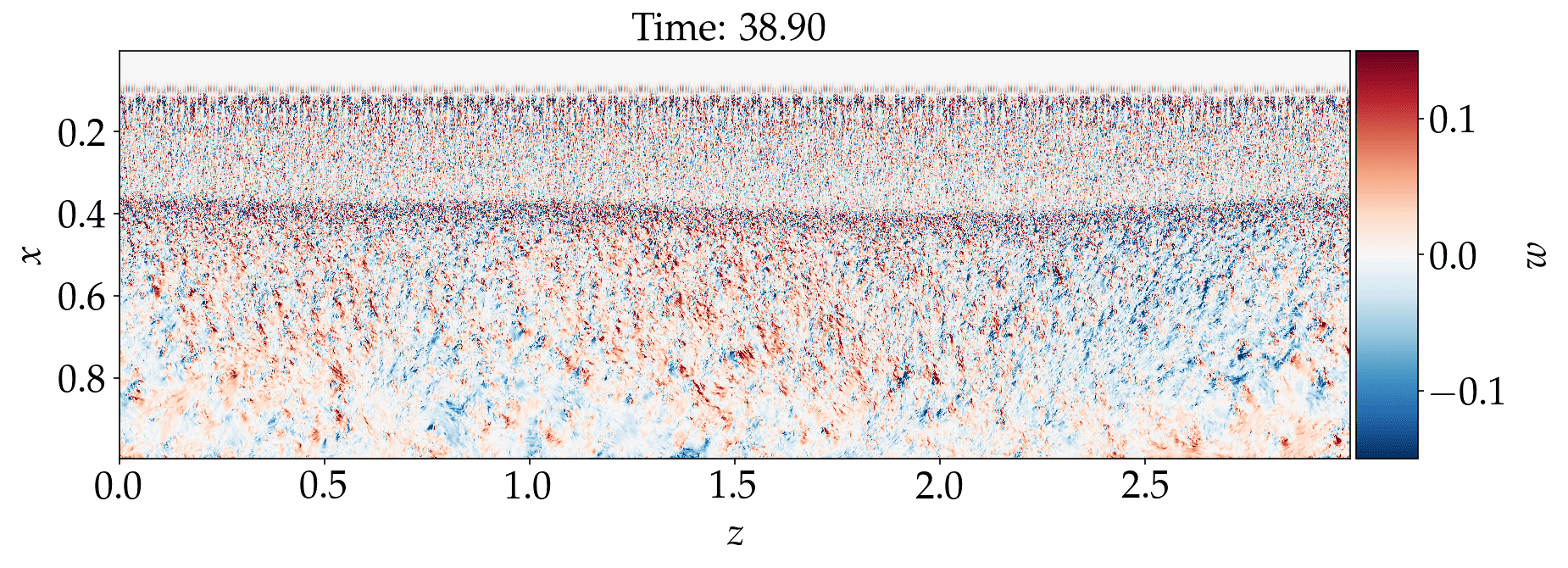}
\includegraphics[width=0.75\textwidth]{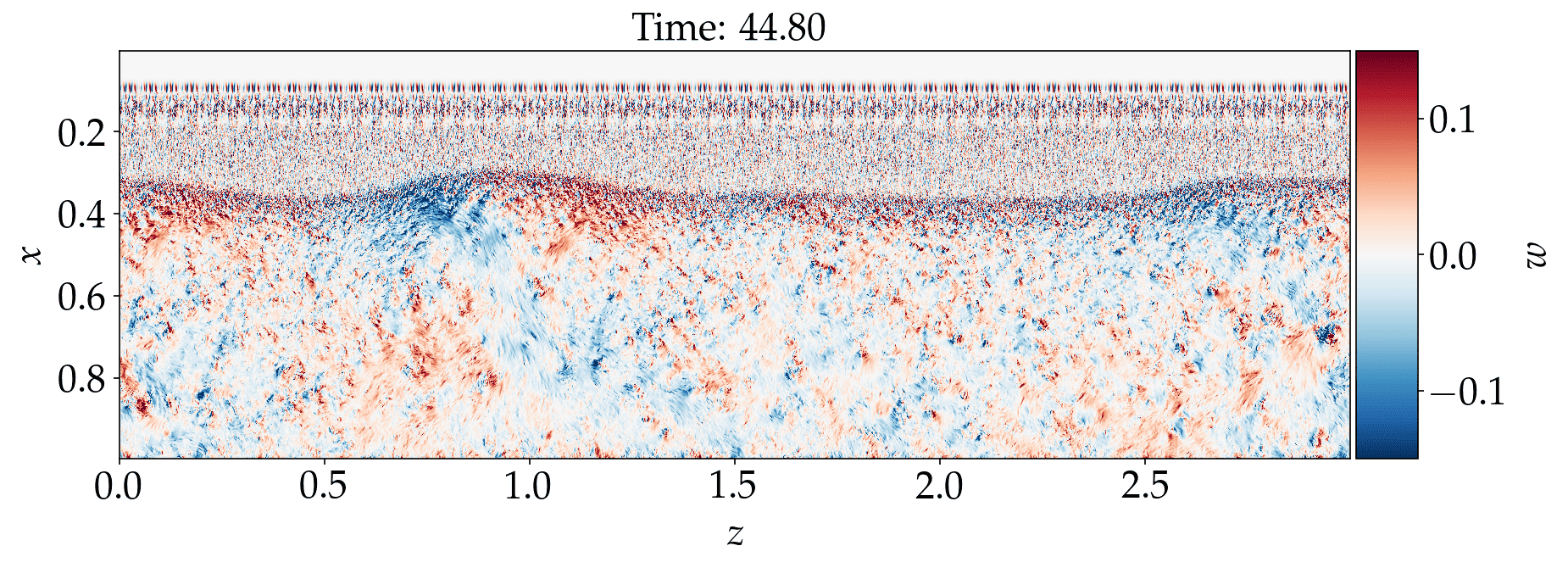}
\includegraphics[width=0.75\textwidth]{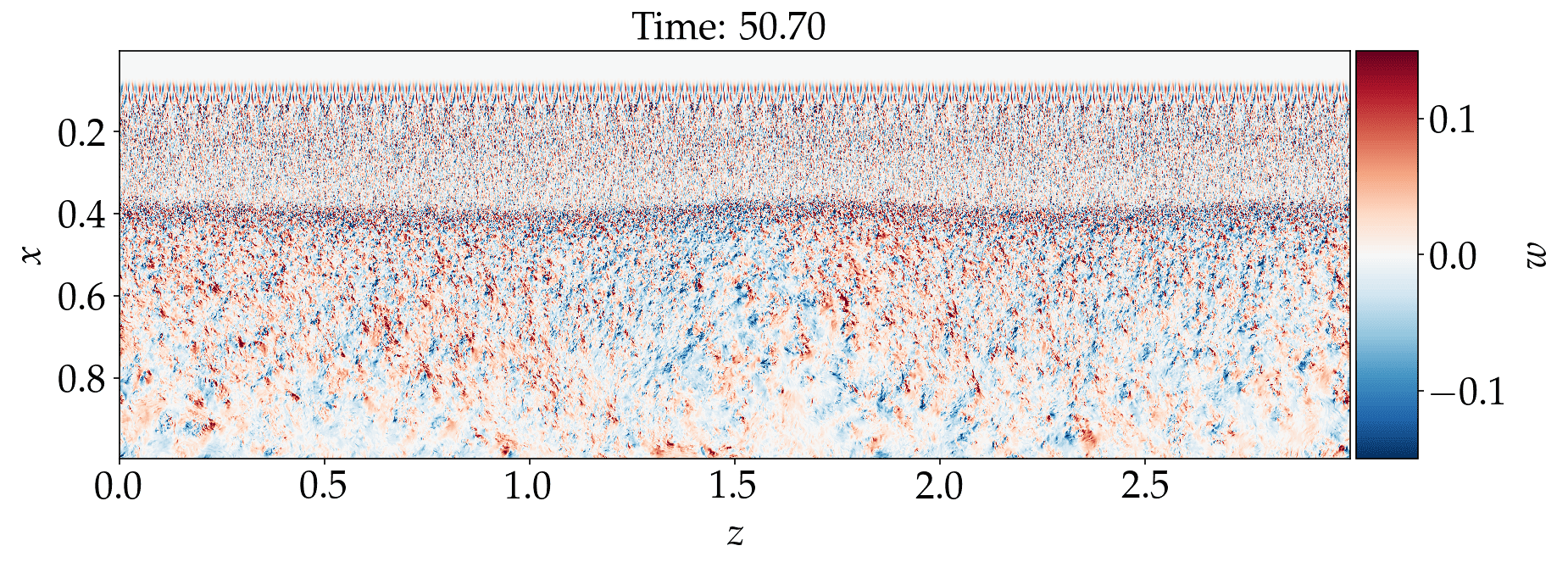}
\caption{Instantaneous spanwise velocity contours on the first point above the wall, for the case at $\AoA{6}$ and $\myAR{3}$ with $t_{\textrm{buffet}} = 11.8$. Showing four time instances equally spaced by $t_\textrm{buffet}/2$, to demonstrate the temporal emergence/cessation of the three-dimensional buffet/stall-cells.}  
\label{fig:deg6_surface_w_contours}
\end{center}
\end{figure}

Finally for this section, we look at how the flow over the suction side varies at different phases of the buffet cycle. Figure~\ref{fig:deg6_surface_w_contours} plots the AR300-AoA6 case over 1.5 buffet periods, equally spaced in time by $t_{\textrm{buffet}}/2$. The four snapshots show instantaneous near-wall spanwise $w$-velocity on the first point above the suction side of the aerofoil. Referring to the lift history in Figure~\ref{fig:deg6_lines_ALL} (a), the starting time of $t=33$ relates to the phase of the cycle where the flow is switching from high- to low-lift, but has not yet reached the minimum. At this AoA, this was found to be the phase of the buffet cycle where the three-dimensionality was at its strongest. At $t=33$ two stall/buffet-cells are visible at $z=0$ and $z=1.4$. The zero-centred spanwise velocity contours show equal and opposite propagating fluid about a central saddle point at the centre of the cell. The recirculating fluid at $z=1.4$ first travels upstream in the $x$ direction (Figure~\ref{fig:deg6_AR3_contours}) before turning left/right at the front of the separation line within the cell. The same behaviour is observed for the cell located across the periodic boundary at $z=0$ and $z=L_z$, with fluid moving in opposite directions away from both spanwise edges of the domain.

Half a buffet period later in the second snapshot where the flow is switching to high-lift phase (Figure~\ref{fig:deg6_lines_ALL} (a)), the buffet/stall-cells disappear and the flow becomes almost two-dimensional. Although there is still some remnant of the three-dimensionality that was convected downstream, the large-scale perturbations on the shock front and separation line vanish almost entirely. Returning to the original phase $t_{\textrm{buffet}}/2$ later in the third snapshot at $44.8$, the cellular structures once again. Due to the lack of sweep angle, there is no preferential location for them to occur and they are shifted left by approximately $z=0.5$ relative to the same phase in the previous period. As before, the shock moves upstream and strong three-dimensionality develops at the point of maximum flow separation just before minimum lift is reached. The cells again have left- and right-moving fluid about the saddle point on the separation line. The upstream recirculating flow is directed in opposite directions at the separation line in a similar fashion seen in other three-dimensional separation patterns that occur within fluid mechanics \citep{tobak1982topology, eagle2014shock, rodriguez2011birth, lusher_sandham_2020}. Half a cycle later at $t=50.7$ the flow reattaches and the three-dimensional separations are again removed. This cycle continues over the numerous periods (Figure~\ref{fig:deg6_ZT}) simulated as part of the buffet instability. 

\section{Sectional evaluation, cross correlations, and modal decomposition of three-dimensional buffet effects}\label{sec:further_analysis}
In this section, further analysis is performed of the $\myAR{2}$ cases at $\alpha=5^{\circ}, 6^{\circ}$, and $7^{\circ}$, and the $\myAR{3}$ case at $\AoA{6}$. In Section~\ref{sec:sectional} three-dimensional effects are investigated by evaluating quantities at individual locations across the span to observe how they deviate from span-averaged quantities. Section~\ref{sec:cross_correlations} calculates cross correlation maps at different chord-wise locations to comment on spanwise correlation/anti-correlation as a result of the intermittent three-dimensional structures. Finally, Section~\ref{sec:SPOD} performs SPOD-based modal decompositions to identify coherent modes and comment on the frequencies at which they occur for both the 2D- and 3D-instability.

\subsection{Sectional span-wise variations of the aerodynamic forces}\label{sec:sectional}

\begin{figure}
\begin{center}
\includegraphics[width=0.45\textwidth]{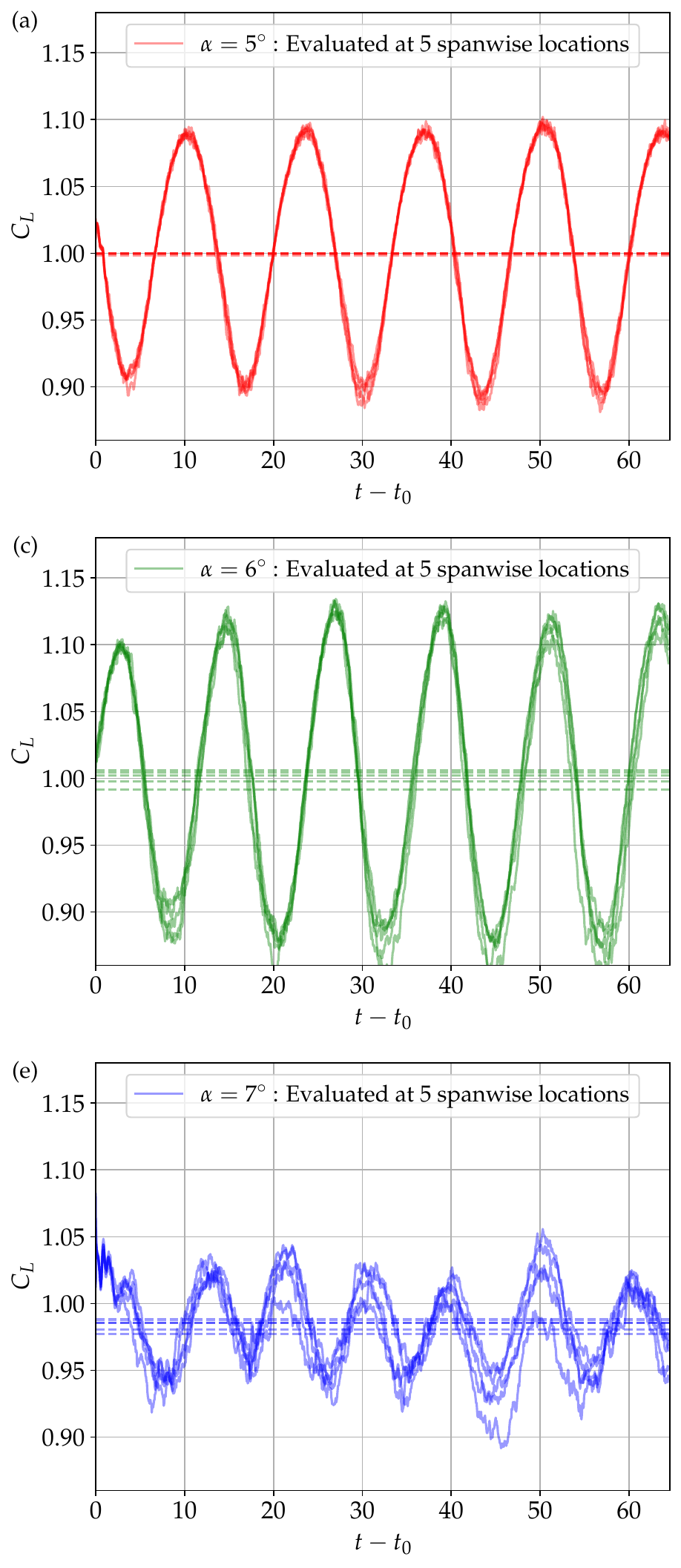}
\includegraphics[width=0.45\textwidth]{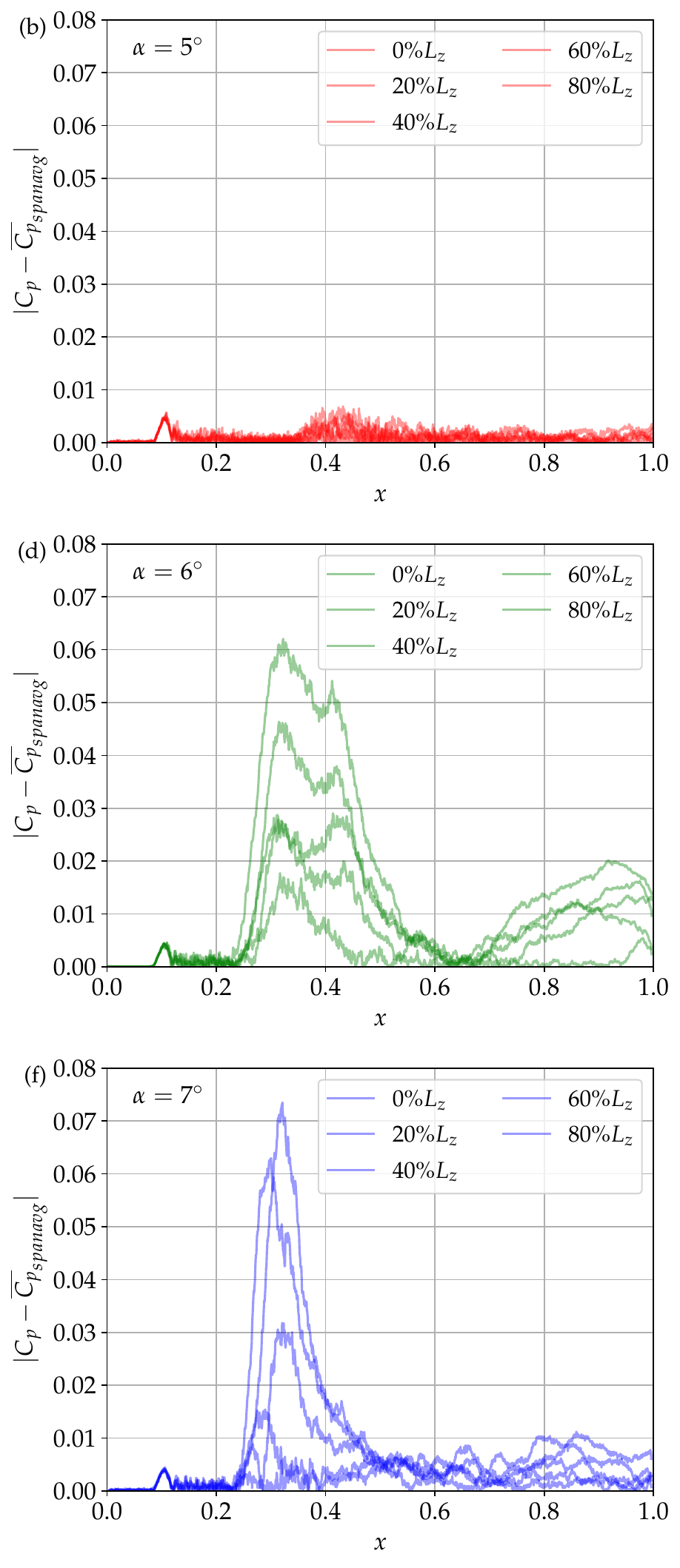}
  \caption{Sectional evaluation of aerodynamic coefficients, for $\myAR{2}$ wide-span buffet cases at $\alpha = 5^{\circ}, 6^{\circ}, 7^{\circ}$. Showing (a,c,e) sectional lift coefficient and (b,d,f) sectional deviation of the surface pressure coefficient from the span-averaged value.}
\label{fig:sectional_AoA567_AR2}
\end{center}
\end{figure}

To further investigate the contrasting behaviour for the buffet phenomenon at moderate and high AoA, it is more illustrative to look at sectional evaluation of aerodynamic coefficients in addition to the span-averaged versions. In the case of the quasi-2D buffet at $\AoA{5}$, evaluation of the aerodynamic coefficients at individual span-wise locations should not show significant deviations from the span-averaged results (Figure~\ref{fig:AR1_AR2_deg5}). Conversely, the cases exhibiting three-dimensional effects should predict different aerodynamic forces at different span-wise stations due to the loss of two-dimensionality of the flow and the finite integration time of the signal.

Figure \ref{fig:sectional_AoA567_AR2} shows the lift coefficient $C_L$ evaluated at single span-wise locations. Five evenly spaced stations are used, located at $z = 0\%, 20\%, 40\%, 60\%, 80\%$ of the span-wise aerofoil width $L_z$. The cases correspond to the wide-span $\myAR{2}$ simulations presented in the previous sections. Figures \ref{fig:sectional_AoA567_AR2} (a,c,e) show the lift coefficient at $\AoA{5}$, $\AoA{6}$ and $\AoA{7}$. In the case of $\AoA{5}$, the results from all of the five stations collapse upon one another, with only very minor deviations observed at the minima and maxima of each low-frequency buffet cycle. All of the five stations are in phase, with very good agreement for the mean $C_L$ prediction between each curve, and also to the span-averaged result (Figure~\ref{fig:AR1_AR2_deg5} (a)). This re-iterates that the buffet phenomena remains quasi-2D at this moderate AoA of $\AoA{5}$, with every span-wise region of the main shock-wave oscillating in phase along the streamwise direction. At $\AoA{6}$ and $\AoA{7}$, there is no longer agreement between the sectional profiles of lift. The five stations diverge all throughout the buffet cycle, with variations in lift magnitude. While the low-frequency trend is similar for each profile, we observe relative lags in reaching minima/maxima depending on the span-wise location used to evaluate the forces. This is due to the three-dimensional effects observed for these higher AoA cases in Figure~\ref{fig:deg6_deg7_contours}, where the flow can either be attached or separated depending on the span-wise probe location relative to the instability at a given time instance.

Figures~\ref{fig:sectional_AoA567_AR2} (b,d,f) show the absolute difference between the (i) time- and span-averaged suction side $\overline{C_p}$ distribution and the (ii) time-averaged $\overline{C_p}$ distribution when evaluated only at single span-wise location. As before, five equally spaced span-wise stations are used across the $\myAR{2}$ span width, to assess whether or not the flow maintains two-dimensionality. Furthermore, we can observe the regions of the chord-wise length that show the strongest three-dimensionality due to the buffet/stall-cells. Figure~\ref{fig:sectional_AoA567_AR2} (b) shows the result for the moderate AoA of $\AoA{5}$. The variation between the individual stations and span-averaged distribution is minimal. This quantity is evaluated over the relatively short $\approx 4.5$ low-frequency cycles shown in Figure~\ref{fig:AR1_AR2_deg5} (a). There is a small rise in the span-deviation around the mean shock position $(x = 0.45)$, but it is minor and of the same order of magnitude as that invoked by the boundary-layer tripping $(x=0.1$). When considering the sectional lift (Figure~\ref{fig:sectional_AoA567_AR2} (a)) and pressure profiles (Figure~\ref{fig:sectional_AoA567_AR2} (b)) relative to the span-averaged results, it is clear that the buffet for the moderate AoA case is essentially two-dimensional for the full length of the chord, despite the wide span-wise domain sizes used.

\begin{figure}
\begin{center}
\includegraphics[width=0.45\textwidth]{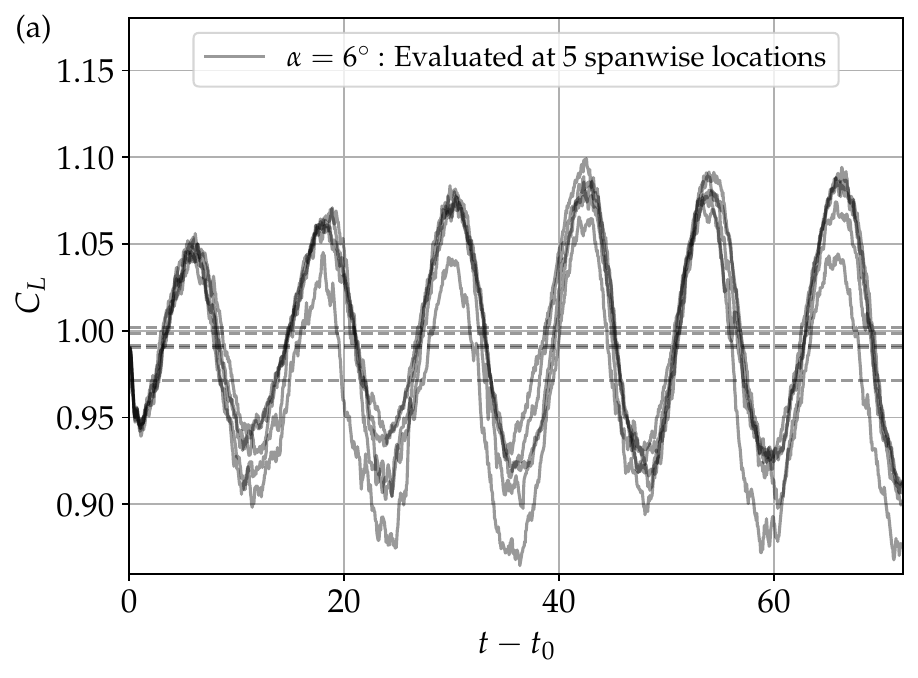}
\includegraphics[width=0.45\textwidth]{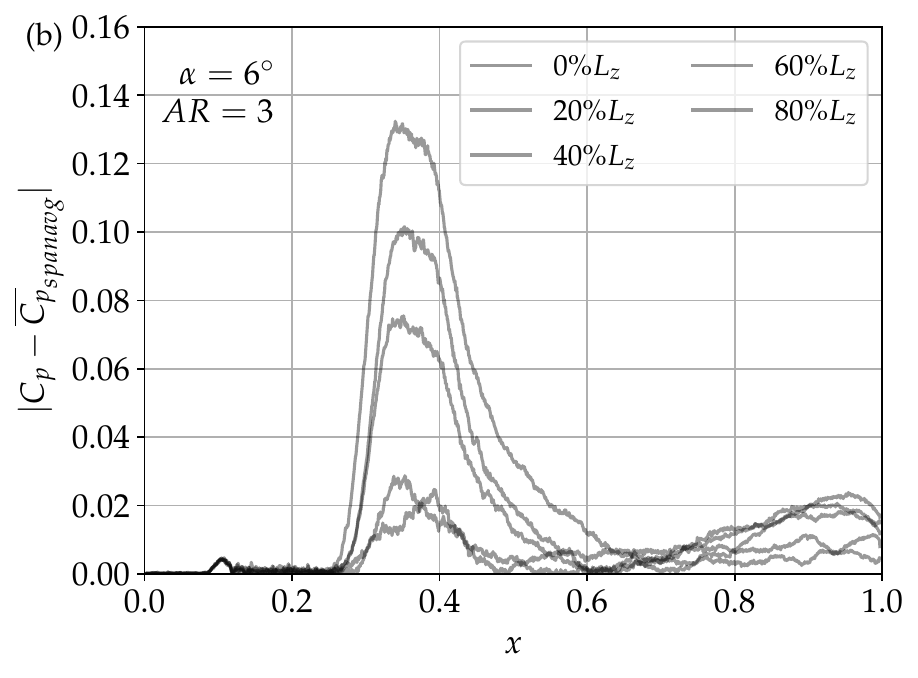}
  \caption{Sectional evaluation of aerodynamic coefficients, for the $\myAR{3}$ wide-span buffet case at $\alpha = 6^{\circ}$. Showing (a) sectional lift coefficient and (b) sectional deviation of the surface pressure coefficient from the span-averaged value.}
\label{fig:sectional_deg6_AR3}
\end{center}
\end{figure}

Figures~\ref{fig:sectional_AoA567_AR2} (b,d) shows the same measure of three-dimensionality again for the higher AoAs of $\AoA{6}$ and $\AoA{7}$. In this case, there are large peaks visible for all of the sectional profiles, indicating strong span-wise deviation from the span-averaged result (Figure~\ref{fig:AR2_deg6_deg7_LINES} (c)) due to the appearance of intermittent buffet/stall-cells. Interestingly, the three-dimensionality is mainly concentrated at the peaks centred at the mean shock locations $(x=0.325, 0.375)$. These are the same chord-wise locations used for the $z-t$ signals in Figure~\ref{fig:ZT_deg5_deg6_deg7_AR2}. For this study with zero sweep angle, the three-dimensional buffet/stall-cells are observed to be somewhat irregular in their span-wise location and this leads to the variation seen between the five equally spaced stations. While the aft region of the aerofoil downstream of the main SBLI ($x > 0.5$) is very two-dimensional in the moderate AoA case (Figure~\ref{fig:sectional_AoA567_AR2} (a)), secondary peaks at higher AoAs show the three-dimensionality persists all the way to the trailing edge in the cases with buffet cells (Figure~\ref{fig:sectional_AoA567_AR2} (d,f)). Finally, the same sectional quantities are plotted in Figure~\ref{fig:sectional_deg6_AR3} for the widest ILES case of AR300-AoA6. Similar trends are observed, with period-to-period variations in the sectional lift and a sharp peak in spanwise pressure deviation at the mean shock location. As previously noted, for a fixed AoA of $\AoA{6}$, the strength of the three-dimensionality increases with increasing aspect ratio. This is further evidenced by noting the doubling in scale for the amplitude of the peak $\left|C_p - \overline{C_p}\right|$ at $AR=3$ (Figure~\ref{fig:sectional_deg6_AR3} (b)) compared to that at $AR=2$ (Figure~\ref{fig:sectional_AoA567_AR2} (d)).

\subsection{Cross correlation analysis of three-dimensional structures}\label{sec:cross_correlations}
\begin{figure}
\begin{center}
\includegraphics[width=0.49\textwidth]{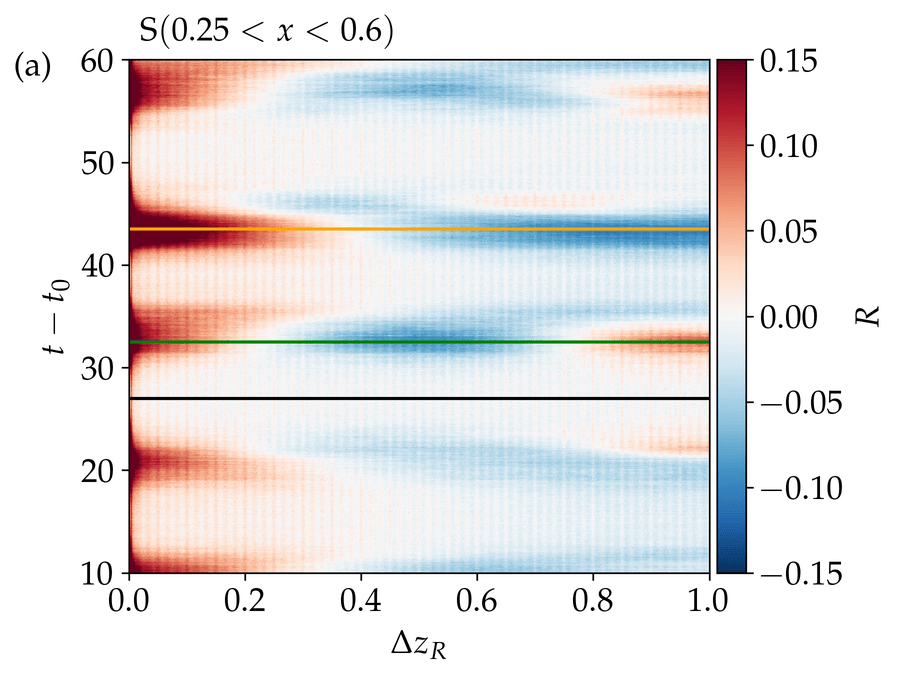}
\includegraphics[width=0.48\textwidth]{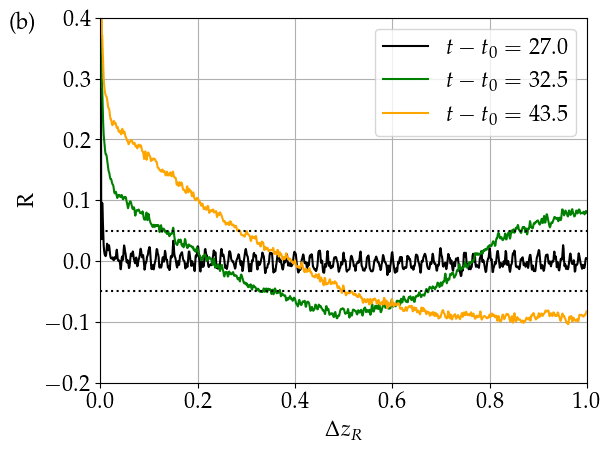}
  \caption{(a) Spanwise cross-correlation $R$ as a function of $\Delta z_R$, as it evolves in time in the chordwise stencil range of $0.25<x<0.6$. Coloured lines correspond to specific selected time instances shown in (b), where $R$ is shown as a function of $\Delta z_R$.}
\label{fig:Corr2D_AoA6_AR2_2D}
\end{center}
\end{figure}

For angles of attack above $\AoA{5}$, we have observed the appearance of large vortical cellular separation structures downstream of the main shock wave (Figure~\ref{fig:deg6_deg7_contours}). These large-scale three-dimensional structures can be intermittent in time. Their intensity and the spanwise location at which they appear varies between cycles, due to the zero sweep angle and subsequent absence of a preferential direction for convection (as in \citet{IR2015,PDL2020,PDBLS2021}). Therefore, in this section, we apply the cross-correlation technique described in Section~\ref{sec:correlation_setup} to identify the presence and size of the structures, plus their phase relation to the global aerodynamic coefficients. Correlations computed on the velocity profiles act as a footprint of the separated flow structures. For the present approach, we select profiles of streamwise-oriented $u$-velocity, taken from one point above the aerofoil suction-side surface.

In order to assess 3D effects by considering the flow at two different spanwise locations, we should look at correlations on the velocity profiles after first subtracting their spanwise-average. This is because if there are no large-scale 3D features, the spanwise fluctuations of the velocity component would be purely due to uncorrelated chaotic turbulent oscillations. If there are large-scale 3D effects, correlations will increase within the scale of such 3D structures.
The present approach is different from correlations of time histories (at a fixed streamwise location) for different spanwise locations, which is often used in the literature to justify sufficient spanwise domain width of span-periodic simulations \citep{jones2008direct}. For our correlations, regular appearance of 3D structures would increase the correlation values over the time signal. We note that correlations in buffet can also increase due to 2D shock oscillations \citep{zauner2019direct}, yet they remain a good indicator of repeated large-scale coherent structures within the flow field. The present approach enables us to analyse only spatial correlation of flow features, independent of any temporal correlation. This also includes identification of intermittent behaviour.

\begin{figure}
\begin{center}
\includegraphics[width=0.49\textwidth]{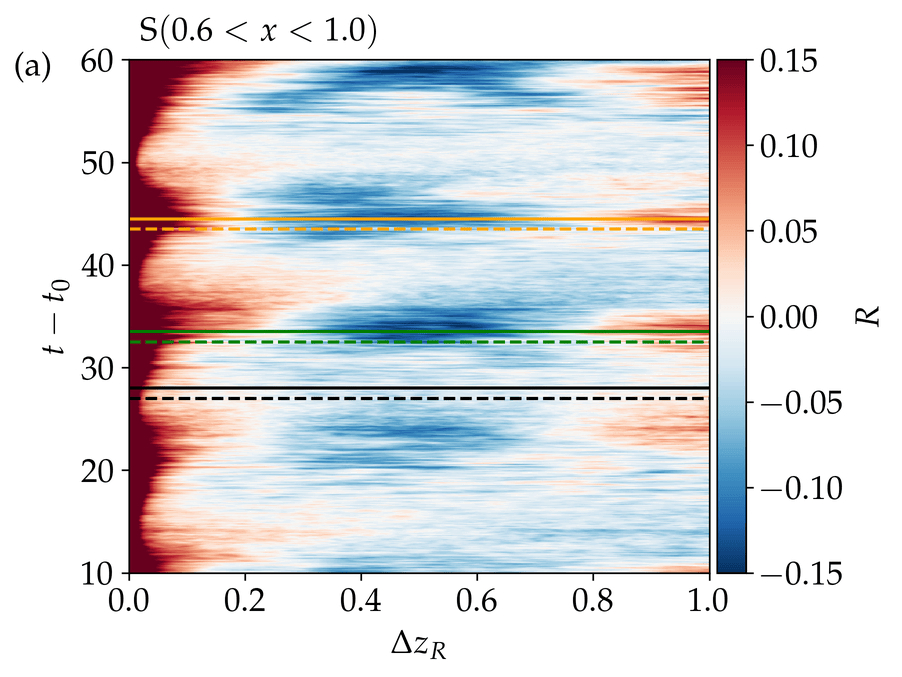}
\includegraphics[width=0.48\textwidth]{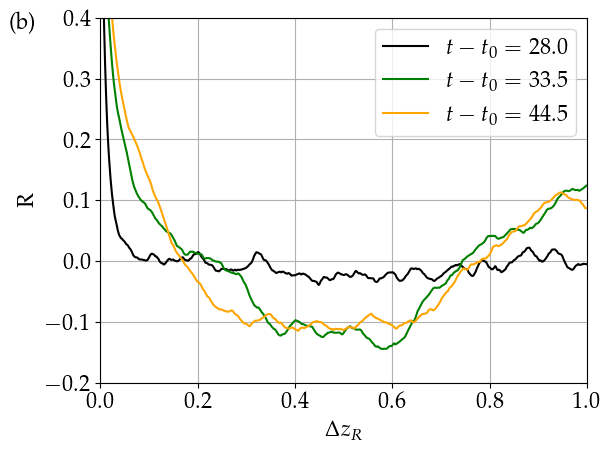}
  \caption{(a) Spanwise cross-correlation $R$ as a function of $\Delta z_R$, as it evolves in time in the chordwise stencil range of $0.6 < x< 1$. Coloured lines correspond to specific selected time instances shown in (b), where $R$ is shown as a function of $\Delta z_R$. Coloured dashed lines in (a) correspond to time-instants extracted in figure \ref{fig:Corr2D_AoA6_AR2_2D}.}
\label{fig:Corr2D_AoA6_AR2_2D_2}
\end{center}
\end{figure}

The stencils are selected to contain data between certain segments of the chordwise length. Based on the sectional evaluation of aerodynamic coefficients in Figure~\ref{fig:sectional_AoA567_AR2} which showed a strong peak of spanwise in-homogeneity around the streamwise shock displacement region when three-dimensionality was present, a first stencil is selected between $0.25 < x < 0.6$ to observe the dynamics in the shocked region. A second region is then selected as the remaining chordwise length downstream to the trailing edge, which is dominated by large turbulent vortex structures and flow separation. Evaluating two regions separately allows us to establish a phase relationship between structures in the shocked region and those appearing downstream. Histories of both sections can be compared to aerodynamic coefficients.
For the case with an angle of attack of $\AoA{6}$ and $\myAR{2}$, Figure~\ref{fig:Corr2D_AoA6_AR2_2D} (a) shows contours of the cross-correlation $R$ as a function of spanwise stencil-displacement $\Delta z_R$ as it evolves in time, for the region $0.25<x<0.6$. Within large-scale coherent flow structures (e.g. vortices, separation bubbles), we expect increased spanwise correlation (denoted by positive red regions) as their foot print (e.g. streamwise velocity profiles) at different spanwise locations is similar. In the event that the distance $\Delta z_R$ between streamwise velocity stencils becomes wider than the spanwise extent of the separation bubble, the correlation decreases as stencils in and outside the flow structure become less similar. For instances where one stencil is located within a separated region (velocity deficit with respect to span-average), and one outside the separation region (velocity surplus), the correlation $R$ may become negative for anti-correlation. In the absence of large-scale 3D structures, the correlation between two streamwise oriented stencils containing velocity data drops rapidly as soon as the $\Delta z_R$ becomes larger than those associated with turbulent length scales.

Figure~\ref{fig:Corr2D_AoA6_AR2_2D} (a) shows the time evolution of the cross correlation $R$ as a function of the stencil separation size $\Delta z_R$, with line data at specific time instances extracted in Figure~\ref{fig:Corr2D_AoA6_AR2_2D} (b). Focusing first on $t-t_{0} \approx 27$ in black, we observe a time instance where the flow is essentially uncorrelated with $R$ fluctuating marginally around zero for all medium-to-large length scales, except at the very small separations $\Delta z_R$ (associated with turbulent scales). Later on ($t-t_{0} \approx 32.5$), we observe increased $R$ up to $ \Delta z_R \approx 0.1$, which suggests the presence of medium-scale correlated 3D structures. The green curve in Figure~\ref{fig:Corr2D_AoA6_AR2_2D} (b) again shows $R$ as a function of $\Delta z_R$ at this time instance. A more gradual (almost linear) correlation decay is observed in the range of scales of $0.1<x<0.3$. A local correlation minimum is reached at a $\Delta z_R \approx 0.5$, corresponding to a quarter of the spanwise domain width for this $\myAR{2}$ case. As we shall see in the other cases also, this region of moderate anti-correlation ($R < 0$) is characteristic of the 3D phenomena shown in the previous sections. The corresponding $\Delta z_R$ is approximately half of the spanwise extent of the buffet-cell structures. Increasing values of $R$ towards a stencil separation of $\Delta z_R = L_z/2$ indicates spanwise periodicity with a spanwise wave length of $\lambda_z = L_z/2$. At that time instance, two 3D separation bubbles appear across the width of the span. At $t-t_{0} \approx 43.5$, we observe positive flow correlations up to $\Delta z_R \approx 0.3$, indicating the appearance of a large 3D structure. In contrast to $t-t_{0} \approx 32.5$, the local minimum appears now at $\Delta z_R \approx 1$, indicating that only a single 3D flow structure is present at that time. It is important to emphasise the clear temporal separation (white uncorrelated bands with $|R|<0.05$) between the appearance of 3D phenomena. Even though the 3D phenomena occurs periodically (at the main buffet frequency), the variation of $\Delta z_R$ associated with the local correlation minima suggests intermittency in their spanwise extent and organisation within the chordwise range of $0.25<x<0.6$.

\begin{figure}
\begin{center}
\includegraphics[width=0.325\textwidth]{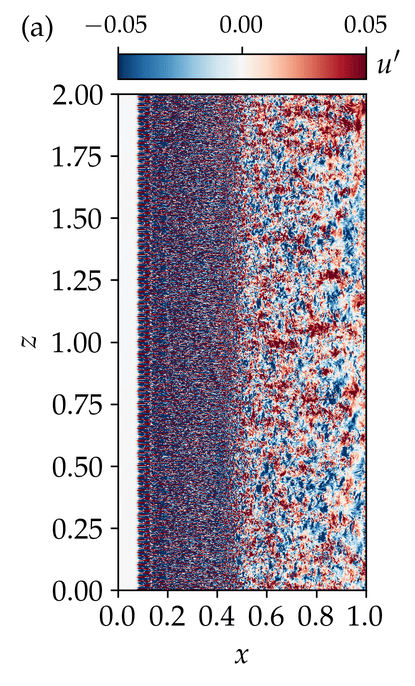}
\includegraphics[width=0.325\textwidth]{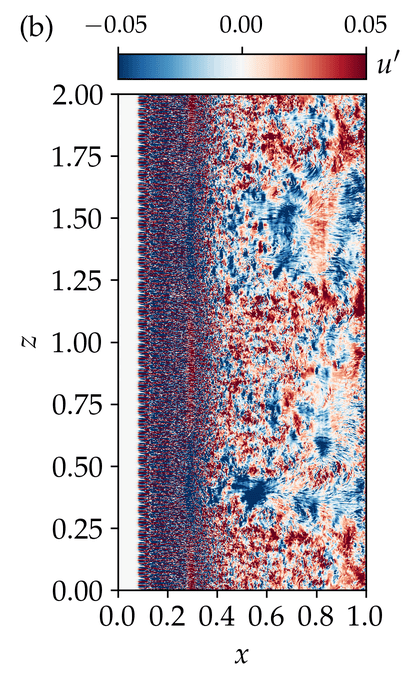}
\includegraphics[width=0.325\textwidth]{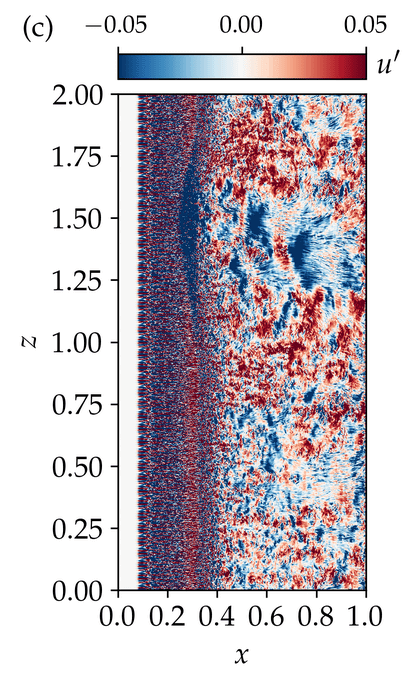}
  \caption{Instantaneous snapshots showing spanwise fluctuations of streamwise velocity component ($u'(x,z,t)=u(x,z,t)-\overline{u_{span}}(x,t)$) as a function of $z$ and $x$ at (a) $t-t_{0} = 28$, (b) $32.5$, and (c) $43.5$ for the $\AoA{6}$ and $\myAR{2}$ case. Data is extracted from an $x/z$ slice at the first grid point off the suction side surface.}
\label{fig:Q_2D_AoA6_AR2}
\end{center}
\end{figure}

Figure \ref{fig:Corr2D_AoA6_AR2_2D_2} shows the same analysis for the stencil centred further downstream over the chordwise section $0.6<x<1$. Dashed horizontal lines in (a) correspond to the time instances marked in Figure~\ref{fig:Corr2D_AoA6_AR2_2D} (a). We can see that contours associated with the 3D phenomena are slightly delayed in time compared to the dashed lines, which indicates downstream convection of the separation cells. To account for this feature, we shifted the horizontal lines for extracting $R$ data by a time interval of $\Delta t = 1$, by approximating the convective structures propagating at around $40\%$ of the speed of the boundary-layer edge velocity. The time instances considered in Figure~\ref{fig:Corr2D_AoA6_AR2_2D} are denoted by the horizontal dashed lines in Figure~\ref{fig:Corr2D_AoA6_AR2_2D} (a).
As in the stencils for the shocked region $(0.25 < x < 0.6$), we observe in Figure~\ref{fig:Corr2D_AoA6_AR2_2D_2} (a) blue regions of anti-correlation at the scale of $\Delta z_R \approx 0.5$. This is true even for $t-t_{0} \approx 44.5$, identified previously in Figure~\ref{fig:Corr2D_AoA6_AR2_2D} (a) (capturing mainly shock-wave characteristics). While the green and orange curves in Figure~\ref{fig:Corr2D_AoA6_AR2_2D_2} (b) appear fairly similar at first glance, we can again identify clear temporal separation by the white uncorrelated bands in Figure~\ref{fig:Corr2D_AoA6_AR2_2D_2} (a). This behaviour is confirmed by the black curve in Figure~\ref{fig:Corr2D_AoA6_AR2_2D_2} (b), where correlations drop rapidly and do not exceed $|R|=0.05$ for $\Delta z_R>0.1$.

\begin{figure}
\begin{center}
\includegraphics[width=0.7\textwidth]{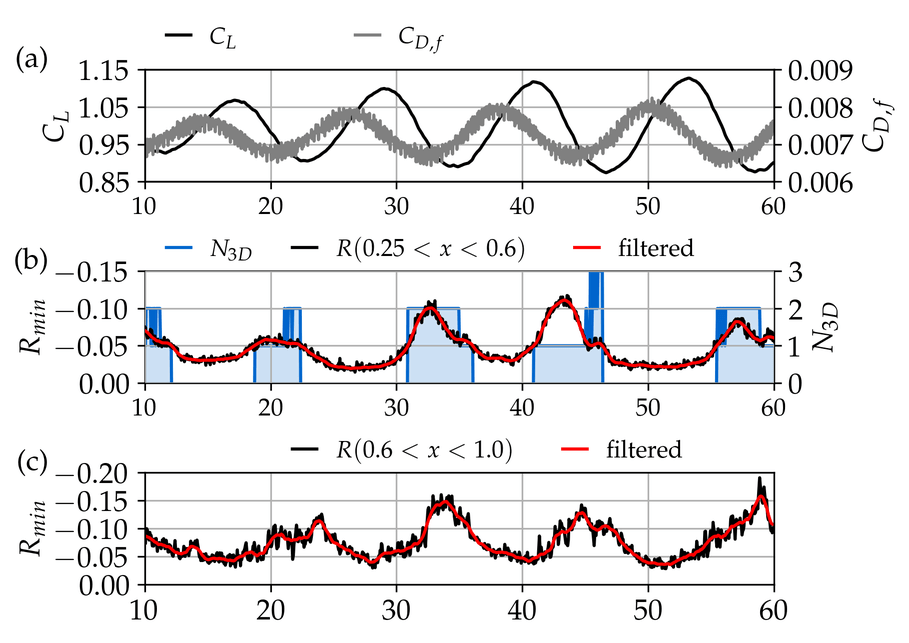}
  \caption{Evolution of anti-correlation for the case with $\alpha=6^{\circ}$ and \myAR{2}. Showing (a) Lift coefficient $C_L$ and skin-friction drag coefficient $C_{D,f}$ as functions of time. The number of 3D structures detected in the spanwise direction $N_{3D}$, is estimated based on the $\Delta z_R$ at the corresponding local cross correlation minima $R_{min}$. The quantities are plotted in time for (b) $0.25<x<0.6$ and (c) $0.6<x<1.0$, respectively.}
\label{fig:R_time_AoA6_AR2}
\end{center}
\end{figure}

To further elucidate the three-dimensionality of the $\AoA{6}$ case at $\myAR{2}$, instantaneous snapshots of streamwise velocity fluctuations $u'(x,z,t)=u(x,z,t)-\overline{u_{span}}(x,t)$ are shown at the same selected time instances as before in Figure~\ref{fig:Q_2D_AoA6_AR2}, for data one point above the aerofoil suction side. When the correlations are close to zero (Figure~\ref{fig:Corr2D_AoA6_AR2_2D}), the surface plot of velocity fluctuations in Figure~\ref{fig:Q_2D_AoA6_AR2} (a) shows no significant large-scale 3D structures and the location of the main shock-wave at $x \approx 0.45$ is essentially-2D and perpendicular to the freestream. However, as we progress in time to $t-t_{0} = 32.5$ and $t-t_{0} = 43.5$ when there are both strong positive and negative correlations present (Figure~\ref{fig:Corr2D_AoA6_AR2_2D} (b)), the plot of fluctuations shows clear alternating red and blue patches along the shock-front at $x\approx 0.3$, reminiscent of the alternating pressure perturbations seen along the shock-front for buffet cells in low-fidelity computations \citep{IR2015,PBDSR2019,PDBLS2021} and experiments \citep{SKNNNA2018,SNKNNA2021,SKK2022}. Looking at the region immediately downstream of the shock-wave ($x>0.4$), we observe in both latter figures two convective cellular 3D structures. These large-scale 3D structures at the shock and downstream in the turbulent region are clearly identified by the fluctuations and cross-correlation technique. We note that the panels in Figure~\ref{fig:Q_2D_AoA6_AR2} (b,c) show that the same flow conditions and aspect ratio can support either one or two buffet-cell structures depending on the time instance, providing further evidence for the intermittency and irregularity associated with 3D buffet cells in the absence of sweep.

\begin{figure}
\begin{center}
\includegraphics[width=0.7\textwidth]{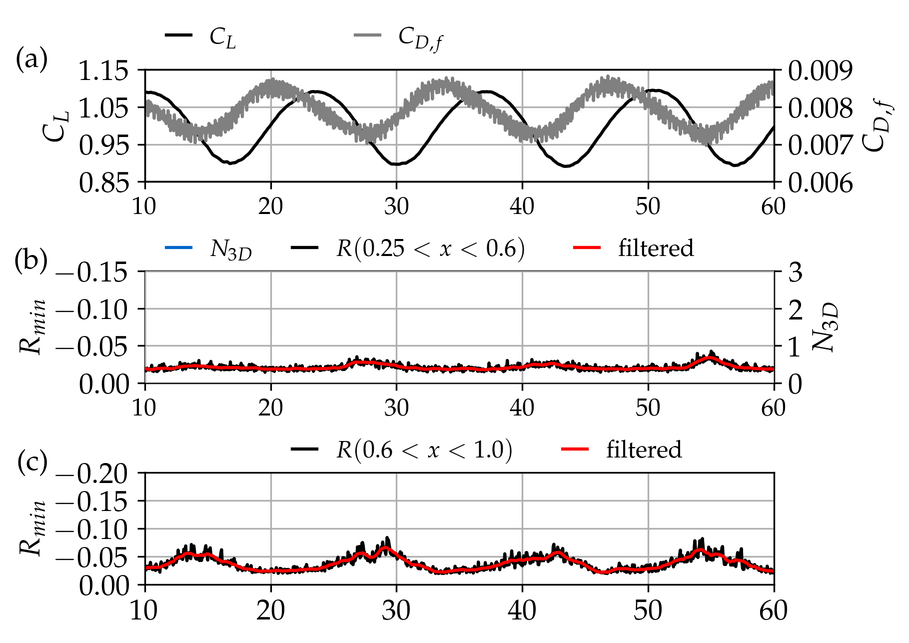}
\includegraphics[width=0.7\textwidth]{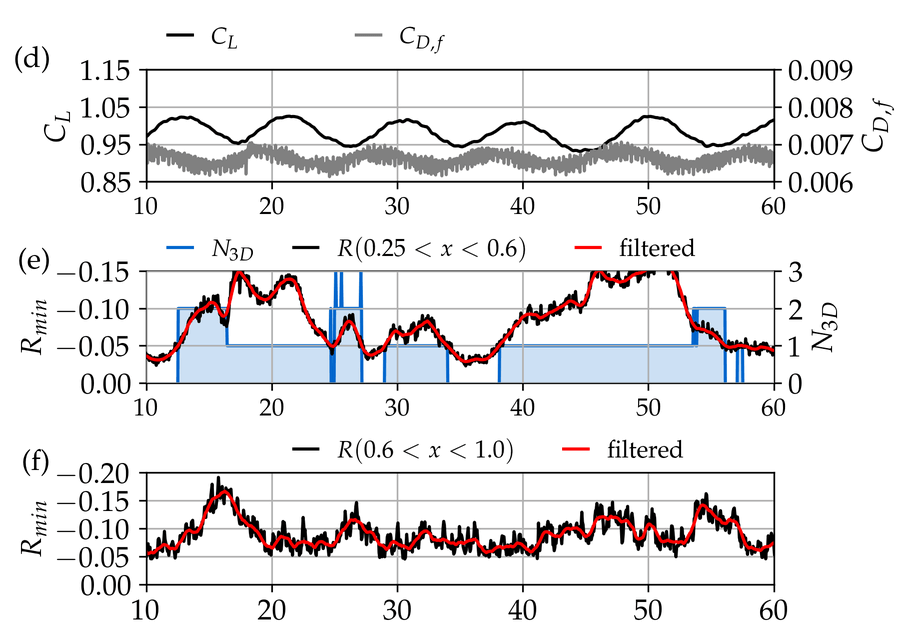}
  \caption{Evolution of anti-correlation for the cases with \myAR{2} and (a-c) $\alpha=5^{\circ}$ and (d-f) $\alpha=7^{\circ}$. Showing (a, d) Lift coefficient $C_L$ and skin-friction drag coefficient $C_{D,f}$ as functions of time. The number of 3D structures detected in the spanwise direction $N_{3D}$, is estimated based on the $\Delta z_R$ at the corresponding local cross correlation minima $R_{min}$. The quantities are plotted in time for (b, e) $0.25<x<0.6$ and (c, f) $0.6<x<1.0$, respectively.}
\label{fig:R_time_AoA5_7_AR2}
\end{center}
\end{figure}

To take a closer look at the maximum anti-correlation behaviour for the same case at $\AoA{6}$ and $\myAR{2}$, Figure \ref{fig:R_time_AoA6_AR2} shows how the minimum correlation value $R_{\textrm{min}}$ evolves in time. This quantity is shown for both the shocked region $(0.25 < x < 0.6)$ and the aft section of the aerofoil $(0.6 < x < 1)$. The corresponding lift coefficient and skin-friction drag components are shown above to make reference to the aerodynamic forces. While $C_L$ and $C_D$ are dominated by pressure forces, $C_{D,f}$ is a more sensitive measure of flow separation. In addition to $R_{\textrm{min}}$, the number of structures identified by the $\Delta z_R$ separation is overlaid as a measure defined as $N_{3D}=\verb|round|(L_z/2\Delta z_{R,min})$. To identify strong 3D structures and remove instances of uncorrelated data, a threshold for three-dimensionality is imposed on the anti-correlation $R_{min}$ equal to $R_{min}= \pm 0.05$. These bounds are used to identify large-scale structures in the shocked region and are marked by the dotted horizontal lines in Figures~\ref{fig:Corr2D_AoA6_AR2_2D} (b). Black curves in the second and third rows of Figure~\ref{fig:R_time_AoA6_AR2} show $R_{min}$ as a function of time. To improve visual clarity, a $3^{rd}$-order filter is applied to the data and plotted in red. The blue regions show the number of 3D structures detected from the smoothed data. In general, the 3D structures are observed at time instances where the global skin-friction drag reaches a local minima (just before the lift starts recovering). In the shock region ($0.25<x<0.6$), it is hard to identify a clear number of 3D structures. This may be even due to the co-existence of multiple structures at different length scales. However, we can clearly see that the correlation levels are dropping also in this region, when the shock wave becomes quasi two-dimensional. In the aft section of the aerofoil the same structures are identified at a slightly later time instance, albeit with increased noise in the correlation signal.

The previous sections have shown that an essentially-2D solution is obtained at $\AoA{5}$, whereas $\AoA{6}$ and $\AoA{7}$ have strong 3D buffet effects. To investigate whether this behaviour can also be identified by the correlation analysis in this section, Figure~\ref{fig:R_time_AoA5_7_AR2} again shows the aerodynamic coefficients in time compared to the correlation measure at (a-c) $\AoA{5}$ and (d-f) $\AoA{7}$. Despite the clear 2D periodic buffet oscillations in the lift coefficient at $\AoA{5}$, the minimum correlation $R_{\textrm{min}}$ is almost flat within the shocked region ($0.25 < x < 0.6$), with no strong anti-correlations observed. In the rear part of the aerofoil, however, weak peaks of anti-correlation are present. The instances of weak anti-correlation at the rear of the aerofoil are seen to be in phase with the local minima of the skin-friction drag component $C_{D,f}$. The same quantities are plotted in Figures~\ref{fig:R_time_AoA5_7_AR2} (d-f) for the 3D case at $\AoA{7}$. Looking at the correlation curve of minimum correlation at each time instance in red, the signal is quite noisy, with additional fluctuations observed at different time scales. Compared to the $\AoA{5}$ case, strong anti-correlations are observed over the majority of the signal both within the shocked region and near the trailing edge, with far less regularity than seen at $\AoA{6}$. Interestingly, less pronounced peaks are observed in the aft region of the airfoil.

\begin{figure}
\begin{center}
\includegraphics[width=0.24\textwidth]{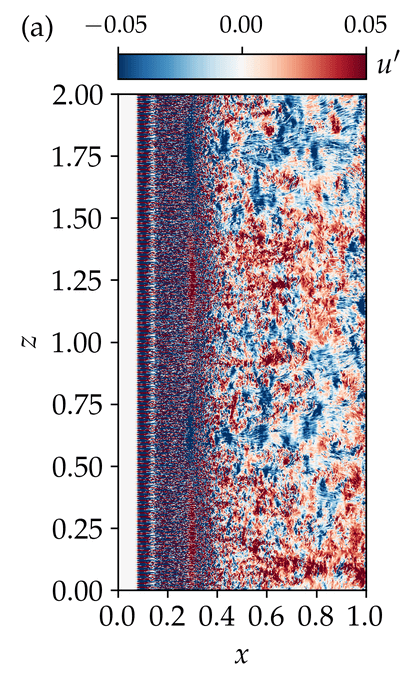}
\includegraphics[width=0.24\textwidth]{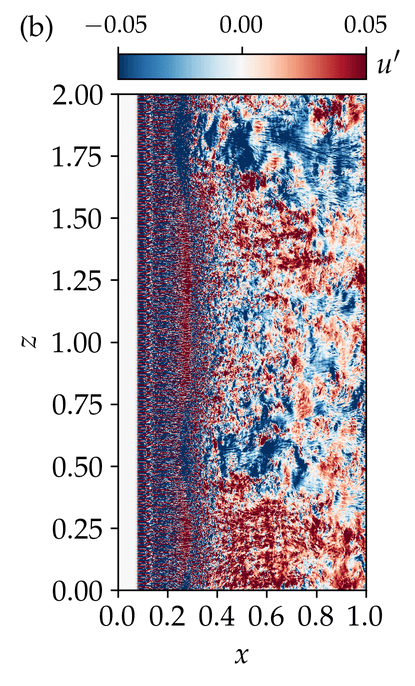}
\includegraphics[width=0.24\textwidth]{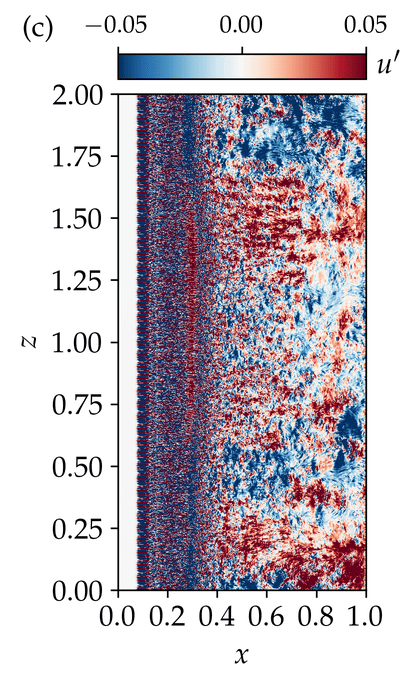}
\includegraphics[width=0.24\textwidth]{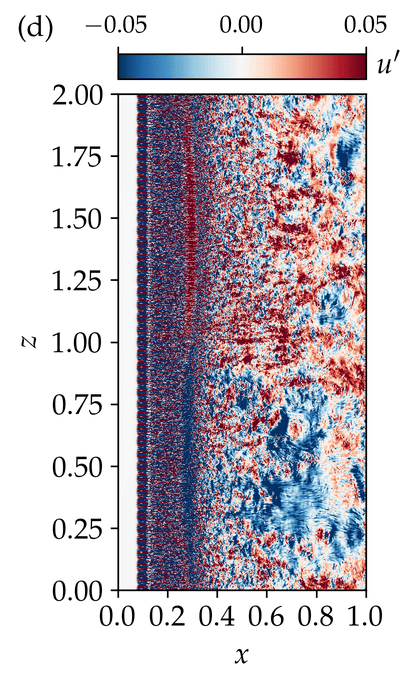}
  \caption{Instantaneous snapshots showing spanwise fluctuations of streamwise velocity component ($u'(x,z,t)=u(x,z,t)-\overline{u_{span}}(x,t)$) as a function of $z$ and $x$ at (a) $t-t_{0} = 14$, (b) $16$, (c) $18$, and (d) $46$ for the $\AoA{7}$ and $\myAR{2}$ case. Data is extracted from an $x/z$ slice at the first grid point off the suction side surface.}
\label{fig:Q_2D_AoA7_AR2}
\end{center}
\end{figure}

To look at the dynamics of the separation cells in time for this case with strong three-dimensionality, Figure~\ref{fig:Q_2D_AoA7_AR2} shows instantaneous snapshots at (a) $t-t_0=14$, (b) $16$, (c) $18$, and (d) $46$. At $t-t_0=14$ we observe two large-scale spanwise structures (in blue) along the undulated imprint of the shock-wave at $x\approx0.3$. A clear alternating velocity pattern is observed with two clear wavelengths across the span, in a similar manner to those observed in Figure~\ref{fig:Q_2D_AoA6_AR2} (b) at $\AoA{6}$. The relative strength of these structures increases along the shock-front at the later time of $t-t_0=16$, before eventually merging into a single wavelength across the span at $t-t_0=18$. This highlights the unsteady and irregular nature of the buffet-cell phenomena without sweep, where different numbers of separation cells can emerge throughout the standard 2D low-frequency buffet cycle if the aspect ratio is sufficiently wide. Separation cells at the SBLI are convected downstream and reattach/merge at later time instances due to turbulent mixing. Figures~\ref{fig:Q_2D_AoA7_AR2} (c,d), show snapshots of the fluctuations at phases where $C_{D,f}$ reaches a maximum (similar phases within the 2D buffet cycles). In both cases, a single wavelength is seen along the position of the shock-front. However, the region downstream of the shock-wave in (c) is dominated by two to three large-scale structures, but Figure~\ref{fig:Q_2D_AoA7_AR2} (d) shows only a single wavelength across the span, aligned with the velocity defect/surplus pattern at the shock-wave. Finally, the same cross correlation analysis is shown for $\AoA{6}$ and $\myAR{3}$ in Appendix~\ref{sec:AR3_correlations}, with similar trends observed to the $\myAR{2}$ case shown in Figure~\ref{fig:R_time_AoA6_AR2}.

%% END OF CORRELATIONS, START OF SPOD
\subsection{Spectral Proper Orthogonal Decomposition (SPOD)}\label{sec:SPOD}
\begin{figure}
\begin{center}
\includegraphics[width=0.325\textwidth]{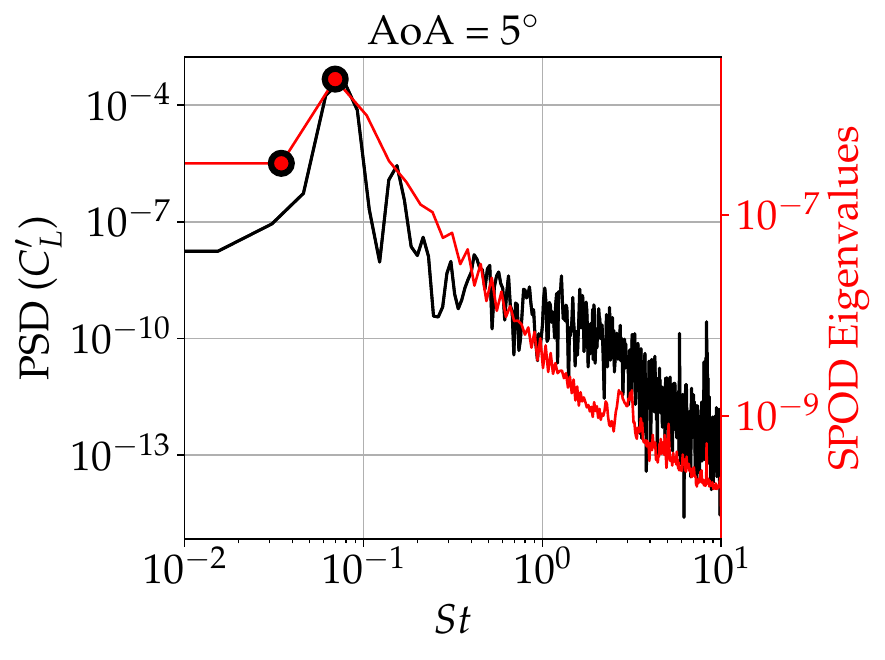}
\includegraphics[width=0.325\textwidth]{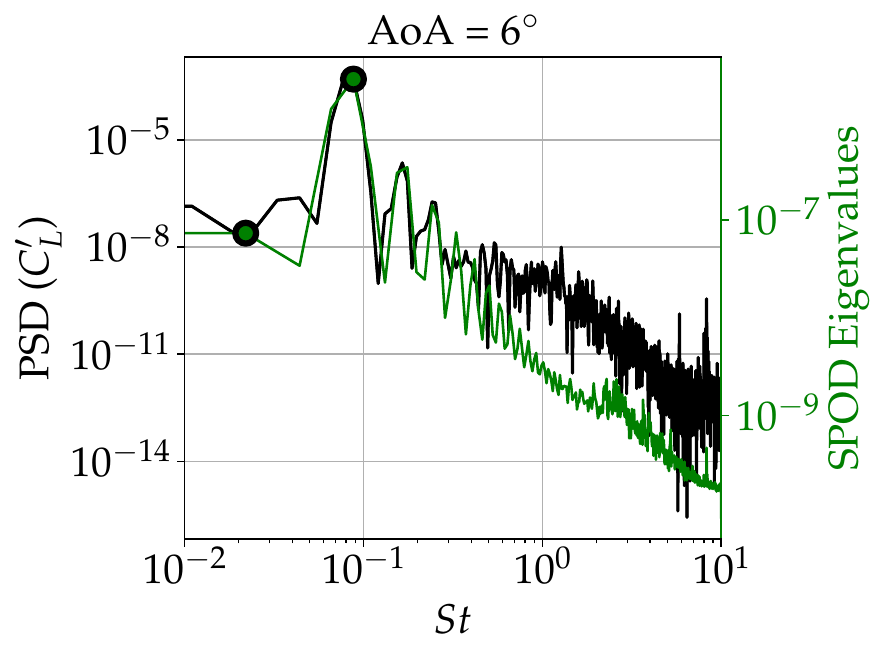}
\includegraphics[width=0.325\textwidth]{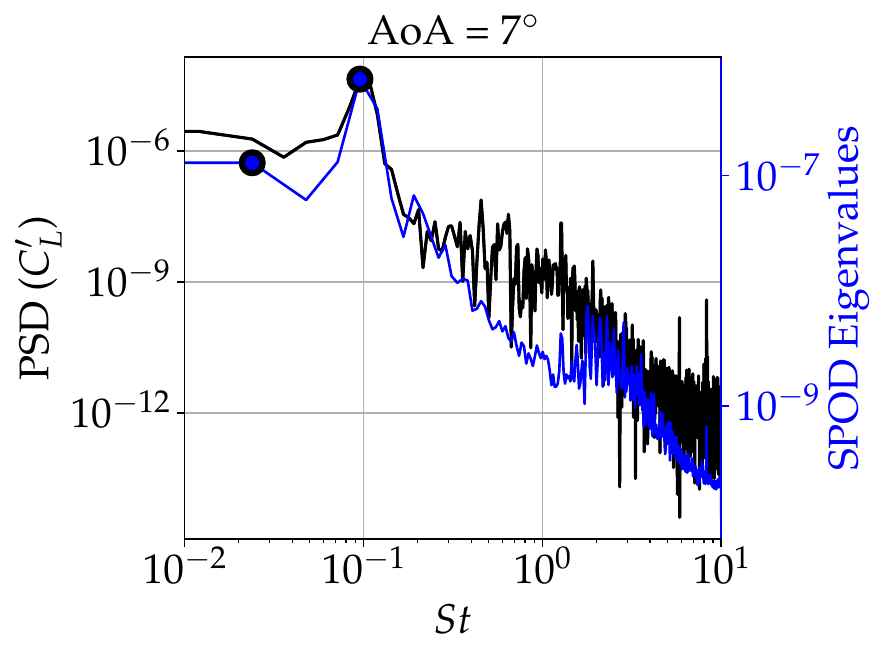}
\includegraphics[width=0.325\textwidth]{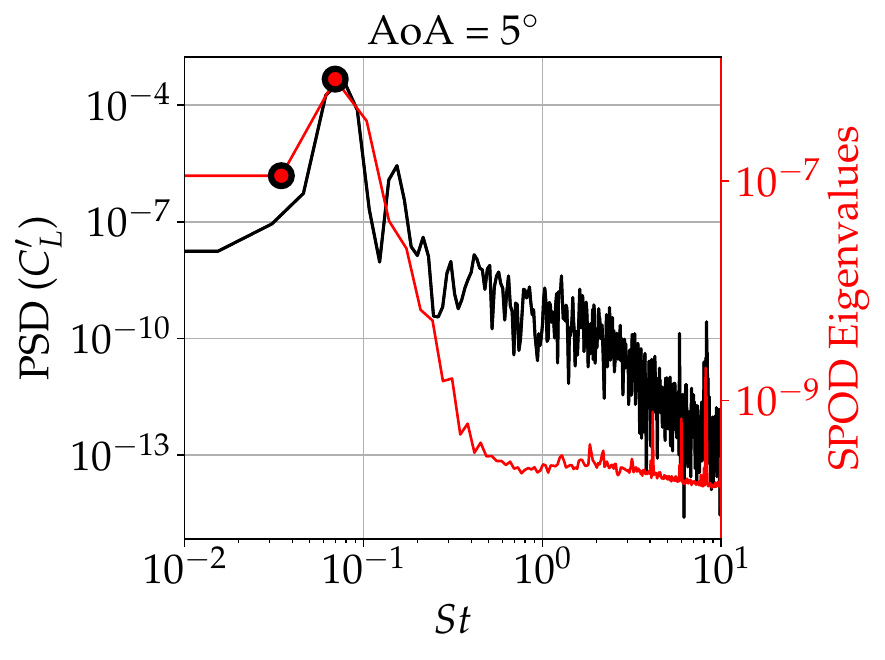}
\includegraphics[width=0.325\textwidth]{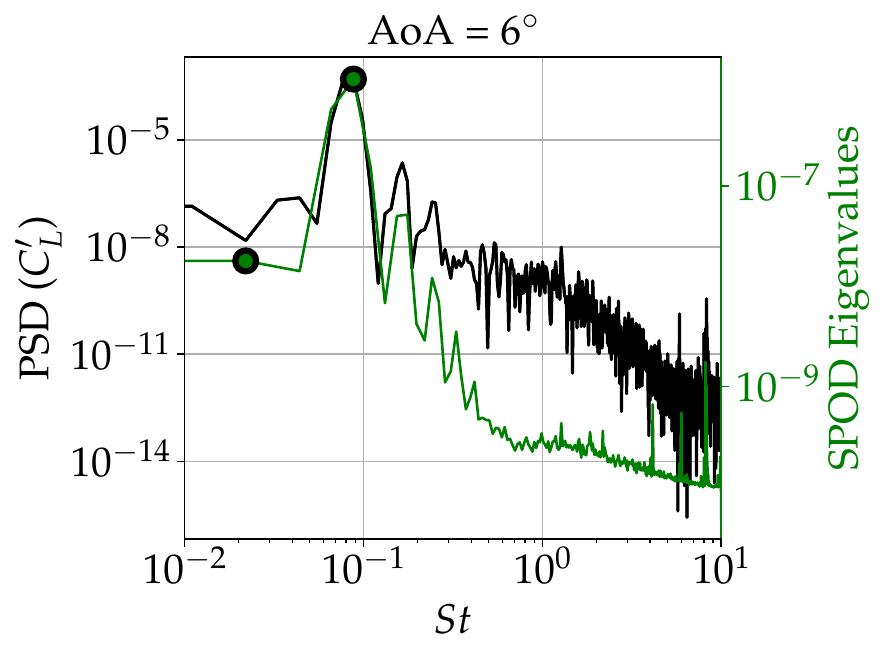}
\includegraphics[width=0.325\textwidth]{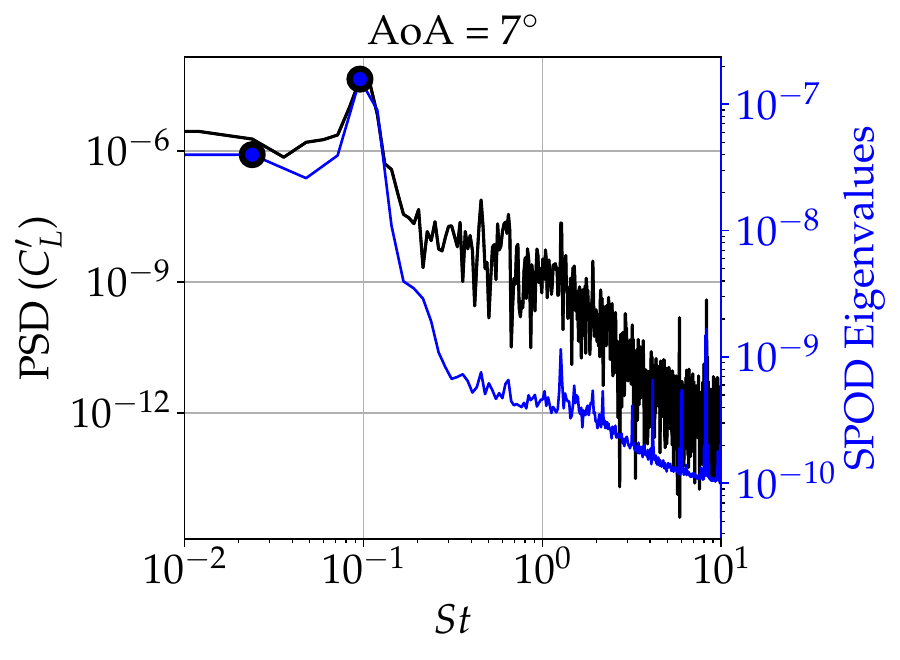}
\caption{SPOD eigenvalue spectra (coloured solid lines) for pressure data on side $x$-$y$ plane at $z=1$ (top plots) and $x$-$z$ suction-side wall (bottom plots) are plotted with the PSD of lift-coefficient fluctuations (black solid lines) for cases $\alpha = 5^{\circ}, 6^{\circ}, 7^{\circ}$ and $\myAR{2}$. Circles correspond to the selected modes for visualization.}
\label{fig:spod_p_spectra}
\end{center}
\end{figure}

In this section, a Spectral Proper Orthogonal Decomposition (SPOD) method \citep{ModalDecomp_Taira2017,towne_schmidt_colonius_2018} is applied to a selection of the cases presented at $\myAR{2}$ and $\myAR{3}$, to further analyse the structure and frequency content of the three-dimensional buffet effects in the absence of sweep. Details of the SPOD configuration are available in Section~\ref{sec:SPOD_setup}. To perform the analysis, two-dimensional snapshots of pressure data are used from both a side ($x$-$y$) plane (at $z=L_z / 2$), and at the wall ($x$-$z$ plane, $y=0$). To further elucidate the three-dimensional behaviour, SPOD is also performed on spanwise $w$-velocity data at the first grid point off the wall for some select cases.

For all angles of attack investigated $(\AoA{5}, \AoA{6}, \AoA{7})$, the eigenvalue spectra of the first SPOD mode are plotted for both side plane and wall-pressured-based data sets in Figure~\ref{fig:spod_p_spectra}. To aid comparison to the global dynamics and aerodynamic forces, the PSD distributions calculated on lift-coefficient fluctuations are also overlaid. For all angles of attack and data sets, the dominant SPOD mode matches the frequency of the dominant peak in the $C'_L$-based PSD very well. While significant energy content can be seen at low-frequencies, the spectra rapidly decays for frequencies above the dominant peak. Although not shown here for brevity, SPOD modes above the dominant one are all either higher harmonics of the dominant frequency (peaks at multiples of the dominant $St$), wake modes ($0.3<St<5$) \citep{Moise2023_AIAAJ,LongWong2024_Laminar_buffet}, or due to the numerical boundary-layer tripping ($3<St<10$). When considering data only from the aerofoil surface wall pressure (i.e. excluding data from the wake blocks), it can be seen that the energy of the wake and tripping modes is significantly reduced in relation to the dominant peak. Based on this observation, we focus the modal analysis on the low-frequency ($0.02<St<0.04$) and dominant ($0.07<St<0.096$) modes, indicated by the colored circles in the SPOD spectra.

\begin{figure}
\begin{center}
\includegraphics[width=0.325\textwidth]{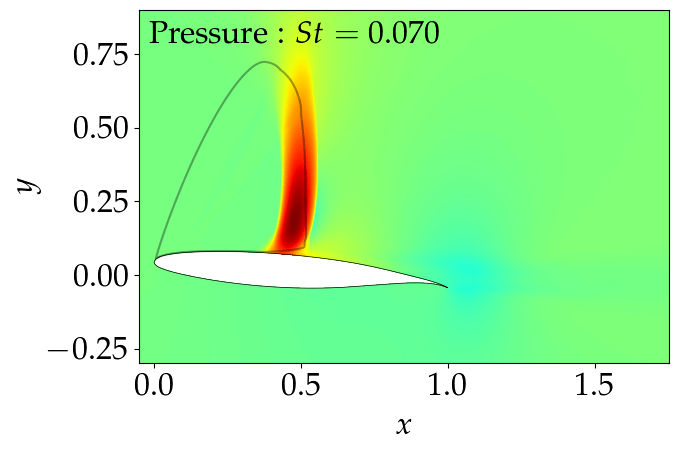}
\includegraphics[width=0.325\textwidth]{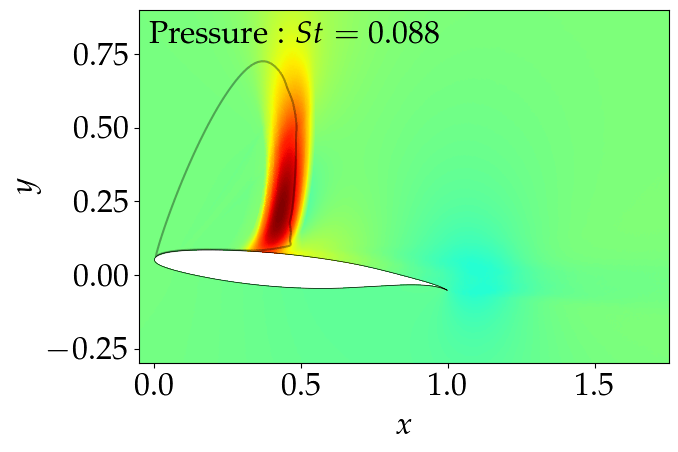}
\includegraphics[width=0.325\textwidth]{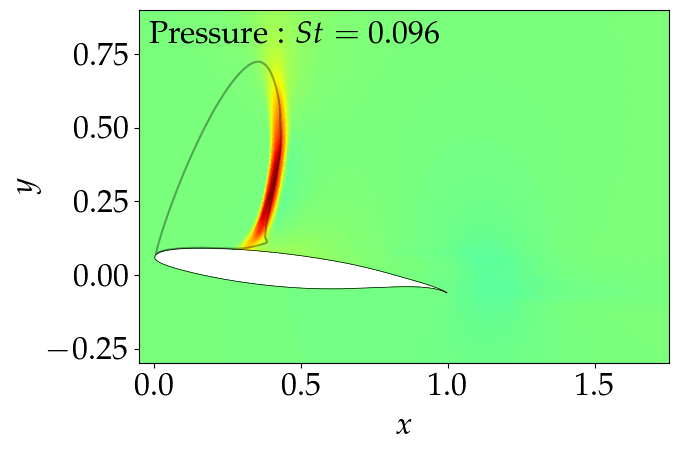}
\includegraphics[width=0.325\textwidth]{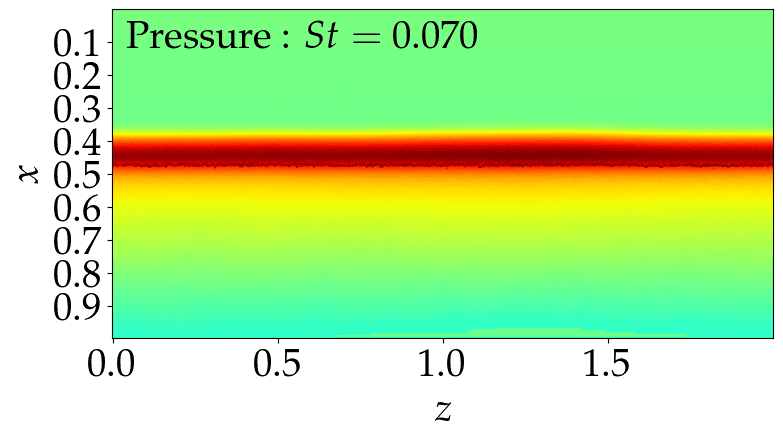}
\includegraphics[width=0.325\textwidth]{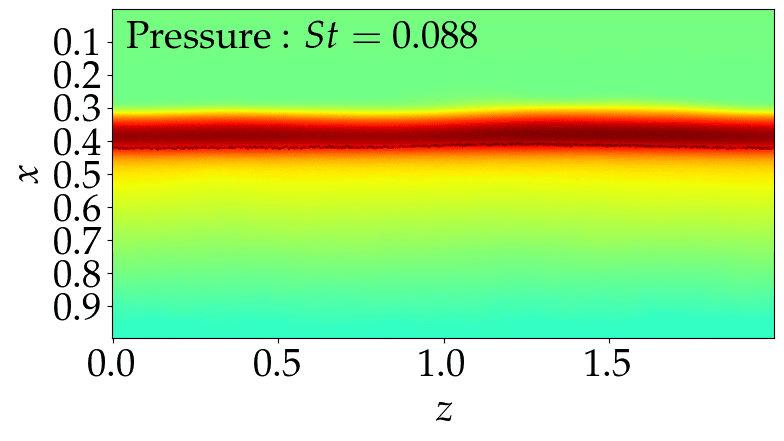}
\includegraphics[width=0.325\textwidth]{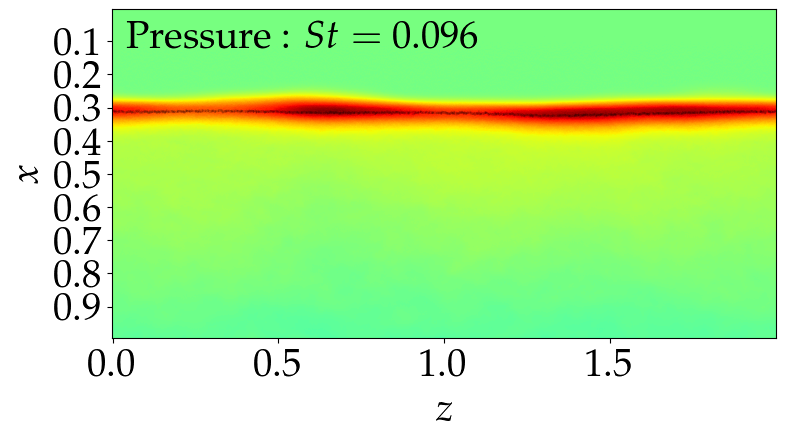}
\caption{SPOD shock-oscillation mode for pressure data on side $x$-$y$ plane at $z=1$ (top plots) and $x$-$z$ suction-side wall (bottom plots) for cases $\alpha = 5^{\circ}, 6^{\circ}, 7^{\circ}$ and $\myAR{2}$. Time-averaged sonic iso-line and separation line are plotted for side plane and wall visualisations, respectively.}
\label{fig:spod_p_bf_ar2}
\end{center}
\end{figure}

The SPOD modes for the dominant peak are plotted in Figure~\ref{fig:spod_p_bf_ar2} for both side ($x$-$y$) plane at $z=L_z/2$ (top plots) and ($x$-$z$, $y=0$) suction-side wall (bottom plots). The Strouhal numbers associated with these mode are $St=0.070$ ($\AoA{5}$), $St=0.088$ ($\AoA{6}$) and $St=0.096$ ($\AoA{7}$). Similar to previous SPOD analysis of transonic buffet in the literature \citep{moise_zauner_sandham_2022,Moise2023_AIAAJ,LongWong2024_Laminar_buffet}, the dominant mode is localised around the main shock-wave. For all angles of attack, this mode is essentially-2D and is associated to the chordwise convection of perturbations that are synchronised with the shock oscillations. The different ranges of shock excursion in the chordwise direction for the three angles of attack can be also seen, and is in agreement with the lift-coefficient oscillation amplitudes shown in Figure~\ref{fig:sectional_AoA567_AR2}. While the modes on the surface become slightly wavy in $z$ at the higher angles of attack, this mode remains in phase across the spanwise direction with the same sign and is essentially-2D (despite the presence of 3D features in the actual flow-fields at $\AoA{6}, \AoA{7}$, (Figure~\ref{fig:sectional_AoA567_AR2})). For all of the above reasons and to be consistent with the literature, we will refer to this mode as the 2D `shock-oscillation mode'. This shock-oscillation mode isolates the chord-wise shock oscillations, without revealing the 3D structures that are also present at $\AoA{6}, \AoA{7}$.

\begin{figure}
\begin{center}
\includegraphics[width=0.325\textwidth]{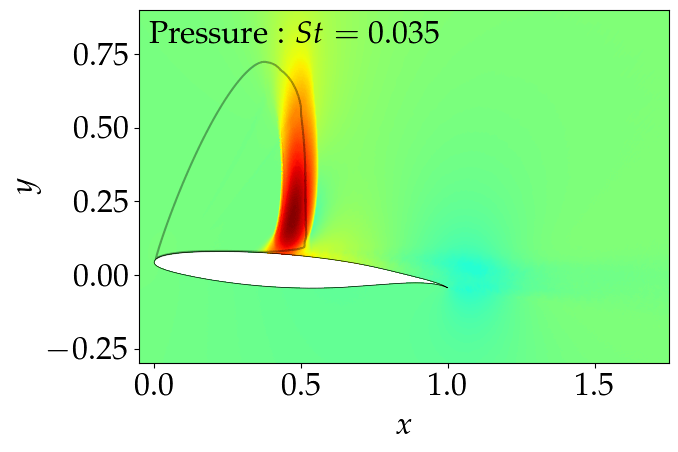}
\includegraphics[width=0.325\textwidth]{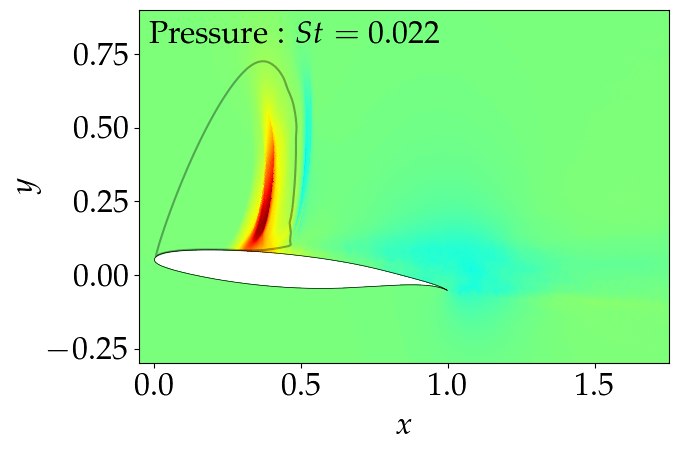}
\includegraphics[width=0.325\textwidth]{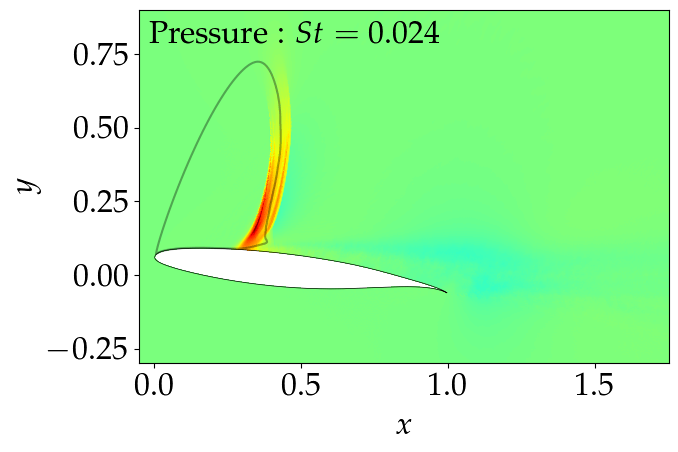}
\includegraphics[width=0.325\textwidth]{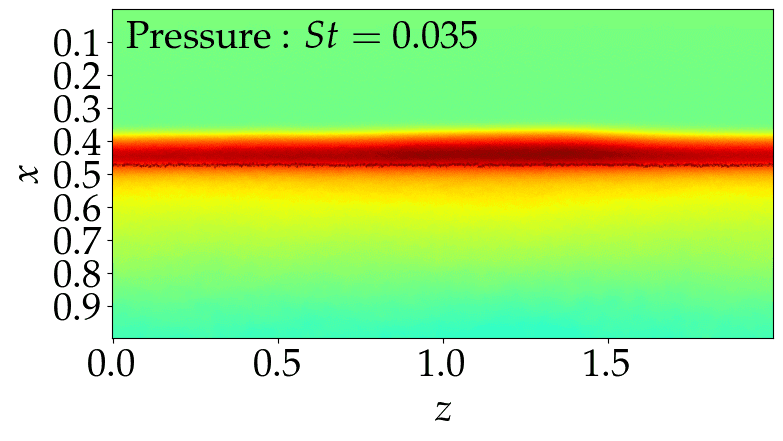}
\includegraphics[width=0.325\textwidth]{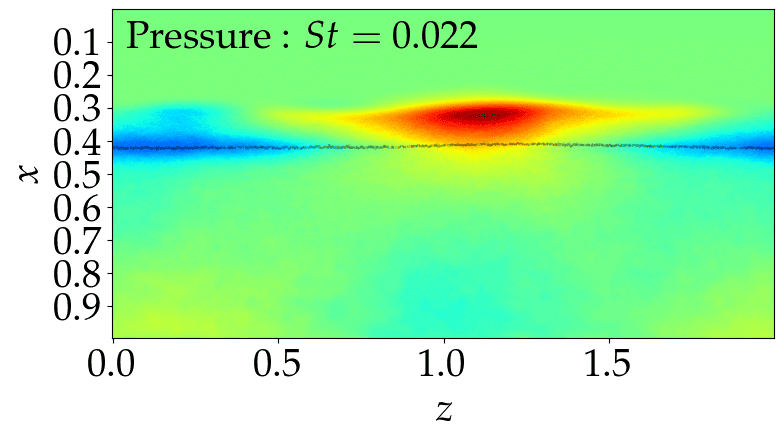}
\includegraphics[width=0.325\textwidth]{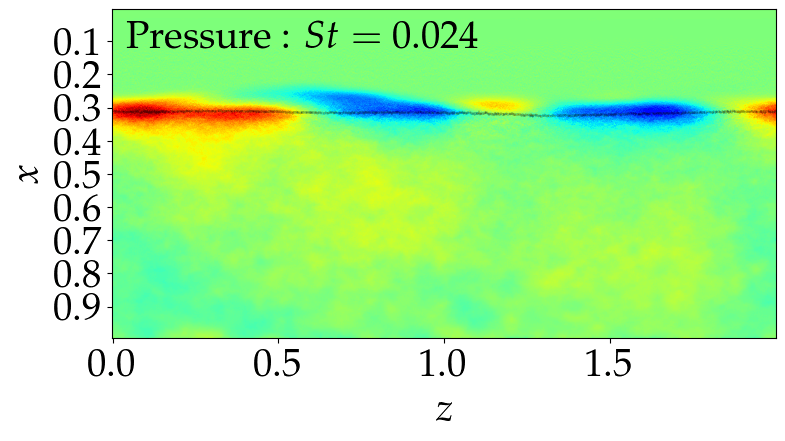}
\caption{SPOD low-frequency mode for pressure data on side $x$-$y$ plane at $z=1$ (top plots) and $x$-$z$ suction-side wall (bottom plots) for cases $\alpha = 5^{\circ}, 6^{\circ}, 7^{\circ}$ and $\myAR{2}$. Time-averaged sonic iso-line and separation line are plotted for side plane and wall visualisations, respectively.}
\label{fig:spod_p_lfm_ar2}
\end{center}
\end{figure}

To investigate the spanwise structure and arrangement of the observed three-dimensionality at certain conditions, the low-frequency SPOD modes corresponding to $St=0.035$ ($\AoA{5}$), $St=0.022$ ($\AoA{6}$) and $St=0.024$ ($\AoA{7}$) are plotted in Figure~\ref{fig:spod_p_lfm_ar2}. Low-frequency in this instance is defined relative to the 2D shock-oscillation mode as above. While at $\AoA{5}$ this mode is still essentially-2D, for $\AoA{6}$ and $\AoA{7}$ this breaks down and the SPOD shows the imprint of the 3D cellular patterns seen previously in the instantaneous flow-fields (Figure~\ref{fig:deg6_deg7_contours}). 

These 3D structures move irregularly along the shock front and can be referred to as buffet cells. As shown in the URANS calculations by \citep{IR2015,PDL2020,PDBLS2021}, in the absence of sweep the amplitude and convection of these cells is irregular, the spanwise location at which they appear is random and the number of buffet cells varies in time. While for swept wings the buffet cells frequency is usually in the $St=0.2-0.3$ range, \citep{PDL2020} showed that for unswept and low sweep-angle wings this frequency can be lower than the one associated with the dominant 2D shock oscillations. As can be seen from the SPOD modes calculated on the suction-side wall pressure data, the spanwise movement of these cellular perturbations leaves slanted bands reminiscent of the buffet-cell mode found in the global stability analyses by \citep{PBDSR2019,PDBLS2021} for swept wings. In contrast to the 2D shock oscillation mode (Figure~\ref{fig:spod_p_lfm_ar2}), the low-frequency 3D mode shows variations in the sign across the span at $\AoA{6}$ and $\AoA{7}$. The lower-AoA ($\AoA{5}$) mode, however, is coherent across the span and visually very similar to the purely 2D shock-oscillation mode shown in Figure~\ref{fig:spod_p_lfm_ar2}. In this sense, the appearance of this low-frequency 3D buffet cells mode seems to be linked to the loss of two-dimensionality observed for certain cases in the present study, as shown in Sections~\ref{sec:deg5_intro} and \ref{sec:deg6_deg7_AR2}. 

\begin{figure}
\begin{center}
\includegraphics[width=0.325\textwidth]{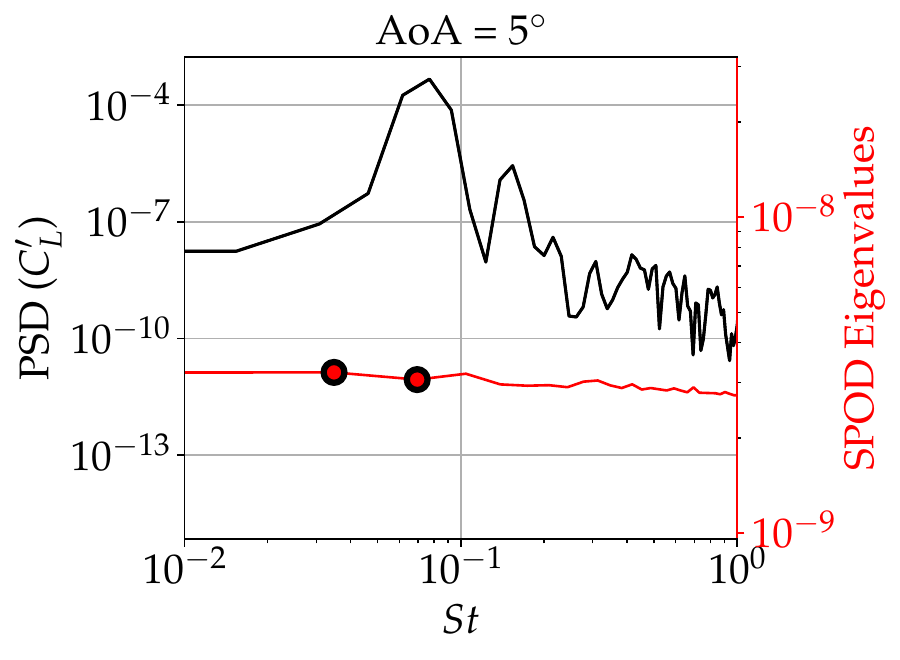}
\includegraphics[width=0.325\textwidth]{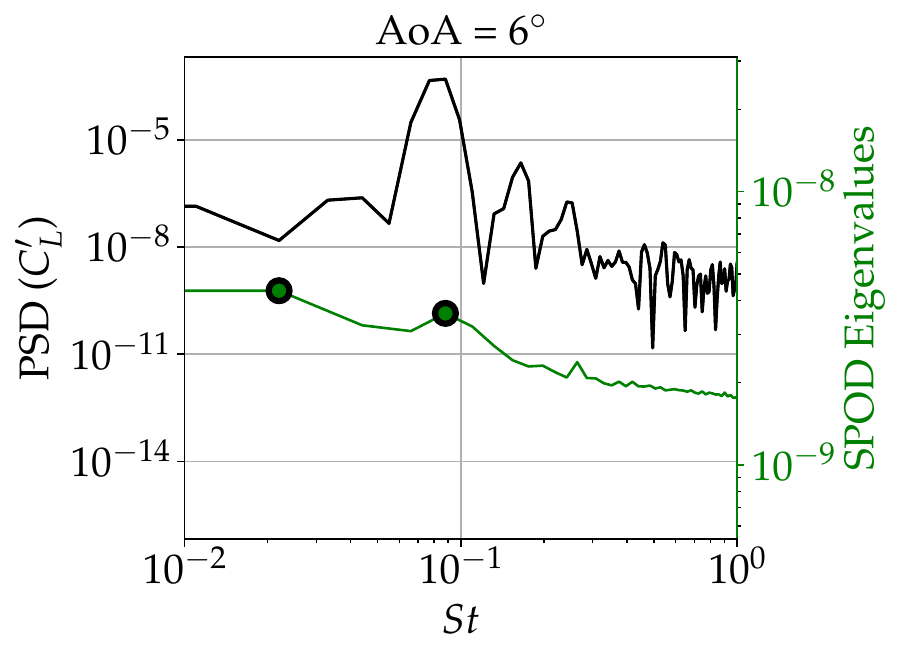}
\includegraphics[width=0.325\textwidth]{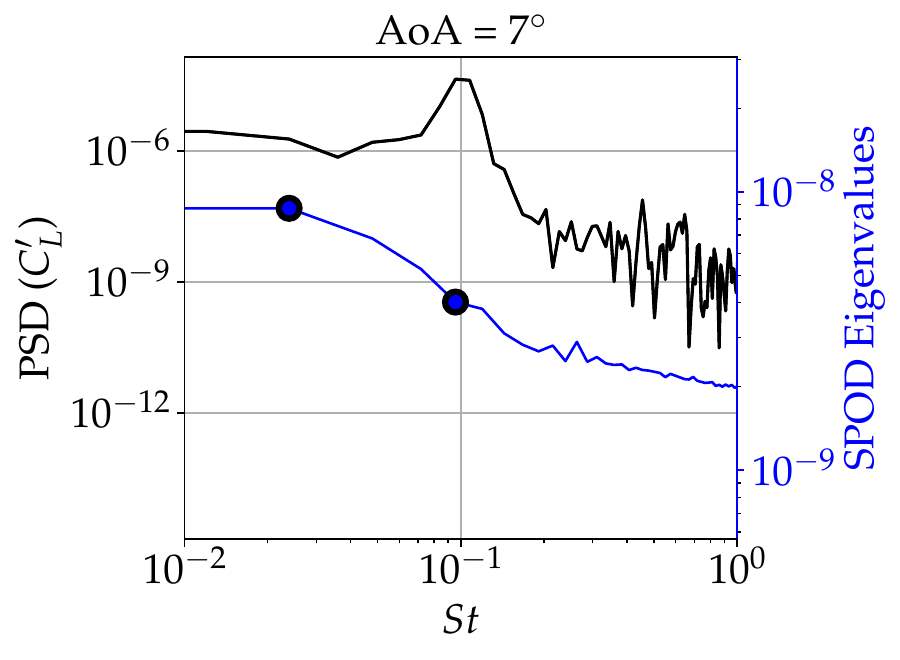}
\includegraphics[width=0.325\textwidth]{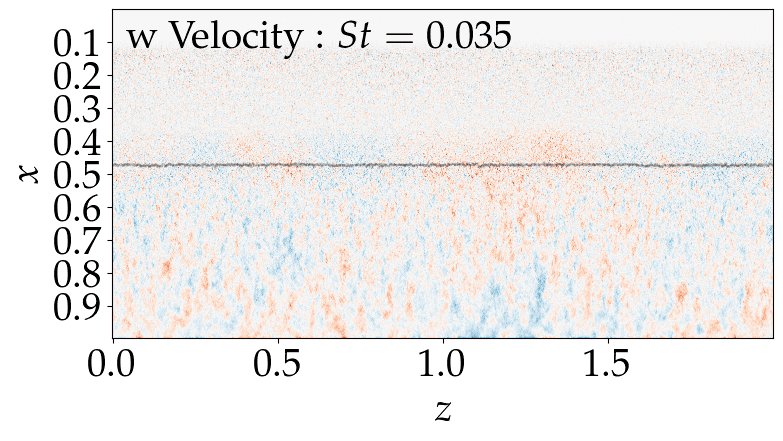}
\includegraphics[width=0.325\textwidth]{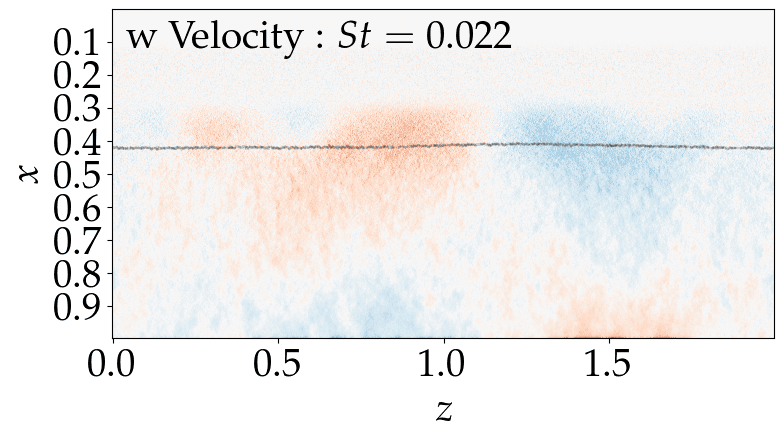}
\includegraphics[width=0.325\textwidth]{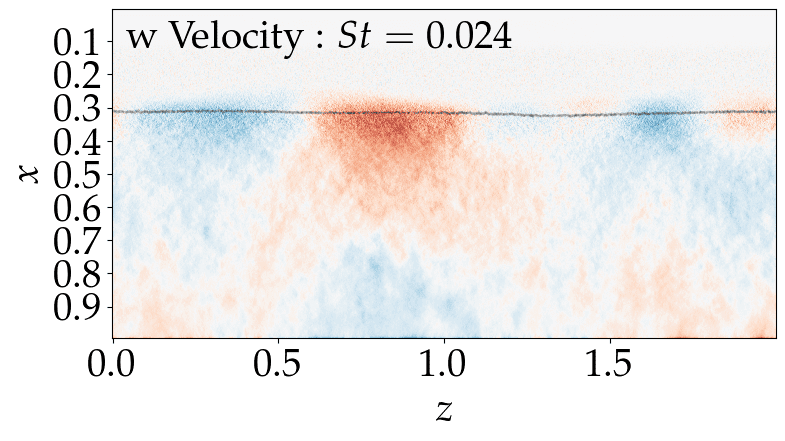}
\includegraphics[width=0.325\textwidth]{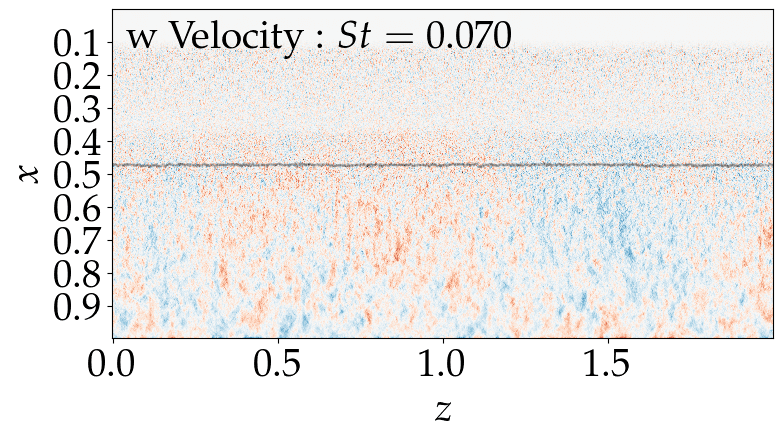}
\includegraphics[width=0.325\textwidth]{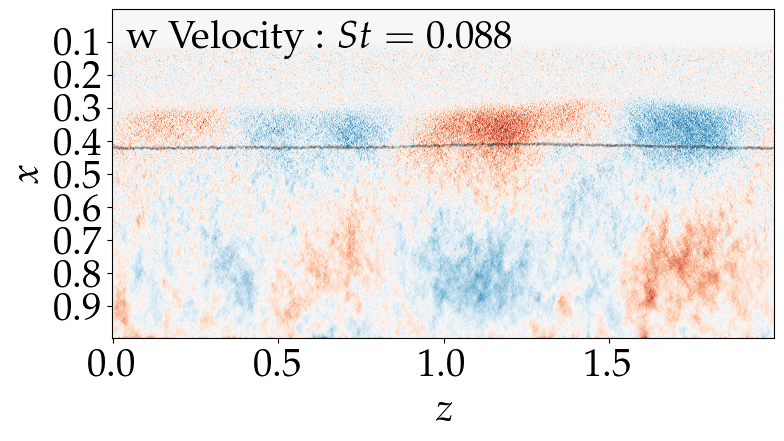}
\includegraphics[width=0.325\textwidth]{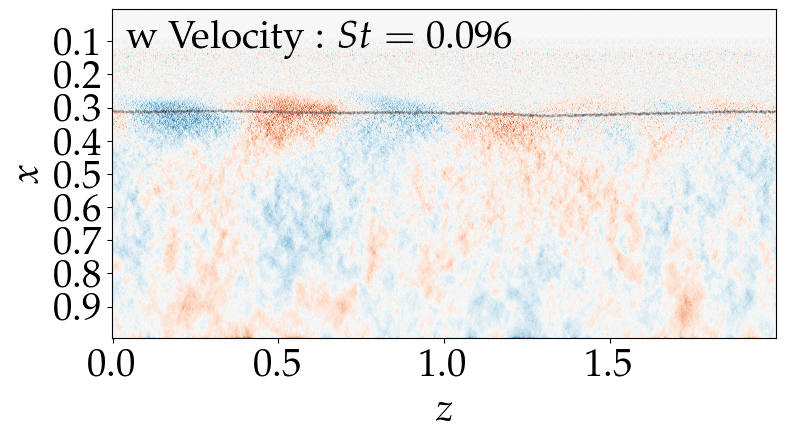}
\caption{SPOD eigenvalue spectra (coloured solid lines) for spanwise velocity data on a $x$-$z$ suction-side near-wall plane (top plots) are plotted with the PSD of lift-coefficient fluctuations (black solid lines) for cases $\alpha = 5^{\circ}, 6^{\circ}, 7^{\circ}$ and $\myAR{2}$. Circles correspond to the SPOD low-frequency (middle plots) and 2D shock-oscillation (bottom plots) modes plotted. Time-averaged sonic iso-line and separation line are plotted for side plane and wall visualisations, respectively.}
\label{fig:spod_w_lfm-bf_ar2}
\end{center}
\end{figure}

In agreement with the GSA studies by \citep{CGS2019,PBDSR2019,PDBLS2021}, buffet cells modes are admitted also for unswept wings. However, these have been reported to become unstable concomitantly with the 2D shock-oscillation modes. To understand whether this is in contradiction with the simulations presented in the current study at $\alpha=5^\circ$ (for which only 2D unsteady modes are present), the SPOD analysis is repeated on spanwise-velocity data at the first grid point above the suction-side wall. This quantity is expected to be more sensitive to 3D effects as, unlike pressure or other thermodynamic state variables tested (not shown for brevity), the $w$-velocity contains additional directional information about the fluid propagation as seen in Figure~\ref{fig:deg6_surface_w_contours}.

For all angles of attack, Figure~\ref{fig:spod_w_lfm-bf_ar2} shows SPOD spectra and modes at the same 2D and 3D mode frequencies visualised previously. The first thing to note is that the SPOD spectrum appears very flat at $\AoA{5}$ and both low-frequency and shock-oscillation modes have similar energy content. For the $\AoA{6}$ and $\AoA{7}$ cases, the low-frequency mode becomes dominant, possibly causing the appearance of the 3D separation structures present in the flow. It is also interesting to see that for all cases, both low-frequency and shock-oscillation modes are no longer strictly two-dimensional. This is surprising especially for the $\AoA{5}$ case, where the main flow dynamics has been shown to be essentially 2D in the previous sections. In agreement with the discussion of \citet{JMDMS2009}, this might be an indication that while 3D effects may be present, until the velocity associated with the shock motion in the streamwise direction is much larger than the velocity associated with the 3D structures, the shock-oscillation related dynamics remain essentially 2D. In this sense, marginal 3D effects may be present, as depicted by the aforementioned global stability analysis studies that predict the appearance of 2D and 3D unstable modes concomitantly. However, these 3D effects might be near the onset and extremely weak, hidden under the prevailing 2D mechanisms of the non-linearly saturated shock-oscillations.

% \subsubsection{Spectral Proper Orthogonal Decomposition (SPOD) $\myAR{3}$}
\begin{figure}
\begin{center}
\includegraphics[width=0.325\textwidth]{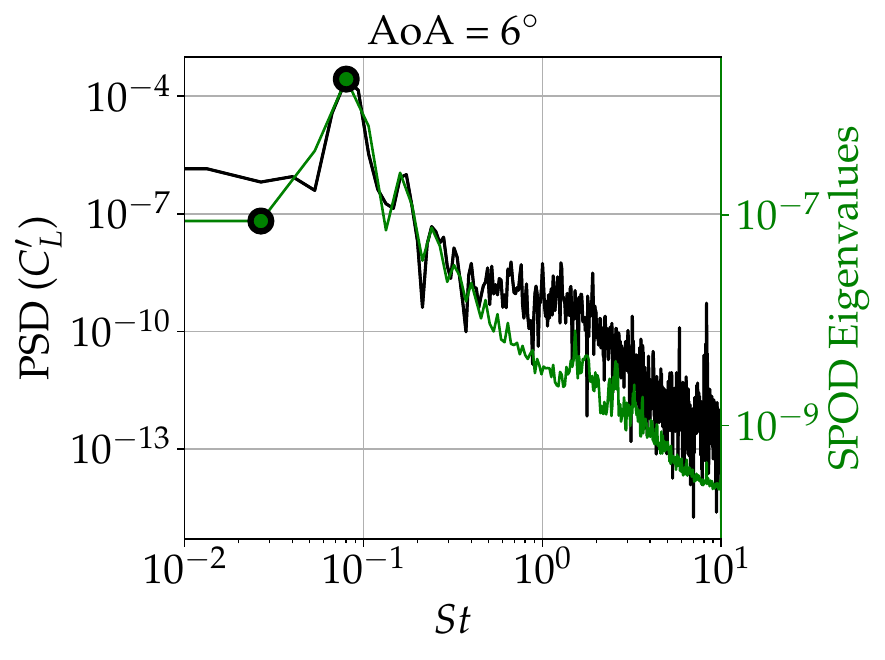}
\includegraphics[width=0.325\textwidth]{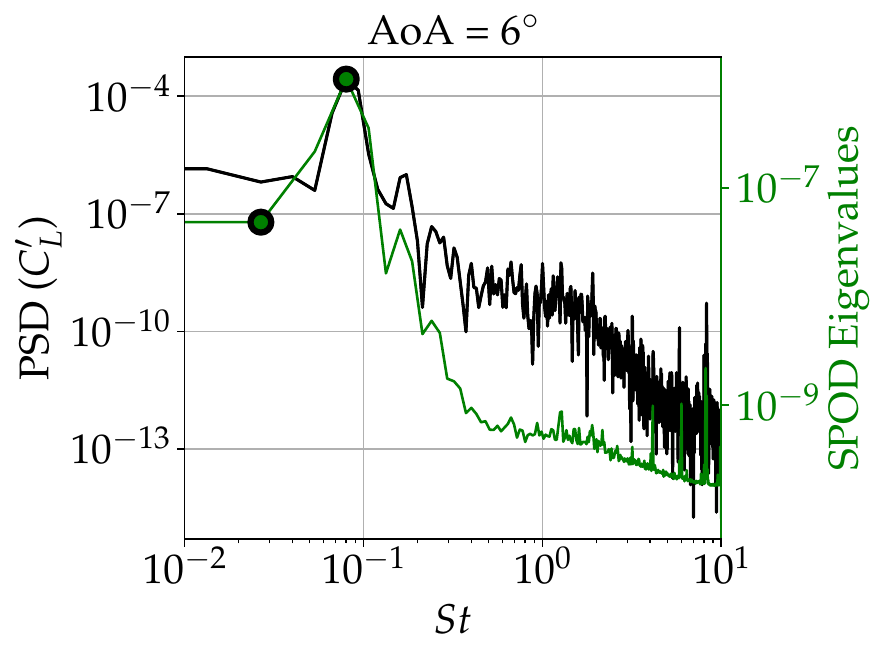}
\includegraphics[width=0.325\textwidth]{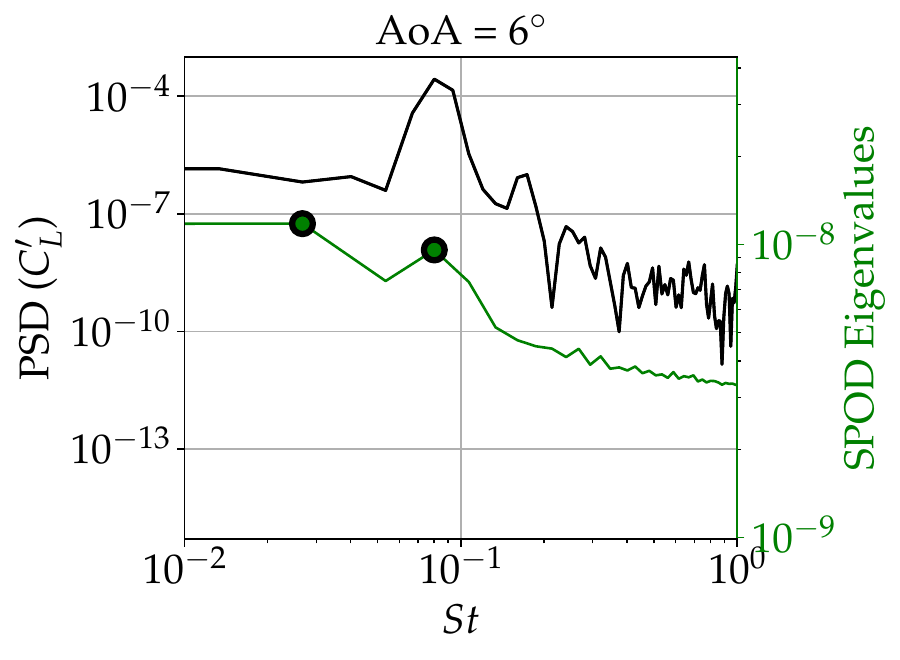}
\caption{SPOD eigenvalue spectra (coloured solid lines) for pressure (left and middle plots) and spanwise velocity (right plot) data on side $x$-$y$ plane at $z=1.5$ (left plot) and $x$-$z$ suction-side wall (middle and right plots) are shown with the PSD of lift-coefficient fluctuations (black solid lines) for case $\alpha = 6^{\circ}$ and $\myAR{3}$. Circles correspond to the selected modes for visualization.}
\label{fig:spod_spectra_ar3}
\end{center}
\end{figure}

\begin{figure}
\begin{center}
\includegraphics[width=0.49\textwidth]{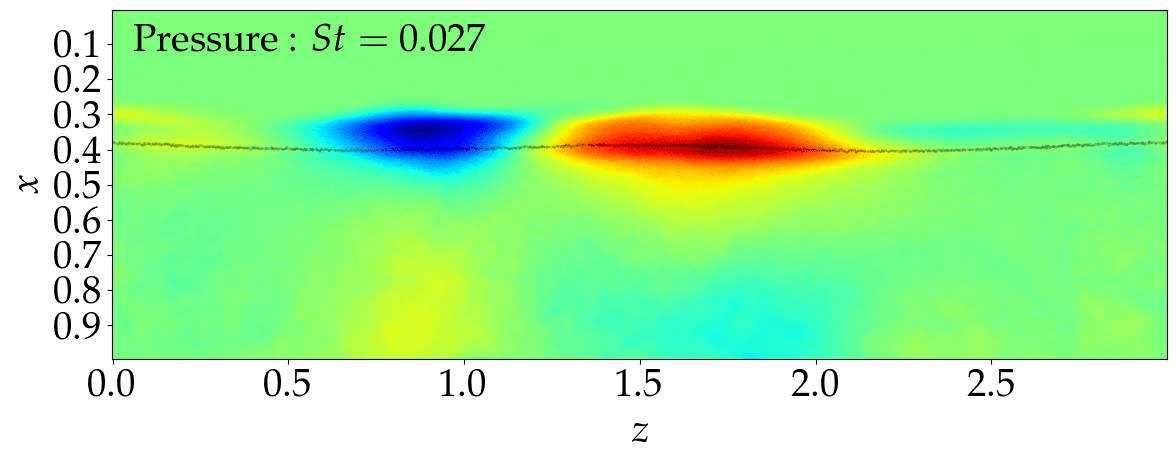}
\includegraphics[width=0.49\textwidth]{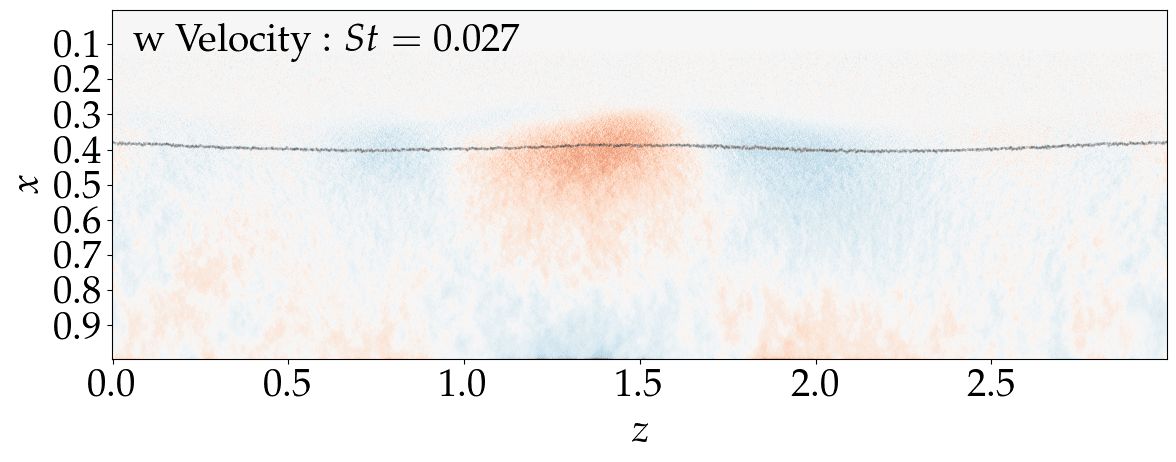}
\includegraphics[width=0.49\textwidth]{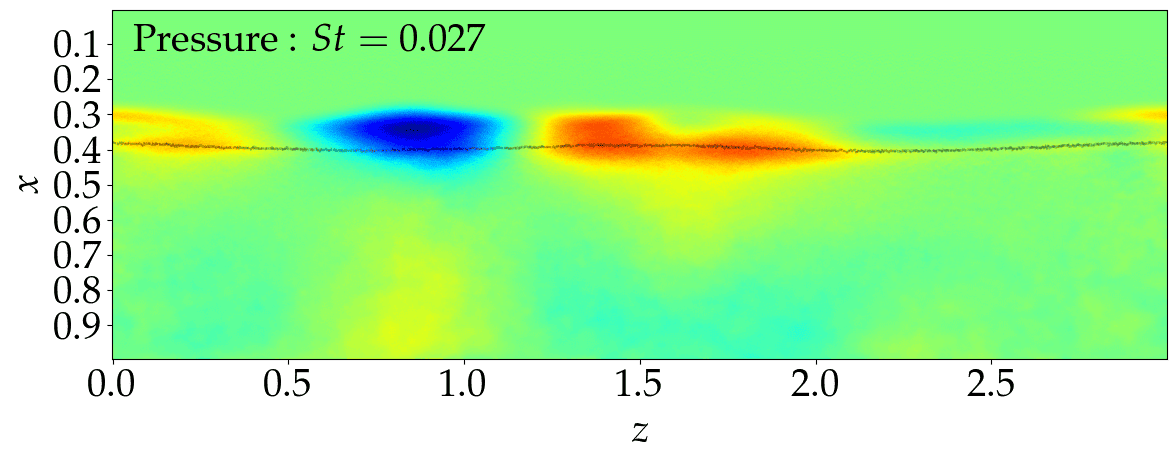}
\includegraphics[width=0.49\textwidth]{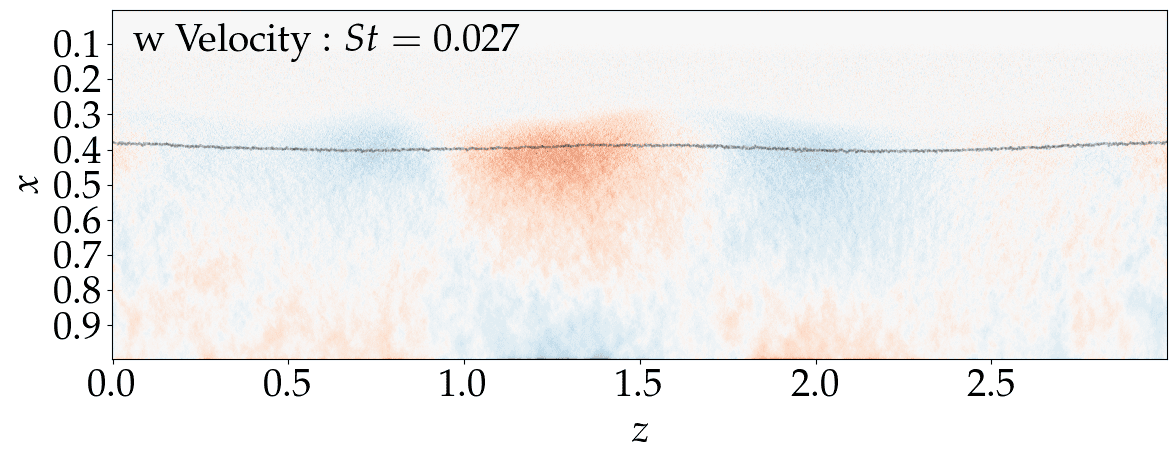}
\includegraphics[width=0.49\textwidth]{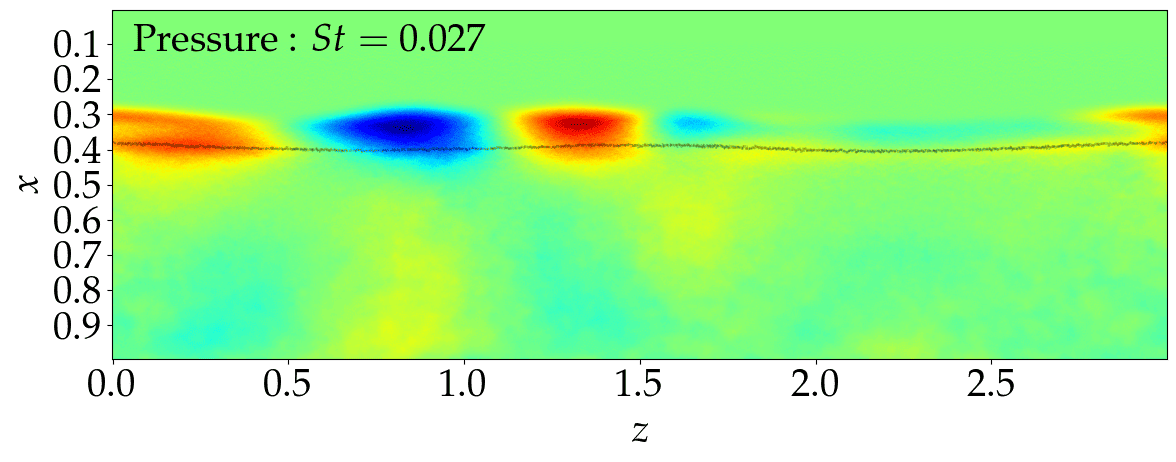}
\includegraphics[width=0.49\textwidth]{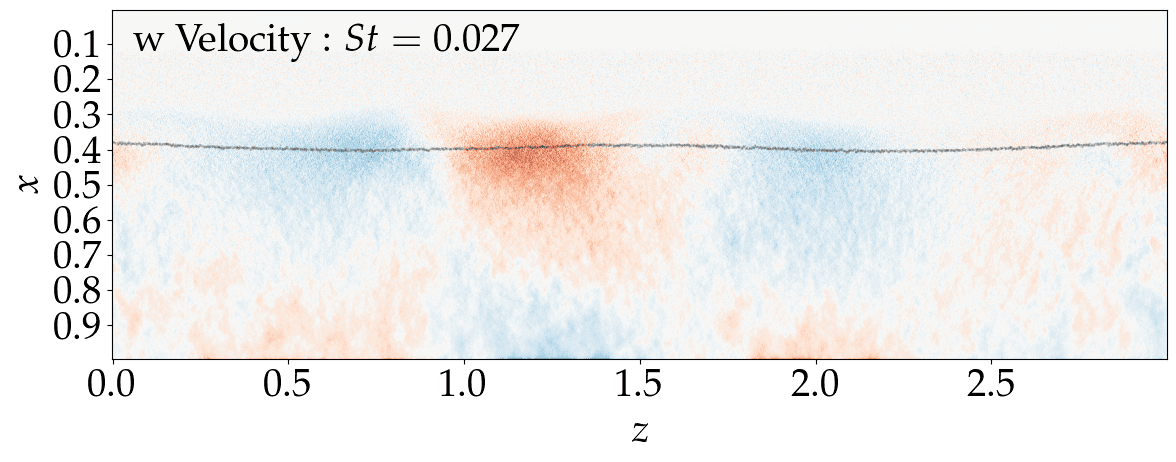}
\includegraphics[width=0.49\textwidth]{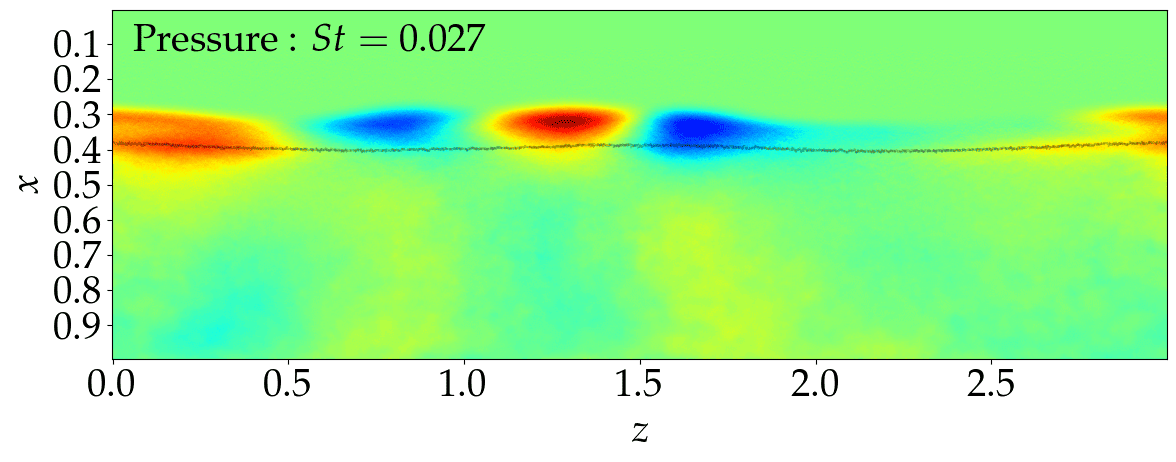}
\includegraphics[width=0.49\textwidth]{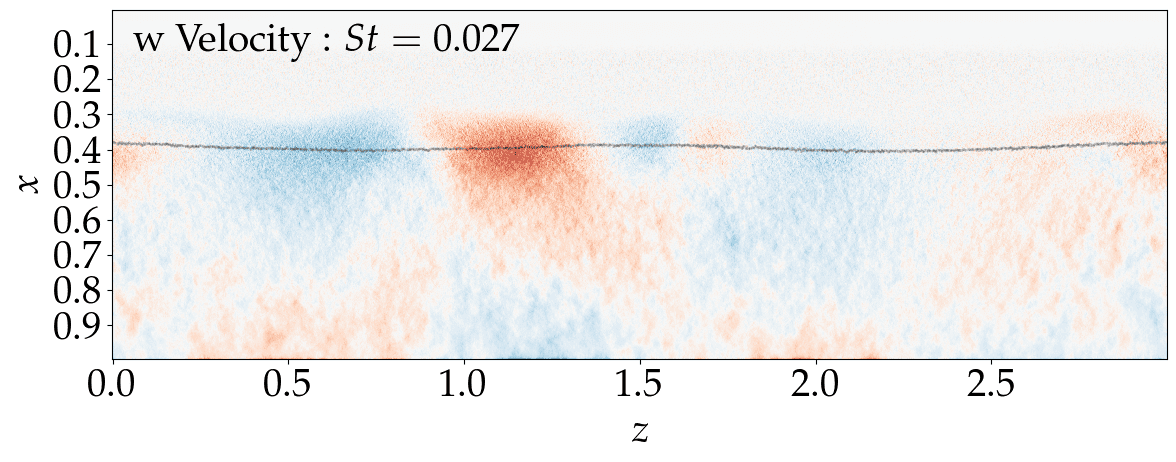}
\caption{SPOD low-frequency mode for pressure (left plots) and spanwise velocity (right plots) data on the $x$-$z$ suction-side wall / near-wall plane for case $\alpha = 6^{\circ}$ and $\myAR{3}$. Rows correspond to $t/T_{period}=\pi/8$ (first row), $\pi/4$ (second row), $3\pi/8$ (third row) and $\pi/2$ (fourth row) time instances. Time-averaged separation line are also plotted.}
\label{fig:spod_w_lfm_ar3}
\end{center}
\end{figure}

\begin{figure}
\begin{center}
\includegraphics[width=0.49\textwidth]{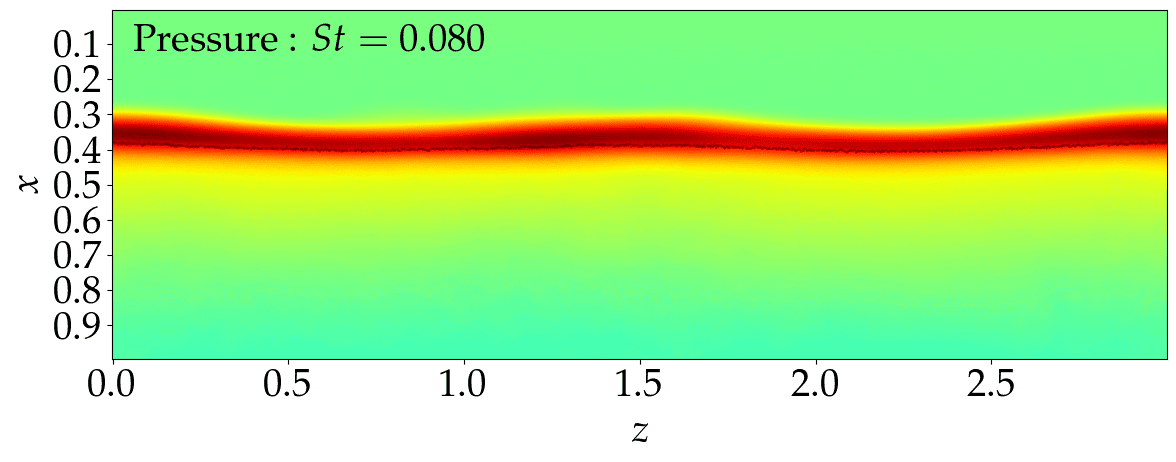}
\includegraphics[width=0.49\textwidth]{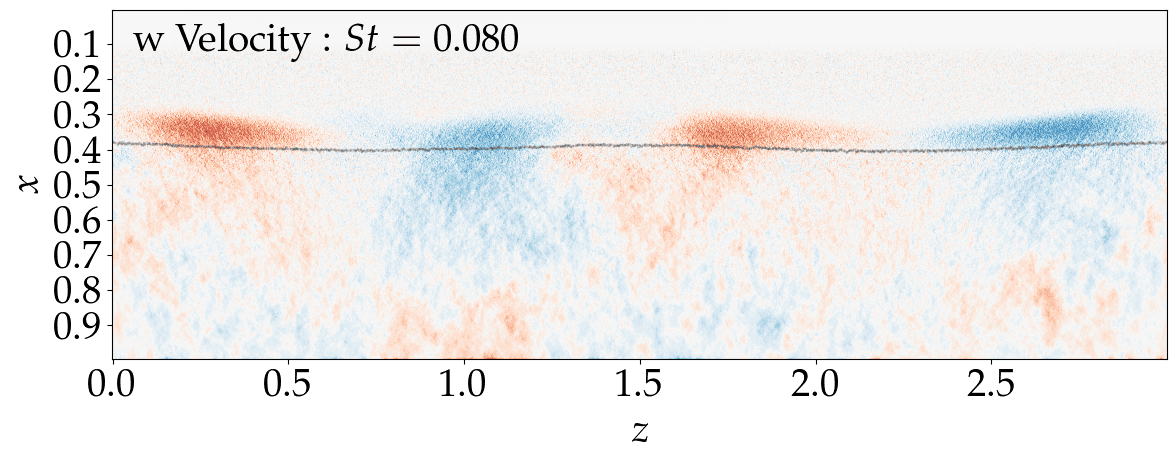}
\includegraphics[width=0.49\textwidth]{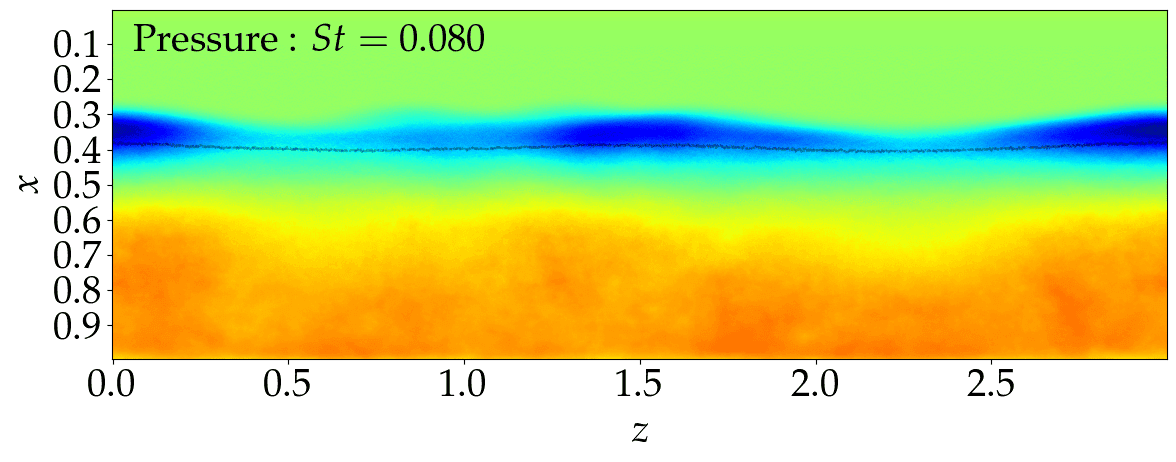}
\includegraphics[width=0.49\textwidth]{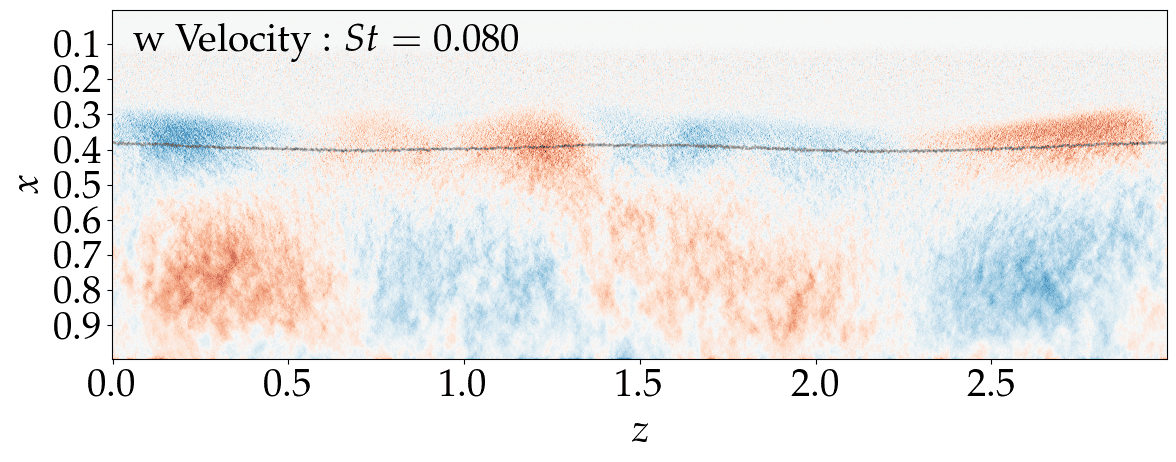}
\includegraphics[width=0.49\textwidth]{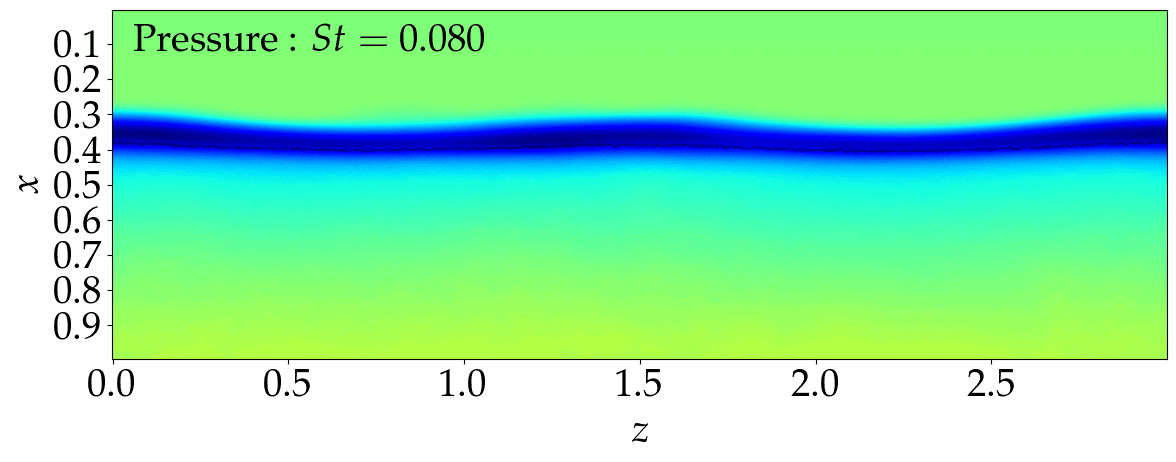}
\includegraphics[width=0.49\textwidth]{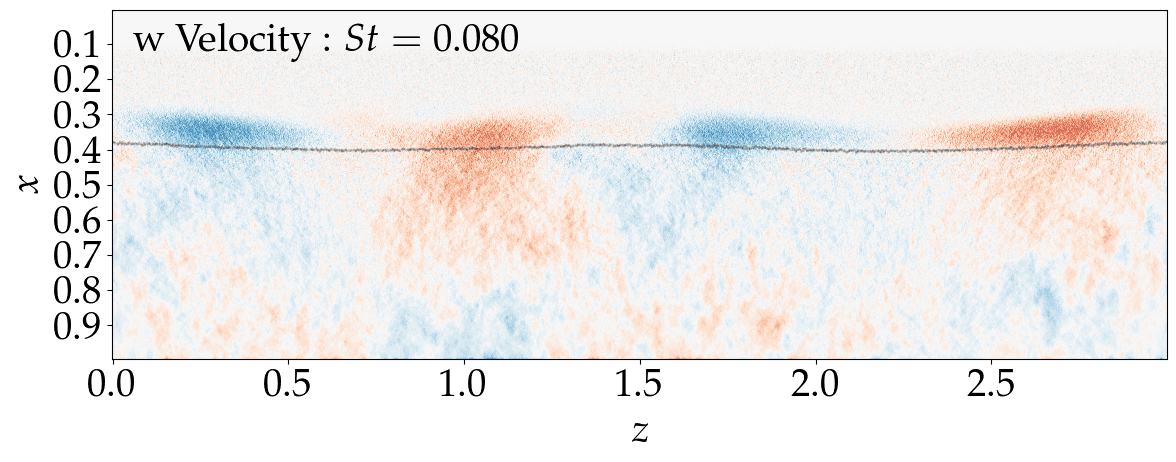}
\includegraphics[width=0.49\textwidth]{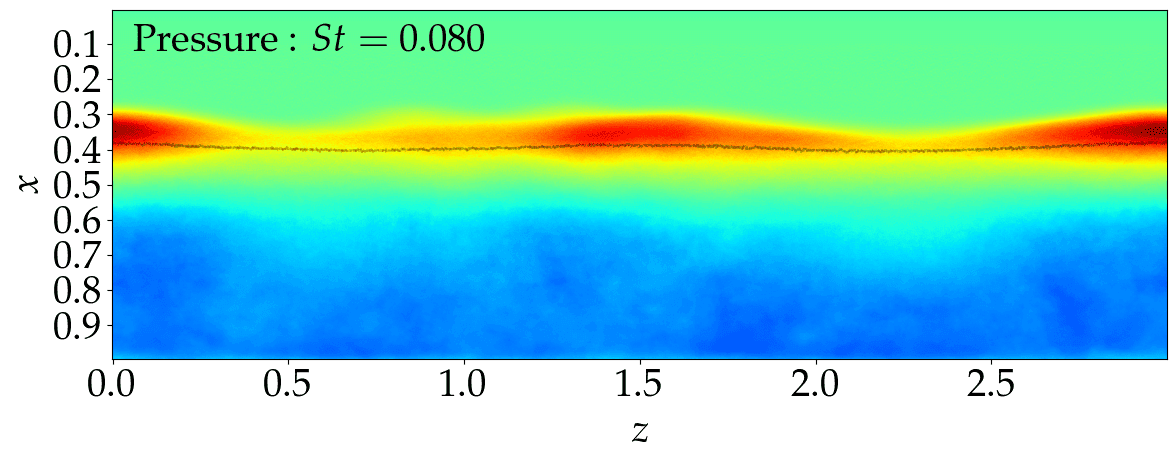}
\includegraphics[width=0.49\textwidth]{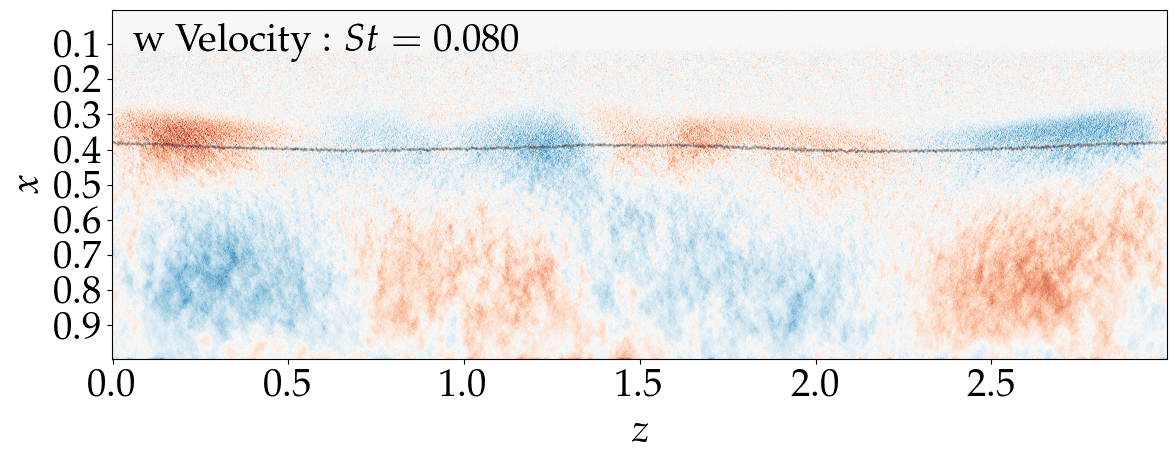}
\caption{SPOD shock-oscillation mode for pressure (left plots) and spanwise velocity (right plots) data on the $x$-$z$ suction-side wall / near-wall plane for case $\alpha = 6^{\circ}$ and $\myAR{3}$. Rows correspond to $t/T_{period}=\pi/2$ (first row), $\pi$ (second row), $3\pi/2$ (third row) and $2\pi$ (fourth row) time instances. Time-averaged separation line are also plotted.}
\label{fig:spod_w_bm_ar3}
\end{center}
\end{figure}

To gain further insight into the aspect ratio dependence and phase behaviour of the low-frequency 3D modes, the SPOD analysis is repeated for the widest $\myAR{3}$ case at $\AoA{6}$. Focusing only on the established 3D effects, the analysis at $\myAR{3}$ is discussed only in the context of the suction-side wall/near-wall data sets on both pressure and spanwise-velocity as above. Firstly, the SPOD spectral content shown in Figure~\ref{fig:spod_spectra_ar3} generally agrees well with the characteristics observed between the side and surface views for the same angle of attack at the lower aspect ratio of $\myAR{2}$. To observe the dynamics at different phase angles, the low-frequency ($St=0.027$) and shock-oscillation modes ($St=0.080$) are reconstructed over a quarter and one period ($T_{period}$) of their corresponding frequencies, respectively. For the low-frequency mode, four time instances at equispaced $\pi / 8$ intervals are reported in Figure~\ref{fig:spod_spectra_ar3}. For the shock-oscillation mode, four snapshots at $\pi / 2$ intervals are given in Figure~\ref{fig:spod_w_lfm_ar3}. The low-frequency mode still shows 3D cellular patterns that are convected (in this case from right to left) along the shock front. As shown in the GSA findings \citep{CGS2019,PBDSR2019,PDBLS2021}, different spanwise wavelengths can emerge as dominant depending on the aspect ratio considered. The shock-oscillation mode is essentially 2D for the pressure-based SPOD modes, although some spanwise modulation is more noticeable than in the equivalent modes at $AR=2$ (Figure~\ref{fig:spod_p_lfm_ar2}). For the SPOD analysis based on spanwise-velocity, the shock-oscillation mode is again 3D and sub-dominant with respect to the low-frequency mode.

% END OF SPOD
\section{Further discussion and conclusions}\label{sec:conclusions}
Wide-span $\left(1 \leq AR \leq 3\right)$ turbulent transonic buffet has been investigated for the first time with high-fidelity scale-resolving simulations. The Implicit Large Eddy Simulations (ILES) were performed for a freestream Mach number of $M_{\infty} = 0.72$ at a moderate Reynolds number of $Re=5 \times 10^5$ on periodic (infinite, unswept) configurations of the supercritical NASA-CRM wing geometry, which was tripped to turbulence. The aspect ratios studied here are between 20 to 60 times wider than typically used for high-fidelity buffet studies ($\AR=0.0365 - 0.073$ \citep{GD2010}, $\AR=0.065$ \citep{FK2018,Nguyen2022}, $\AR=0.05$ \citep{moise_zauner_sandham_2022,Moise2023_AIAAJ}, and $\AR=0.25$ \citep{LongWong2024_Laminar_buffet}). Building upon our previous recent work on this configuration \citep{LSH2024_narrow_buffet} which investigated domain sensitivity of the two-dimensional buffet phenomenon on narrow-to-moderate aspect ratios $\left(0.025 \leq \AR \leq 0.5\right)$, this study instead focused on domain widths expected to be wide enough to observe three-dimensional buffet effects $\left(1 \leq \AR \leq 3\right)$. Initial results at $\AoA{5}$ and $\myAR1,2$ were cross-validated against low-fidelity URANS predictions, with excellent agreement observed for both aerodynamic forces and buffet frequencies. At a moderate AoA of $\AoA{5}$ with mostly attached flow (in a time-averaged sense), buffet was found to remain essentially 2D with no spanwise modulation observed. The low-frequency buffet oscillations agreed well with narrow-span ILES predictions \citep{LSH2024_narrow_buffet} and URANS. Widening the domain from $\myAR{1}$ to $\myAR{2}$ had no effect on the main aerodynamic quantities and buffet characteristics (Figure~\ref{fig:AR1_AR2_deg5}). A very-wide URANS at $\myAR{6}$ confirmed this quasi-2D buffet behaviour at $\AoA{5}$ (Figure~\ref{fig:URANS_AR6}).

However, at higher angles of attack ($\AoA{6}, \AoA{7}$), significant 3D buffet effects were observed, which persisted through multiple low-frequency buffet cycles. The 3D effects were similar to those found in lower-fidelity RANS-based studies in the literature \citep{IR2015, PBDSR2019, PDBLS2021}. In addition to the chord-wise 2D shock oscillations found at lower AoA, large cellular 3D separation bubbles were observed on the suction side of the wing once the angle of incidence was raised to $\AoA{6}, \AoA{7}$ (Figure~\ref{fig:deg6_deg7_contours}). The 3D separations were primarily localised to the shock-foot region and result in long-wavelength spanwise perturbations of the shock front. The three-dimensionality was observed to occur for aspect ratios of $\myAR{1}$ and above, with wavelengths in the range $\lambda = 1-1.5$, depending on the applied aspect ratio. Due to the absence of imposed sweep angle ($\Lambda = 0^{\circ}$), the buffet/stall-cells were observed to be irregular and intermittent in their spanwise location and strength between consecutive buffet cycles (Figure~\ref{fig:ZT_deg5_deg6_deg7_AR2}). The three-dimensionality at the shock front was found to be linked to the level of flow separation present, with the strongest 3D effects seen during low-lift phases of the buffet cycle where the shock reaches its farthest upstream position and the flow is highly-separated. 

Span- and time-averaged aerodynamic quantities showed minimal sensitivity to the appearance of the three-dimensional structures, generally agreeing well to 2D predictions. In contrast, sectional evaluation of the same quantities at different individual spanwise stations showed large deviations from span-averaged values. These deviations were only observed for the cases showing three-dimensionality, with the sectional evaluations for the quasi-2D cases at $\AoA{5}$ largely agreeing well with the span-averaged result. The three-dimensionality was found to be located mainly at the main shock-wave, as demonstrated by the peaks in the sectional evaluation of aerodynamic quantities in Figure~\ref{fig:sectional_AoA567_AR2} and Figure~\ref{fig:sectional_deg6_AR3}. For a fixed AoA of $\AoA{6}$, increasing the aspect ratio from $\myAR{2}$ to $\myAR{3}$ modified the wavelength of the cellular separations, increased the number of buffet/stall-cells accommodated by the span-wise width, and strengthened the three-dimensionality (Figure~\ref{fig:sectional_AoA567_AR2}, Figure~\ref{fig:sectional_deg6_AR3}). Instantaneous spanwise velocity contours above the suction-side wall (Figure~\ref{fig:deg6_surface_w_contours}) showed the buffet/stall-cell separations appear as perturbations along the shock-front, with left- and right-moving fluid either side of the saddle point. The 3D effects were mostly seen during low-lift buffet phases where the separation reaches a maximum. The configuration reverts to quasi-2D topologies during high-lift phases as the shock moves downstream and the flow reattaches. 

Cross-correlation analysis was performed on the wide-span aerofoil data to detect coherent large-scale structures present within the flow due to 3D buffet. The correlation approach was applied both to the shocked region of the aerofoil $(0.25 < x < 0.6)$, and the remainder of the chord length $(0.6 < x < 1)$, separately. While there exists time instances where the flow is essentially all uncorrelated turbulence, at other phases of the buffet cycle we observed strong patches of correlation and anti-correlation for the cases showing 3D buffet effects. These were found to occur for length scales on the order of $\lambda_z = L_z/2$, consistent with the observed instantaneous flow structures. Tracking the evolution of the maximum anti-correlation in time showed repeated phases of three-dimensionality overlaid on the main 2D buffet oscillations. The number of 3D structures identified by the correlation metric varied between cycles, highlighting the irregular and intermittent nature of 3D buffet on unswept infinite wings. The appearance of these separation cells was related to the phases of the aerodynamic lift and skin-friction drag. Meanwhile, as observed in the other sections, the $\AoA{5}$ case at buffet conditions was found to remain essentially-2D, with only very minor levels of correlation observed in the region downstream of the shock-wave.

To further analyse the structure and frequency content of the three-dimensional structures, a modal SPOD method was applied to cases at $\myAR{2}$ and $\myAR{3}$. For all cases considered, a 2D shock-oscillation mode was observed both when considering side-views and surface-pressure data. The Strouhal numbers associated with this mode occurred in the range $St = \left[0.07, 0.1\right]$, consistent with those commonly reported in the buffet literature \citep{FK2018,moise_zauner_sandham_2022,Moise2023_AIAAJ}. Higher frequency harmonics and wake modes ($0.3<St<5$, \citep{Moise2023_AIAAJ,LongWong2024_Laminar_buffet}) were also present. The 2D shock-oscillation mode from SPOD was shown to remain essentially-2D even in the presence of the strong 3D separation effects demonstrated throughout this work. Instead, the separated 3D structures were associated with SPOD modes occurring at frequencies ($St \sim 0.002 - 0.004$), lower than that of the 2D shock-oscillation mode. This finding is in good agreement with the URANS-based analysis of \cite{PDL2020}, who showed that the frequency of the 3D mode tends towards lower-frequencies in the unswept ($\Lambda = 0^{\circ}$) limit relevant here.

Further SPOD analysis of the spanwise velocity component ($w$) instead of pressure showed that, while the $\AoA{5}$ case was observed to remain essentially two-dimensional in the instantaneous flow-fields and pressure-based SPOD modes, the SPOD mode on $w$ did show weak traces of three-dimensionality at the same scale as expected for buffet-cell phenomena ($\lambda = 1-1.5$). This subtlety could be a possible explanation as to why GSA-based studies identify the onset of 2D- and 3D-buffet occurring at the same conditions, whereas we demonstrate with high-fidelity wide-span simulations that 2D shock-oscillations can be active on infinite-wings at transonic buffet conditions without noticeable 3D buffet/stall-cell phenomena present. Analysis of the phase-dependence of the SPOD surface modes showed spanwise convection occurring only for the low-frequency 3D mode, and not for the 2D shock-oscillation one. Future work of ILES on swept configurations is required to assess whether or not the irregular 3D separation patterns observed here become regular buffet-cells with a fixed convection velocity as the sweep angle is increased. This will be the topic of a future study on the same configuration.

\backsection[Supplementary data]{\label{SupMat}Supplementary material and movies are available at \\https://doi.org/10.1017/jfm.2019...}

\backsection[Acknowledgements]{Computational time was provided by the JAXA JSS3 supercomputing facility and associated support staff, and the Fugaku supercomputer at RIKEN on projects hp220195, hp220226.}

\backsection[Funding]
{Dr. David J. Lusher is funded by the Japan Society for the Promotion of Science (JSPS), on a postdoctoral fellowship awarded to the JAXA Chofu Aerospace Center. Additional funding was provided by a JSPS KAKENHI grant award (22F22059).}

\backsection[Declaration of interests]{The authors report no conflict of interest.}

\backsection[Data availability statement]{The data that support the findings of this study are openly available in [repository name] at http://doi.org/[doi], reference number [reference number]. See JFM's \href{https://www.cambridge.org/core/journals/journal-of-fluid-mechanics/information/journal-policies/research-transparency}{research transparency policy} for more information}

\backsection[Author ORCIDs]{D.J. Lusher, https://orcid.org/0000-0001-8874-5290; A. Sansica, https://orcid.org/0000-0003-1950-2636; M. Zauner, https://orcid.org/0000-0002-6644-2990; A. Hashimoto, https://orcid.org/0000-0003-4428-0406;}

\backsection[Author contributions]{All authors that made contributions are present. \textbf{Writing}: D.L. and A.S. wrote the original manuscript. M.Z., and A.H. contributed to revisions of the manuscript. \textbf{Simulations}: D.L. performed all of the ILES simulations. A.S. performed the URANS simulations, M.Z. assisted with URANS setup. \textbf{Analysis}: A.S. and D.L. implemented the modal SPOD using the PySPOD library, M.Z. and D.L. implemented the two-point correlations. \textbf{Software}: D.L. is the lead developer of the OpenSBLI solver (University of Southampton \& JAXA), A.H. and A.S. contributed to the development of the FaSTAR solver (JAXA). \textbf{Meshes}: A.S. created the meshes in Pointwise. \textbf{Funding}: D.L., A.S. and A.H. obtained project funding from the Japan Society for the Promotion of Science (JSPS). All authors contributed to discussions, proposals for computational time, and post-processing tools.}

% \appendix

\appendix{}\label{Appendix}

\section{URANS up to $AR=6$}\label{sec:URANS_AR6}
%% URANS AR=6 comparison
\begin{figure}
\begin{center}
\includegraphics[width=0.7\textwidth]{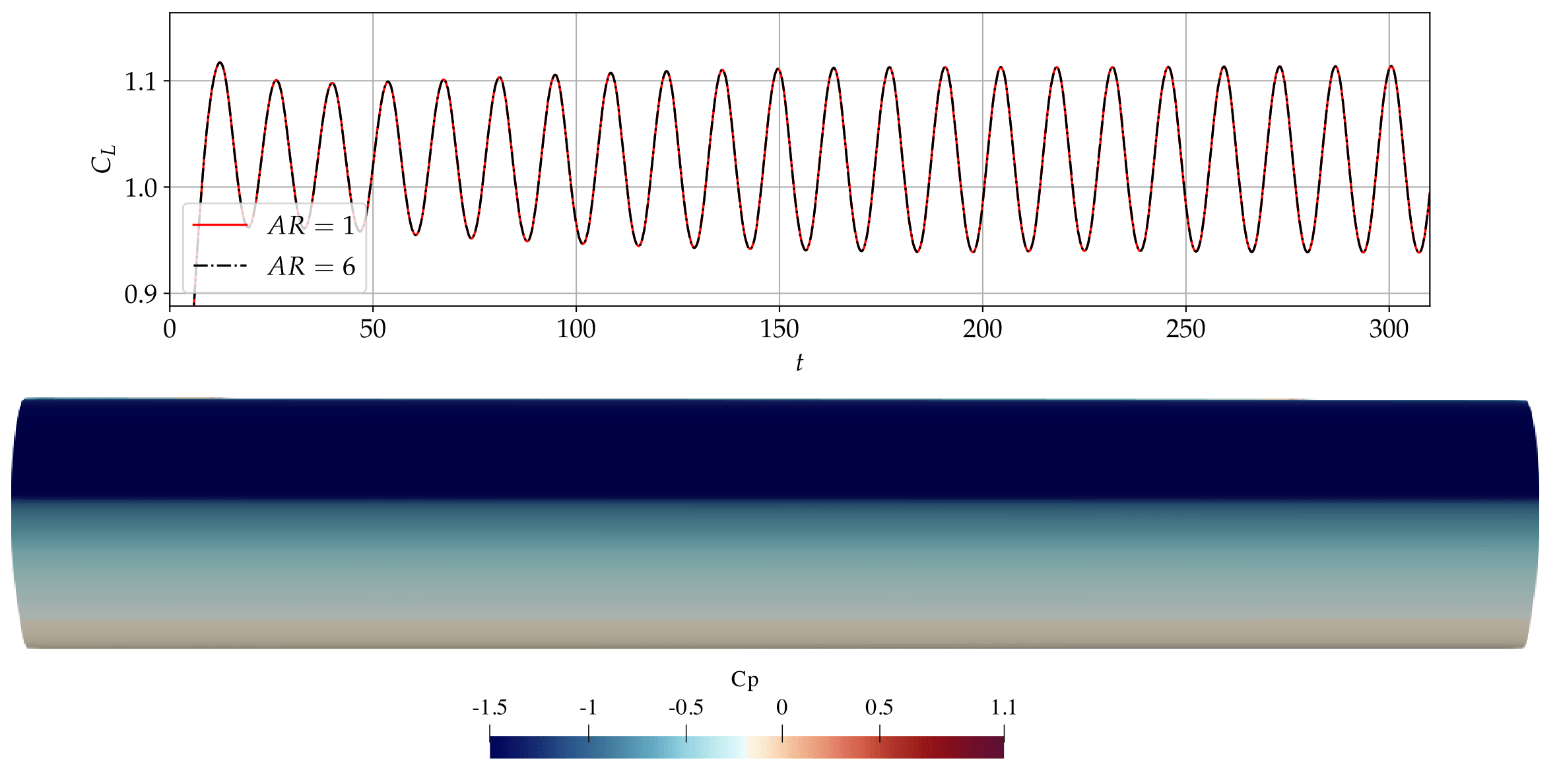}
\caption{Low-fidelity URANS solutions of the $\AoA{5}$ baseline configuration, showing (top) lift coefficient history at $\myAR{1}$ and $\myAR{6}$. The bottom panel shows an instantaneous plot of pressure coefficient contours on the suction side of the $\myAR{6}$ aerofoil, demonstrating the essentially two-dimensional behaviour of this 3D URANS buffet simulation.}
\label{fig:URANS_AR6}
\end{center}
\end{figure}

To further check the two-dimensionality of the $\AoA{5}$ solution, the $\myAR{1}$ URANS (Figure \ref{fig:URANS_validation}) is extended to $\myAR{6}$ to assess whether the $\myAR{2}$ ILES domain is still too narrow. The much cheaper computational cost of URANS compared to ILES allows us to simulate much wider aspect ratios and integrate for far more periods of the buffet cycle. Figure~\ref{fig:URANS_AR6} shows a comparison of lift coefficient for $\myAR{1}$ and $\myAR{6}$, with an instantaneous top-down pressure coefficient snapshot of the flow at $\myAR{6}$. Despite observed chord-wise low-frequency buffet shock oscillations, the surface plot shows no three-dimensionality and the solution is still essentially two-dimensional even at $\myAR{6}$. Numerous individual URANS snapshots were observed at each point in the buffet cycle, but no buffet/stall-cells were identified. Similarly, the lift coefficient shows perfect overlap between $\myAR{1}$ and $\myAR{6}$, suggesting that there are no span-dependent three-dimensional effects occurring.

\section{Sensitivity to aspect ratio with strong 3D effects}\label{sec:appendix:deg7_AR1}

\begin{figure}
\begin{center}
\includegraphics[width=1\textwidth]{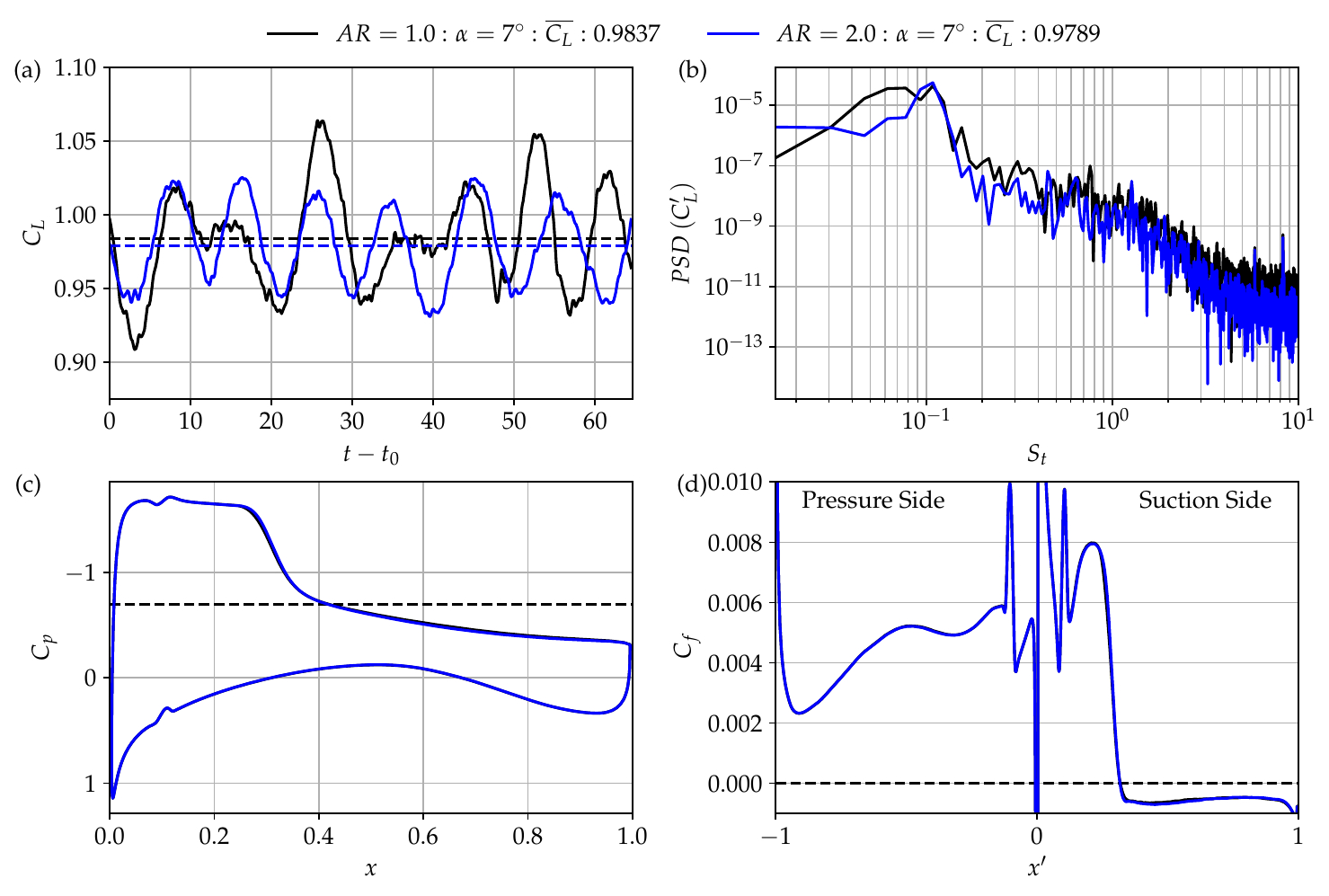}
\caption{ILES cases of wide-span buffet at $\myAR{1}$ and $\myAR{2}$ for an angle of attack of $\AoA{7}$. Showing (a) lift coefficient, (b) PSD of lift fluctuations, (c) time- and span-averaged pressure coefficient, and (d) time- and span-averaged skin-friction distributions.}
\label{fig:appendix_AR1_AR2_deg7}
\end{center}
\end{figure}

Figure~\ref{fig:appendix_AR1_AR2_deg7} shows the aerodynamic coefficients for the $\AoA{7}$ cases with aspect ratios of $\myAR{1}$ and $\myAR{2}$. While the mean values of lift in Figure~\ref{fig:appendix_AR1_AR2_deg7} (a) are similar between the two aspect ratios, the unsteady histories do not collapse in the same manner as at the lower AoA that was essentially two-dimensional (Figure~\ref{fig:AR1_AR2_deg5} (a)). This is due to the span-wise domain width now setting the wavelength of the three-dimensional instability, and not simply being a widened version of an essentially two-dimensional simulation with quasi-2D shock oscillations. Nevertheless, despite the irregular lift histories, the span-averaged distributions of pressure and skin-friction in Figure~\ref{fig:appendix_AR1_AR2_deg7} (c,d) agree very well between $\myAR{1}$ and $\myAR{2}$, with only minor deviations observed on the suction side of the aerofoil. Elsewhere on the aerofoil the profiles collapse exactly. This highlights that although there are three-dimensional effects which have span-wise dependence, the length of the averaging period used ($t = 80$, Figure~\ref{fig:appendix_AR1_AR2_deg7} (a)) is sufficiently long to obtain collapsed profiles in a span-averaged sense. In Section~\ref{sec:sectional} the same quantities are computed on individual span-wise grid locations to show that the three-dimensional buffet effects observed in Section~\ref{sec:deg6_deg7_AR2} do persist and the collapsed profiles in Figure~\ref{fig:appendix_AR1_AR2_deg7} (a) are only due to span-averaging and a sufficiently long time signal.

From the PSD of lift fluctuations in Figure~\ref{fig:appendix_AR1_AR2_deg7} (b), the main buffet shock-oscillation frequency of $St=0.1086$ is found at both aspect ratios. However, there is deviation between the two spectra and additional content at lower frequencies below the shock-oscillation mode. This is consistent with the modal decomposition SPOD findings in Section~\ref{sec:SPOD}, which identified that at zero sweep, the three-dimensionality associated with buffet/stall-cells appears at lower frequencies than the shock-oscillation mode. This also agrees with the findings of \citep{PDL2020,PDBLS2021}, who showed that while at higher sweep angles the three-dimensionality occurs at frequencies higher than the shock oscillation mode, as the sweep angle is decreased the 3D mode tends towards lower frequencies. In this paper a zero sweep angle is applied and the three-dimensionality is expected to occur at the lowest frequencies, below the shock-oscillation mode.

\section{Cross correlations at $\myAR{3}$ and $\AoA{6}$}\label{sec:AR3_correlations}

\begin{figure}
\begin{center}
\includegraphics[width=0.8\textwidth]{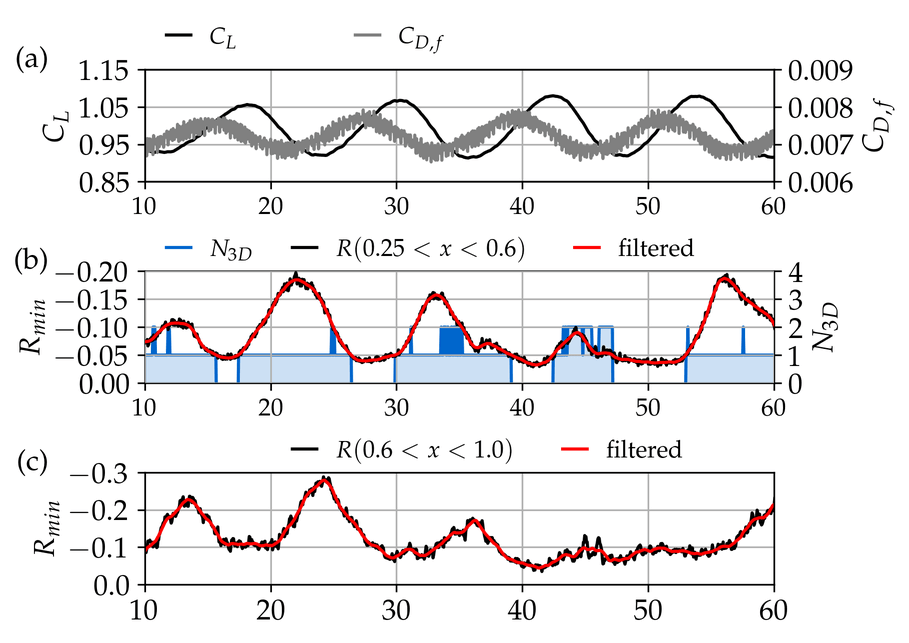}
  \caption{For $\alpha=6^{\circ}$ and \myAR{3}, showing (a) Lift coefficient $C_L$ (black curve - left-hand-side scale) and skin-friction drag coefficient $C_{D,f}$ (grey curve - right-hand-side scale) as functions of time. The number of 3D structures detected in the spanwise direction $N_{3D}$ (left-hand-side scale), estimated based on the $\Delta z$ associated with local minima in cross-correlations $R_{min}$ (right-hand-side scale) as functions of time for (b) $0.25<x<0.6$ and (c) $0.6<x<1.0$, respectively.}
\label{fig:R_time_AoA6_AR3}
\end{center}
\end{figure}

The cross-correlation analysis at $\AoA{6}$ and $\myAR{2}$ in Section~\ref{sec:cross_correlations} (Figure~\ref{fig:R_time_AoA6_AR2}) is extended here to the wider aspect ratio of $\myAR{3}$. Figure~\ref{fig:R_time_AoA6_AR3} again shows the (a) lift coefficient and skin-friction drag histories in time, plus the evolution of the correlation measure in the (b) shocked-region $(0.25 < x < 0.6)$ and (c) rear of the aerofoil $(0.6 < x < 1)$. Strong anti-correlation is observed during certain phases of the buffet cycle, slightly before the instance where the global lift approaches its minimum. The strong anti-correlations are repeatable in their appearance between consecutive buffet cycles, however, the amplitude of the peak anti-correlation varies from cycle-to-cycle. The magnitude of the anti-correlations is stronger than those observed at $\myAR{2}$ (Figure~\ref{fig:R_time_AoA6_AR2}), suggesting more pronounced large-scale 3D flow structures at this wider aspect ratio for a fixed angle of attack. This is consistent with the instantaneous flow-field visualisations shown in Figure~\ref{fig:deg6_AR3_contours}. At $\myAR{3}$, the correlation threshold detects both single and double separation cell patterns, associated with an undulation of the shock front in this region ($0.25 < x < 0.6)$. Multiple structures are also observed in the region downstream of the shock-wave. Comparing the red correlation curves in Figures~\ref{fig:R_time_AoA6_AR3} (b,c) associated with the upstream ($0.25<x<0.6$) and downstream regions ($0.6<x<1.0$) of the aerofoil, we can now clearly see a time lag between anti-correlation peaks of $\Delta t \approx 2-3$ for \myAR{3}, as the buffet-cell structures are convected downstream. 

% \clearpage
\bibliographystyle{jfm}
%\bibliography{jfm2esam}

% \begin{thebibliography}{99}

% \expandafter\ifx\csname natexlab\endcsname\relax
% \def\natexlab#1{#1}\fi
% \expandafter\ifx\csname selectlanguage\endcsname\relax
% \def\selectlanguage#1{\relax}\fi

\bibliography{jfm.bib}

% \end{thebibliography}

%% End of file `jfm2esam.bib'.

\end{document}